\documentclass[%
 reprint,
 amsmath,amssymb,
 aps,
]{revtex4-1}

\usepackage{graphicx}
\usepackage{dcolumn}
\usepackage{bm}
\usepackage{latexsym}
\usepackage{siunitx}
\usepackage{float}
\usepackage{color}

\begin{document}

\preprint{APS/123-QED}

\title{Effects of Oxidation of Top and Bottom Interface  on the Electric, Magnetic, and Spin-Orbit Torque Properties of Pt/Co/AlO$_{\textrm{x}}$ Trilayers}

\author{Junxiao Feng}
\author{Eva Grimaldi, Can Onur Avci, \\Manuel Baumgartner, Giovanni Cossu, Antonella Rossi}
\author{Pietro Gambardella}
\affiliation{%
	Department of Materials, ETH Zurich, 8093 Zurich, Switzerland
}%

\date{\today}

\begin{abstract}
Oxidation strongly influences the properties of magnetic layers employed in spintronic devices. We studied the effect of oxidation on the structural, magnetic, and electrical properties as well as current-induced spin-orbit torques (SOT) in Pt/Co/AlO$_{\textrm{x}}$, Pt/CoO$_{\textrm{x}}$/Co/AlO$_{\textrm{x}}$, and PtO$_{\textrm{x}}$/Co/AlO$_{\textrm{x}}$ layers. We show how the saturation magnetization, perpendicular magnetic anisotropy, anomalous Hall resistance, and SOT are systematically affected by the degree of oxidation of both the Pt/Co and Co/Al interfaces. Oxidation of the Co/Al interface results in a 21\% and 42\% variation of the damping-like and field-like SOT efficiencies, which peak at 0.14 and 0.07, respectively. The insertion of a paramagnetic CoO$_{\textrm{x}}$ layer between Pt and Co maintains a very strong perpendicular magnetic anisotropy and improves the damping-like and field-like SOT efficiencies, up to 0.26 and 0.20, respectively.
In contrast with recent reports, we do not find that the oxidation of Pt leads to a significant enhancement of the torques. Rather, we find that oxygen migrates from Pt to the Co and Al layers, leading to a time-dependent oxidation profile and an effective spin Hall conductivity that decreases with increasing oxygen concentration.
Finally, we studied current-induced switching in Pt/Co/AlO$_{\textrm{x}}$ with different degrees of oxidation and found a linear relationship between the critical switching current and the effective magnetic anisotropy controlled by the oxidation of Al. These results highlight the importance of interfaces and oxidation effects on the SOT and magnetotransport properties of heavy metal/ferromagnet/oxide trilayers, and provide information on how to improve the SOT efficiency and magnetization-switching characteristics of these systems.
\end{abstract}

\maketitle

\section{\label{INTRO}Introduction}
Heavy metal/ferromagnet/oxide (HM/FM/Ox) trilayers are a fundamental component of spintronic devices~\cite{Mtjcofebperpen2010, Brataasnm2012, Cubukcuapl2014,MRAM-IE-2016} due to their tunable perpendicular magnetic anisotropy (PMA)~\cite{Monso-2002}, relatively high Curie temperature~\cite{Sokalski2015}, fast dynamics~\cite{Manuel-nn}, and compatibility with back-end-of-line CMOS processes~\cite{Jan2014,Garello-ie2018}. In recent years, current-induced spin-orbit torques (SOT) have emerged as a powerful tool to manipulate the magnetization of these systems~\cite{Miron-rashba-2010,Miron-switching2011,Avci2012, Liu-science2012, Garello-nn-2013, Hayashi-nm2013, Cubukcu2018, Manchon-rmp-2019}. SOT rely on the spin Hall effect of the HM as well as on Rashba-type interfacial effects in order to convert an in-plane charge current into a non-equilibrium spin accumulation, which diffuses into the FM and transfers angular momentum to the local magnetization via the $s$-$d$ exchange interaction~\cite{Manchon-rmp-2019}. Symmetry considerations show that the current-induced spin accumulation allows for two types of torques, namely the damping-like and field-like SOT~\cite{Zhang2002a, Garello-nn-2013}.

\begin{figure*}[t]
\centering
\includegraphics[width=160mm]{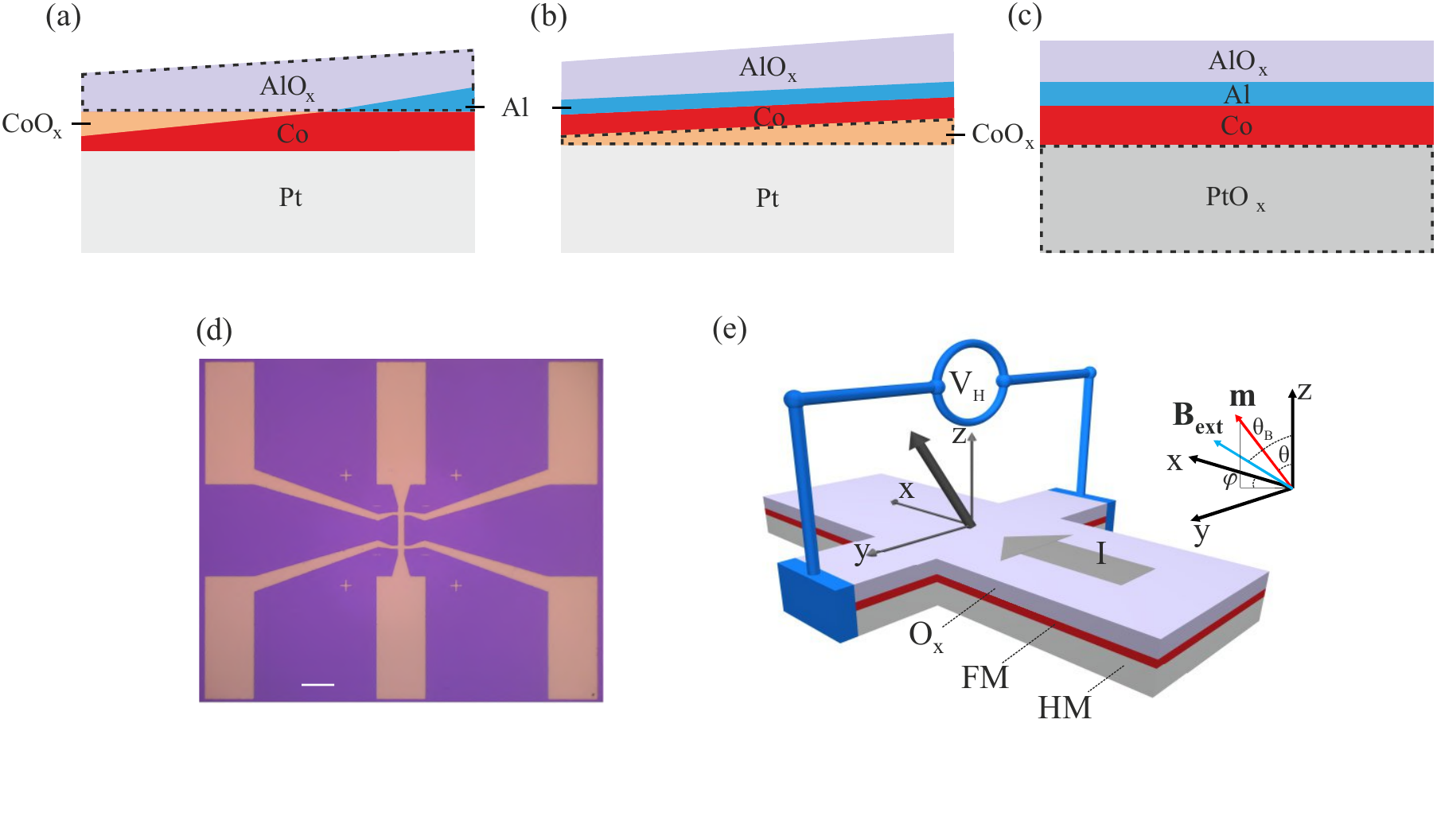}
\caption{\label{Config}Schematic cross-section of the samples employed in this study: (a) Pt/Co/AlO$_{\textrm{x}}$, (b) Pt/CoO$_{\textrm{x}}$/Co/AlO$_{\textrm{x}}$ and (c) PtO$_{\textrm{x}}$/Co/AlO$_{\textrm{x}}$. (d) Micrograph of a patterned Hall bar after photolithography. The scale bar is 50~$\mu$m. (e) Schematic diagram of the experimental geometry and coordinate system. A positive current is injected along the $\textbf{x}$ direction, $\textbf{m}$ is the unitary magnetization and $\textbf{B}_{\textrm{ext}}$ is the external field.}
\end{figure*}

Several studies have shown that modifying the interface oxidation in HM/FM/Ox trilayers leads to substantial changes of the magnetic properties. Early reports focused on the effect of interface oxidation on the magnetic anisotropy~\cite{Monso-2002,Manchon-jap-2008,Rodmacq2009,Chiba2011,Garad2013}. Following the discovery of SOT, the focus has shifted towards using oxidation as a tool to modify the SOT and current-induced switching characteristics by either engineering the oxidation of the top layer during growth~\cite{Miron-switching2011,ZFS-wedge-nn-2014,Yu2014b,Qiu2015,Pai2015,Akyol2015,Hibino2017,Gweon2019} or post-growth dynamic modification of the oxidation profile by applying a gate voltage~\cite{Bauer2014,Emori2014,Bi2014}.

Very recently, the oxidation of the HM layer has been found to dramatically improve the SOT efficiency of trilayer structures such as W/CoFeB/TaN~\cite{Demasius2016}, Pt/Ni$_{81}$Fe$_{19}$/SiO$_2$~\cite{An2018}, and Pt/CoTb/MgO~\cite{An2018a}. In PtO$_{\textrm{x}}$/Ni$_{81}$Fe$_{19}$/SiO$_2$, the SOT efficiency reaches a maximum of 0.9 for fully oxidized Pt, which implies that most of the charge current is converted into a spin current, even though no current actually flows through Pt~\cite{An2018}. All of these studies focused on the oxidation of
one interface only, leaving open questions in regard to the effects of oxidation on either interface of the same system or on the FM itself.
Further, the strong increase of the SOT efficiency reported for oxidized Pt is very promising as a means to achieve switching at low current density, but needs to be confirmed in other systems, particulary in trilayers exhibiting PMA.

In this article, we report a systematic study of the effects of oxidation on the structural, electrical, and magnetic properties of Pt/Co/AlO$_{\textrm{x}}$ trilayers, as well as of the current-induced SOT and magnetization switching characteristics. We find that oxidation of the Co/Al and Pt/Co interfaces, and of the Co layer itself, influences the above properties in distinct ways. More specifically, we find that oxidation of either the Co/Al or Pt/Co interface has a sizeable effect on the SOT. The strongest effects is observed upon oxidation of the Pt/Co interface, pinpointing the critical role of this interface in the spin currents giving rise to the SOT. The insertion of a thin CoO$_{\textrm{x}}$ layer next to Pt leads to a 100\% increase of the SOT efficiency for both damping-like and field-like torques, whereas structures with oxidized Pt present significantly lower efficiencies compared to Pt/Co/AlO$_{\textrm{x}}$ and recent studies~\cite{An2018, An2018a}. Additionally, we find that, while a single layer of PtO$_{\textrm{x}}$ remains stable after oxidation, PtO$_{\textrm{x}}$ is strongly reduced upon deposition of the Co/Al overlayers due to oxygen diffusion towards Co and Al. Finally, we show that the oxidation of Al can be used to tune the critical current for switching of the Co layer, which is directly proportional to the strength of the perpendicular magnetic anisotropy controlled by oxidation. These results highlight the importance of tuning and preserving the oxidation profiles in HM/FM/Ox trilayers in order to control the SOT efficiency and magnetic anisotropy of these systems.

\begin{figure*}[t]
	\includegraphics[width=160mm]{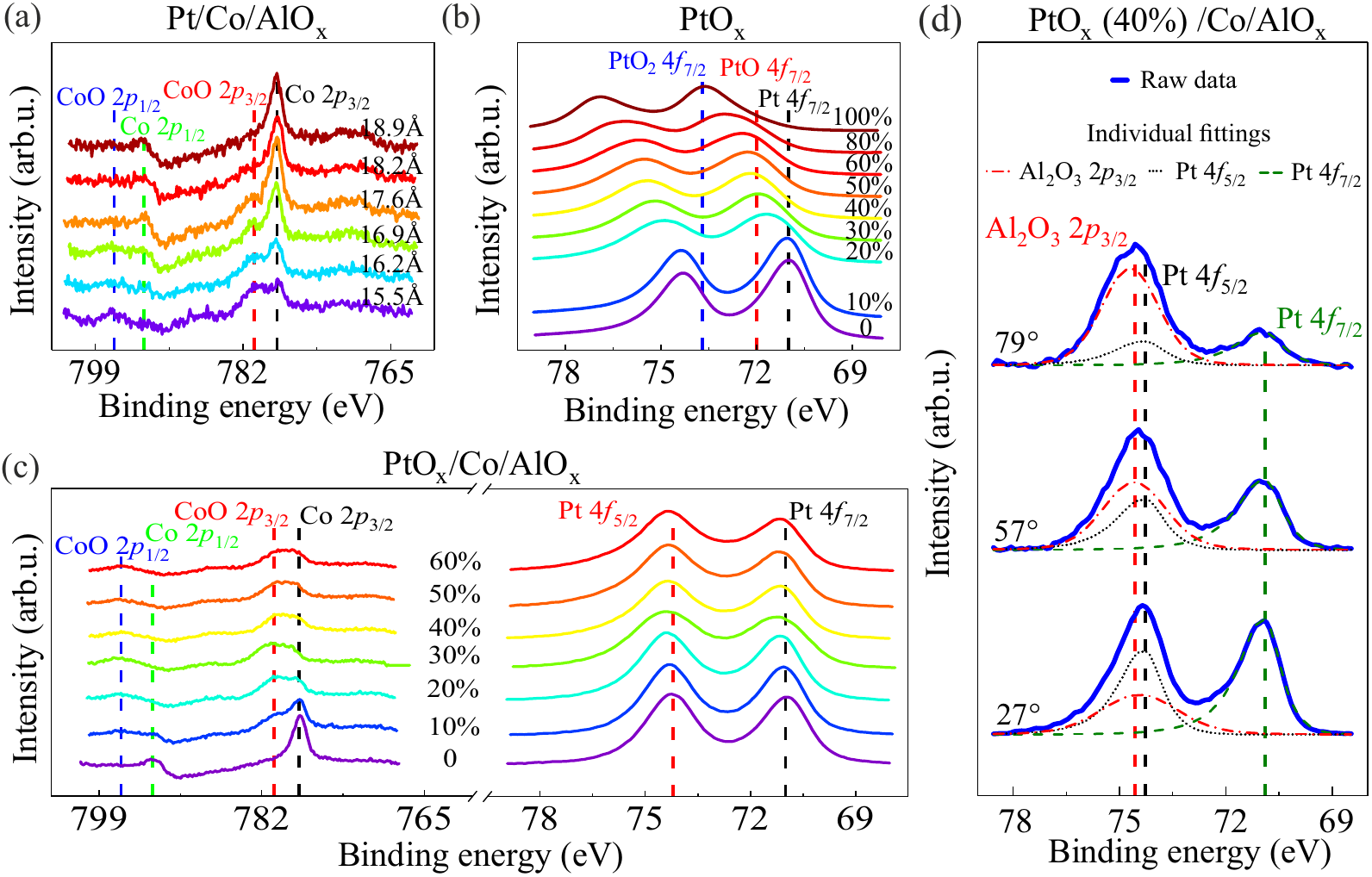}
	\caption{\label{XPS}XPS spectra of (a) Co in Pt/Co/AlO$_{\textrm{x}}$($t_{\rm{Al}}$), (b) Pt in PtO$_{\textrm{x}}$ single layers with different O$_2$:(Ar+O$_2$) ratio, and (c) Co and Pt in PtO$_{\textrm{x}}$/Co/AlO$_{\textrm{x}}$ with different O$_2$:(Ar+O$_2$) ratio. (d) Angle-resolved XPS of the Al 2\textit{p} and Pt 4\textit{f} peaks in PtO$_{\textrm{x}}$(40\%)/Co/AlO$_{\textrm{x}}$ recorded for three incident angles with respect to the sample's surface normal. The dashed and dotted lines show the intensity of the individual peaks derived from a fit of the experimental spectrum (solid line). The dotted (green) and dashed (red) lines show the model Gaussian/Lorentzian functions used to resolve the contributions of the overlapping photoelectron signals.}
\end{figure*}

\section{\label{EXP} Sample Characterization}
\subsection{Sample fabrication}\label{FAB}
We grew three series of samples on Si$_3$N$_4$ wafers using dc magnetron sputtering with base pressure \textless  8.0 $\times 10 ^{-8}$ Torr and Ar pressure between 0.3 and 1.1~mTorr. The first series, which we refer to as Pt/Co/AlO$_{\textrm{x}}$, was obtained by depositing an Al wedge with thickness 12.8~-~19.5~\AA ~on a Pt(5)/Co(1) bilayer (the numbers between parentheses represent the thickness in nm). After deposition of the full metallic stack, an oxygen plasma with 10.0~mTorr pressure and 53~W power was used to oxidize the Al wedge with the purpose of obtaining over- and under-oxidized regions of the top Co interface [Fig. 1(a)]. We note that the easy magnetization axis is perpendicular to the sample plane for an Al thickness $t_{\rm{Al}} = 13.4 - 18.9$~\AA, whereas beyond this range the easy axis is in-plane, in agreement with previous reports~\cite{Monso-2002, Manchon-jap-2008, Rodmacq2009}.
The second series is Pt(5)/CoO$_{\textrm{x}}$($t_{\rm{CoO_{\textrm{x}}}}$)/Co(0.65)/AlO$_{\textrm{x}}$(1.85), which was grown by depositing a CoO$_{\textrm{x}}$ wedge layer with thickness $t_{\rm{CoO_{\textrm{x}}}}= 2.1 - 4.9$~\AA ~on Pt by reactive sputtering of Co with an O$_2$:(Ar+O$_2$) ratio of 50\% [Fig. 1(b)]. In this series, $t_{\rm{Al}}$ was fixed at 1.85~nm in order to ensure PMA for all the samples. The third series is PtO$_{\textrm{x}}$(5)/Co(1)/AlO$_{\textrm{x}}$(2.5) [Fig. 1(c)]. Unlike the first two series, which have a wedge structure, this series includes seven different samples grown with an O$_2$:(Ar+O$_2$) ratio ranging between 0 and 60\% during deposition of the Pt layer. All the samples of the third series have in-plane magnetic anisotropy (IMA) resulting from the under-oxidation of the relatively thick Al cap layer. This choice was made in order to have the same type of magnetic anisotropy in the series, since we found that the oxidation of Pt reduces the PMA of the PtO$_{\textrm{x}}$/Co interface relative to Pt/Co. As reference for the XPS and electrical measurements, we also deposited single PtO$_{\textrm{x}}$(5) layers on Si$_3$N$_4$ with the same O$_2$:(Ar~+~O$_2$) ratios as used for the full PtO$_{\textrm{x}}$/Co/AlO$_{\textrm{x}}$ trilayers. These reference samples are not capped in order to avoid oxygen migration effects. Contact with air is unlikely to influence the oxygen distribution in single PtO$_{\textrm{x}}$ layers deposited by reactive sputtering, consistently with the observation that the resistance remains constant after deposition.

The Pt layers are polycrystalline with [111] preferred orientation, as confirmed by transmission Kikuchi diffraction.
After deposition, all the samples were diced in two pieces, one for the characterization of the structure and magnetization properties and the other for device patterning and SOT measurements. The devices are 10 $\mu m$ wide and 50 $\mu m$ long Hall bars defined by optical lithography followed by Ar milling [Fig. 1(d)]. The coordinate system employed throughout this study is depicted in Fig. 1(e).

\subsection{XPS}\label{XPS}
We performed x-ray photoemission spectroscopy (XPS) in order to characterize the oxidation state of the first and third series described above. Figure 2(a) shows the XPS spectra of Pt/Co/AlO$_{\textrm{x}}$ recorded by measuring photoemission from the Co 2\textit{p}$_{1/2}$ and 2\textit{p}$_{3/2}$ core levels of samples with different $t_{\rm{Al}}$. The spectra of samples with $t_{\rm{Al}}> 17.6$~\AA ~present two peaks at 778.1 and 793.4~eV, which are characteristic of metallic Co. On the other hand, the spectra of samples with $t_{\rm{Al}}< 17.6$~\AA ~ show high energy shoulders at 780.7 and 796.7~eV characteristic of CoO~\citep{Manchon-jap-2008}, the intensity of which increases with decreasing Al thickness. These measurements show that the Co/Al interface is underoxidized and that Co remains fully metallic down to $t_{\rm{Al}}=17.6$~{\AA}, after which CoO forms. We designate the sample with $t_{\rm{Al}}=17.6$~{\AA} as the ``optimized'' sample for which the Co/Al interface is fully oxidized, but Co is not, leading to maximum PMA.

In order to characterize the third series, we first measured as a reference the XPS spectra of single PtO$_{\textrm{x}}$(5) layers.
The spectra, reported in Fig. 2(b), show the evolution of the Pt 4\textit{f}$_{7/2, 5/2}$ doublet with increasing O$_2$ partial pressure. As the peak of the Pt 4\textit{f}$_{5/2}$ level overlaps with the Al 2\textit{p} level around 74.6~eV, typical of Al$_2$O$_3$~\cite{Dua1988163}, we focus on the behavior of the 4\textit{f}$_{7/2}$ peak. This peak shifts from 71~eV below 10\% O$_2$, as typical of Pt, to 72~eV around 40\% O$_2$, characteristic of PtO, and finally to 73.7~eV at 100\% O$_2$, indicating the formation of PtO$_2$~\citep{An2018}. These spectra confirm that a single layer of Pt can be stably oxidized by controlling the amount of O$_2$ during sputtering. However, the measurements of the full PtO$_{\textrm{x}}$/Co/AlO$_{\textrm{x}}$ stack that were performed 6 days after deposition, shown in Fig. 2(c), reveal quite a different picture. In these samples, the Pt 4\textit{f}$_{7/2}$ peak does not shift with increasing O$_2$ partial pressure, showing that Pt is not oxidized. On the other hand, a strong shift of the Co 2$p$ peaks is observed, indicating the formation of CoO upon diffusion of oxygen into Co from the Pt side.

The progressive oxidation of the top layers resulting from oxygen diffusion was further confirmed by angle-resolved XPS, as shown in Fig. 2(d) for PtO$_{\textrm{x}}$(40\%)/Co/AlO$_{\textrm{x}}$. The spectra recorded with an increasing emission angle relative to the surface normal, from 27$^\circ$ to 79$^\circ$, have a higher degree of surface sensitivity. These spectra reveal a progressive increase of the Al$_2$O$_3$ intensity relative to Pt, showing that Al is strongly oxidized, whereas the position of the Pt peak does not change as a function of incidence angle. Note that $t_{\rm{Al}}=25$~{\AA} in this series, such that Al was supposed to be underoxidized. Overall, these observations indicate that PtO$_{\textrm{x}}$ layers are not stable when embedded in a multilayer structure, as oxygen tends to diffuse away from Pt and oxidize both the Co and Al layers. Such a behavior can be rationalized in terms of the enthalpy of formation of PtO$_{\textrm{x}}$ (-101.3 kJ/mol)~\cite{Nagano2002}, which is significantly lower than that of AlO$_{\textrm{x}}$ (-1675.7 kJ/mol)~\cite{enthalpy} and CoO$_{\textrm{x}}$ (-237.4 kJ/mol)~\cite{Holmes1986}.

\subsection{Electrical resistance}\label{ELECTRICAL}
Figure 3 summarizes the electrical resistance of devices belonging to the three series and of PtO$_{\textrm{x}}$ single layers. The typical device resistance of Pt/Co/AlO$_{\textrm{x}}$ is 0.33-0.34~k$\Omega$, which corresponds to a resistivity of 40-41~$\mu\Omega$cm by assuming that only Pt and Co are conductive. This assumption is corroborated by the weak dependence of the resistance on $t_{\rm{Al}}$ [Fig. 3(a)], which shows that the oxidation state of the top Co/Al interface does not influence significantly the conductivity of Pt/Co/AlO$_{\textrm{x}}$.

Devices belonging to the Pt/CoO$_{\textrm{x}}$/Co/AlO$_{\textrm{x}}$ series have a resistance of 0.35-0.36~k$\Omega$, which corresponds to a resistivity of 40-41~$\mu\Omega$cm. Also in this case we assume that only Pt and Co are conductive. Note that the resistance is approximately constant as a function of $t_{\rm{CoO}_{\textrm{x}}}$, suggesting that CoO$_{\textrm{x}}$ is weakly conducting or effectively an insulator [Fig. 3(b)].
From these two sets of data we conclude that the oxidation of the Co/Al and Pt/Co interfaces does not have a major impact on the resistivity, indicating that conduction is dominated by the 5~nm-thick Pt channel.

\begin{figure}[t]
	\includegraphics[width=85mm]{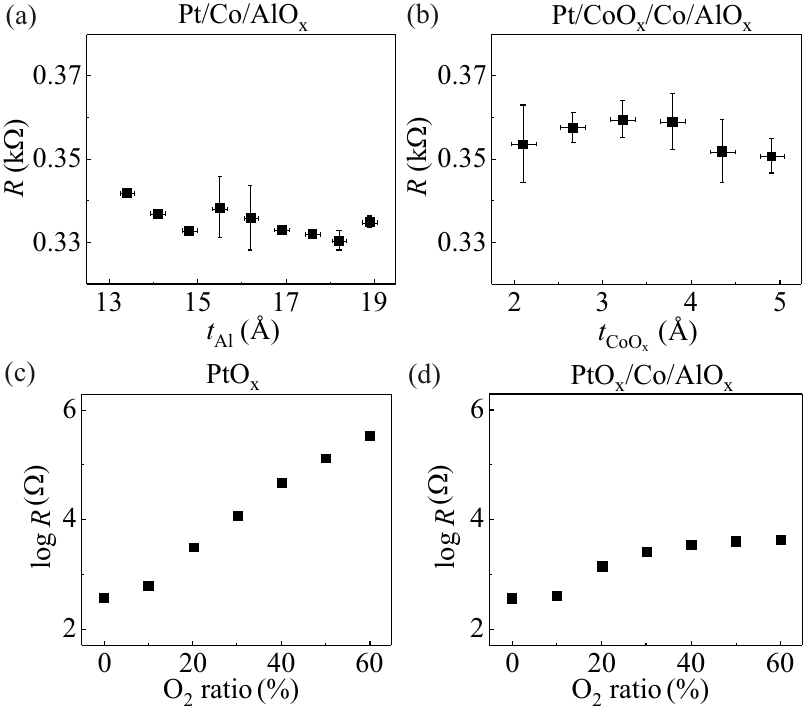}
	\caption{\label{R} 4-point resistance as a function of (a) $t_{\rm{Al}}$ in Pt/Co/AlO$_{\textrm{x}}$, (b) $t_{\rm{CoO_{\textrm{x}}}}$ in Pt/CoO$_{\textrm{x}}$/Co/AlO$_{\textrm{x}}$, and O$_2$:(Ar+O$_2$) ratio in (c) PtO$_{\textrm{x}}$ single layers and (d) PtO$_{\textrm{x}}$/Co/AlO$_{\textrm{x}}$. Vertical error bars represent the standard deviation of the average of different devices from the same series. Horizontal error bars represent the uncertainty of the thickness calibration.}
\end{figure}

The resistances of the reference PtO$_{\textrm{x}}$ layers are shown in Fig. 3(c) as a function of O$_2$:(Ar+O$_2$) ratio. The resistance of PtO$_{\textrm{x}}$ increases dramatically with increasing O$_2$ ratio (note the log scale of the plot), indicating that PtO$_{\textrm{x}}$ becomes nearly insulating for 50-60\% of O$_2$. These results are consistent with the conclusions drawn from the XPS spectra presented in Fig. 2(b).
In contrast with the reference PtO$_{\textrm{x}}$ layers, the resistance of PtO$_{\textrm{x}}$/Co/AlO$_{\textrm{x}}$ saturates around 40\% of O$_2$ rather than increasing monotonically with O$_2$ ratio, to a value that is about two orders of magnitude lower compared to PtO$_{\textrm{x}}$ single layers. These findings corroborate the oxygen diffusion scenario outlined in Sec.~\ref{XPS}.

\section{\label{MAG}Magnetic properties}
\subsection{\label{MS1}Saturation magnetization}
\begin{figure*}[t]
	\includegraphics[width=160mm]{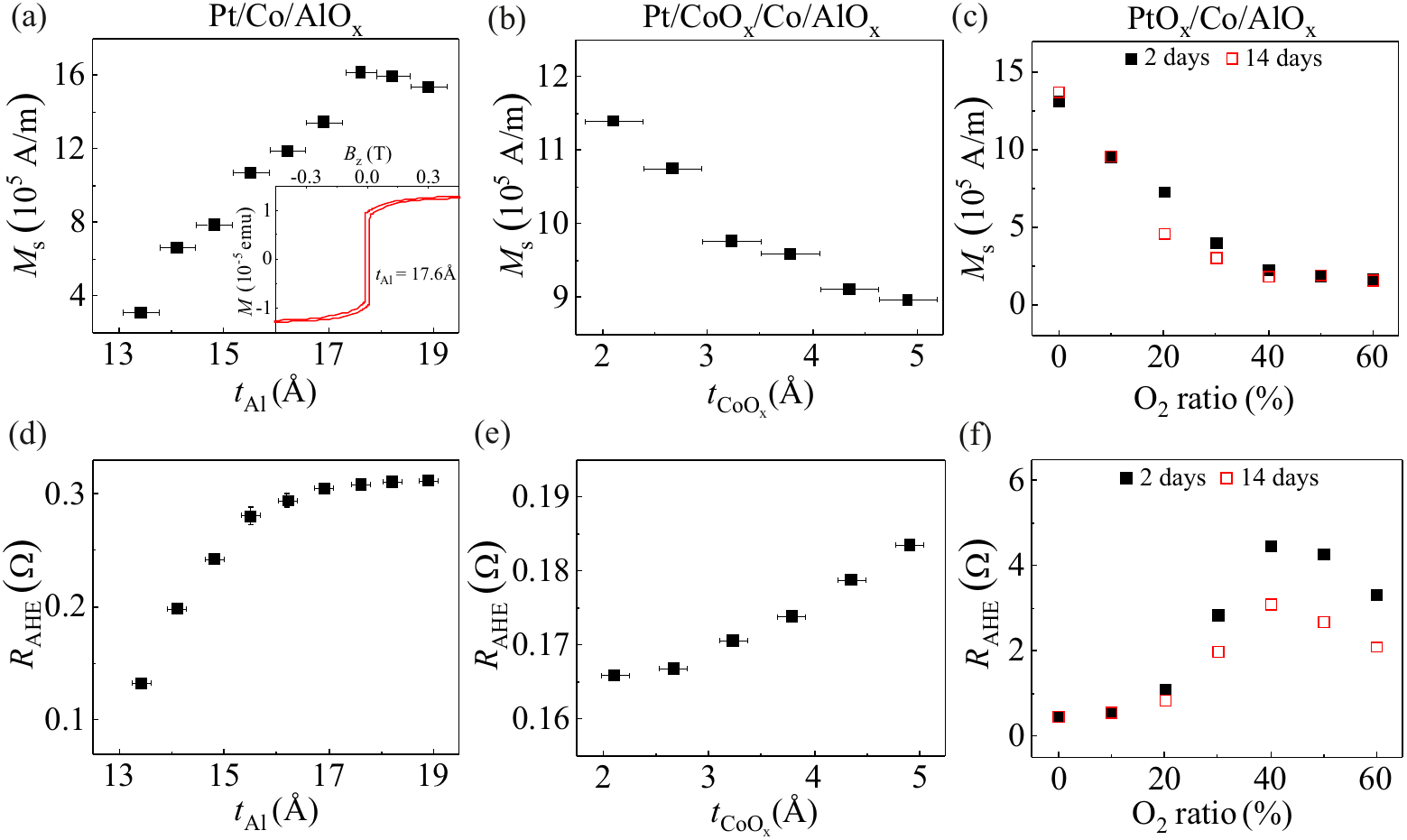}
	\caption{\label{MS} Saturation magnetization as a function of (a) $t_{\rm{Al}}$ in Pt/Co/AlO$_{\textrm{x}}$, (b) $t_{\rm{CoO_{\textrm{x}}}}$ in Pt/CoO$_{\textrm{x}}$/Co/AlO$_{\textrm{x}}$, and (c) O$_2$:(Ar+O$_2$) ratio in PtO$_{\textrm{x}}$/Co/AlO$_{\textrm{x}}$. The inset in (a) shows the total magnetic moment of Pt/Co/AlO$_{\textrm{x}}$  with $t_{\rm{Al}}$ = 17.6~\AA ~ as a function of out-of-plane external field. (d-f) Same as (a-c) for the anomalous Hall resistance. The data shown as solid and open symbols in (c) and (f) were obtained, respectively, 2 and 14 days after the deposition of PtO$_{\textrm{x}}$/Co/AlO$_{\textrm{x}}$. Vertical error bars represent the standard deviation of the average of different devices from the same series. Horizontal error bars represent the uncertainty of the thickness calibration.}
\end{figure*}

We measured the total magnetic moment of each sample using superconducting quantum interference device (SQUID) magnetometry in an out-of-plane external field [see inset of Fig. \ref{MS}(a) for an example of a magnetization curve]. The saturation magnetization $M_s$ is obtained by dividing the saturation magnetic moment by the nominal volume of the Co layer.
Figures \ref{MS}(a-c) show how $M_s$ varies in the three series.

In Pt/Co/AlO$_{\textrm{x}}$, $M_{s}$ increases with increasing $t_{\rm{Al}}$ until it saturates around $t_{\rm{Al}}$ = 17.6~\AA ~[Fig. \ref{MS}(a)]. This behavior is consistent with the formation of CoO with antiferromagnetic correlations for $t_{\rm{Al}}$ $<$ 17.6~\AA ~ and the metallic character of Co evidenced by XPS for $t_{\rm{Al}}$ $\geq$ 17.6~\AA .

In Pt/CoO$_{\textrm{x}}$/Co/AlO$_{\textrm{x}}$, $M_s$ decreases monotonically as a function of $t_{\rm{CoO_{\textrm{x}}}}$ [Fig. \ref{MS}(b)], which we ascribe to several factors. First, some of the oxygen initially present in CoO$_{\textrm{x}}$ likely diffused into the metallic Co part, thus reducing $M_s$. Since the total amount of oxygen is proportional to $t_{\rm{CoO_{\textrm{x}}}}$ and $t_{\rm{Co}}$ is fixed, we expect that this effect becomes more pronounced with increasing $t_{\rm{CoO_{\textrm{x}}}}$. Second, the deposition of a CoO$_{\textrm{x}}$ layer between Pt and Co strongly reduces the proximity magnetization in Pt, thus reducing $M_s$. 
Based on measurements of $M_s$ in Pt/Co bilayers separated by ultrathin metal spacers, we estimate the latter effect to account for a loss of about 
$2\times 10^{5}$~A/m~\cite{Avci2019}.

In PtO$_{\textrm{x}}$/Co/AlO$_{\textrm{x}}$, $M_s$ decreases by about a factor 7 as the O$_2$ ratio increases from 0 to 40\% and beyond~[Fig. \ref{MS}(c)]. This trend is consistent with the significant diffusion of oxygen into the metallic Co layer evidenced by XPS as well as with the resistance measurements presented above. Measurements of PtO$_{\textrm{x}}$/Co/AlO$_{\textrm{x}}$ performed 2 and 14 days after growth [solid and open symbols in Fig. \ref{MS}(c)] further show that $M_s$ changes with time for the intermediate oxidation ratios (20-40\%), suggesting that oxygen diffusion is a relatively slow process.

\begin{figure*}[t]
	\includegraphics[width=170mm]{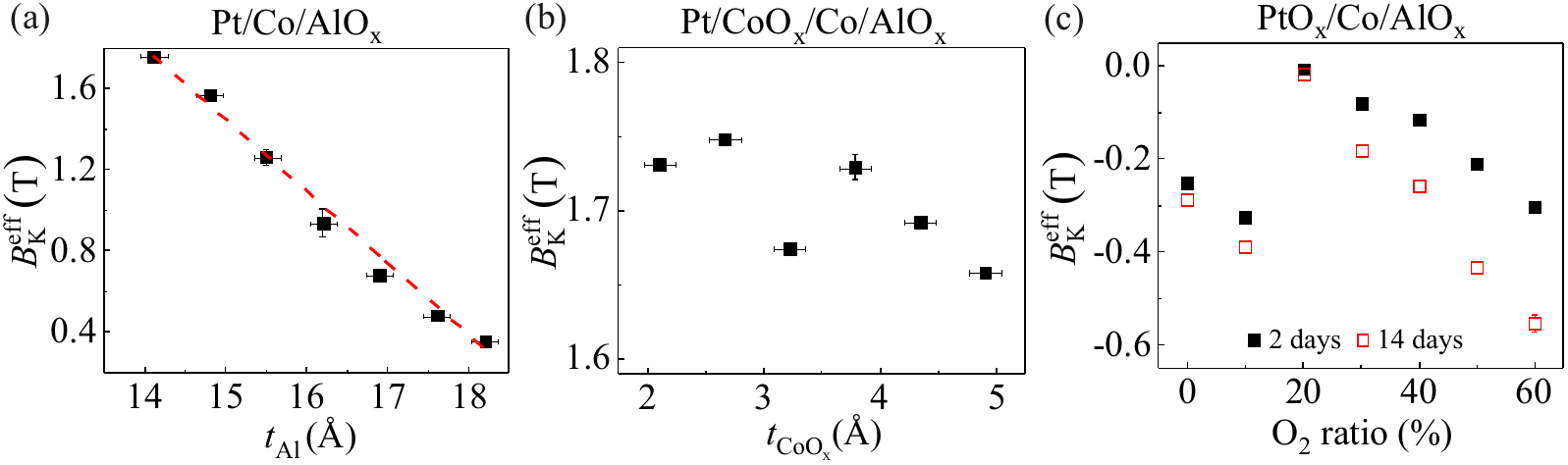}
	\caption{\label{BK} Effective magnetic anisotropy field as a function of (a) $t_{\rm{Al}}$ in Pt/Co/AlO$_{\textrm{x}}$, (b) $t_{\rm{CoO_{\textrm{x}}}}$ in Pt/CoO$_{\textrm{x}}$/Co/AlO$_{\textrm{x}}$, and (c) O$_2$:(Ar+O$_2$) ratio in PtO$_{\textrm{x}}$/Co/AlO$_{\textrm{x}}$. The dashed line in (a) is a linear fit to the data. The solid and open symbols in (c) refer to measurements carried out 2 and 14 days after the deposition, respectively. Vertical error bars represent the standard deviation of the average of different devices from the same series. Horizontal error bars represent the uncertainty of the thickness calibration.}
\end{figure*}

\subsection{\label{AHE} Anomalous Hall resistance}
The anomalous Hall effect (AHE) was measured by using the harmonic Hall voltage method described in Refs.~\onlinecite{Garello-nn-2013,Can-thermal}. Representative AHE measurements for the different samples are reported in the Appendix. Figures \ref{MS}(d-f) show the anomalous Hall resistance coefficient $R_{\textrm{AHE}}$ in the three sample series. In the Pt/Co/AlO$_{\textrm{x}}$ series, we find that $R_{\textrm{AHE}}$ increases and then saturates as a function of $t_{\rm{Al}}$, similar to $M_s$ [Fig. \ref{MS}(d)]. This behavior shows that $R_{\textrm{AHE}}$ depends on the volume fraction of metallic Co, which increases with increasing $t_{\rm{Al}}$. Additionally, we observe that $R_{\textrm{AHE}}$ approaches saturation faster than $M_s$, indicating that the AHE increases upon interfacial oxidation of Co, in agreement with previous work~\cite{Zhang2010}, which we attribute to enhanced electron scattering at the Co/AlO$_{\textrm{x}}$ interface.
In the Pt/CoO$_{\textrm{x}}$/Co/AlO$_{\textrm{x}}$ series, we find that $R_{\textrm{AHE}}$ increases monotonically with $t_{\rm{CoO_{\textrm{x}}}}$, opposite to the behavior of $M_s$ [Fig. \ref{MS}(e)]. Although the relative change of $R_{\textrm{AHE}}$ is small, this observation corroborates the conclusion that moderate oxidation of the Co interfaces, and possibly of the Co grain boundaries, enhances the AHE. These results are consistent with a previous study of the AHE in MgO/[Co/Pt]$_3$/MgO multilayers, which evidenced an increase of $R_{\textrm{AHE}}$ upon insertion of a CoO$_{\textrm{x}}$ layer between MgO and Co~\cite{Zhang2013}.

The strongest increase of $R_{\textrm{AHE}}$, by more than one order of magnitude, is found in PtO$_{\textrm{x}}$/Co/AlO$_{\textrm{x}}$. In this series, $R_{\textrm{AHE}}$ reaches a maximum of 4.5~$\Omega$ around 40\% O$_2$ ratio, compared to 0.3~$\Omega$ in Pt/Co/AlO$_{\textrm{x}}$, and decreases afterwards [Fig. \ref{MS}(f)]. This nonmonotonic behavior suggests that, in the lower oxidation range (0-40\%), the higher resistivity of Pt causes a larger current flow through Co, thereby increasing $R_{\textrm{AHE}}$. The turning point at 40\% and successive decrease of $R_{\textrm{AHE}}$ can then be attributed to the strong reduction of $M_s$ that occurs with the progressive oxidation of Co [Fig. \ref{MS}(c)]. In this case, the difference between samples measured 2 and 14 days after deposition is substantial, which shows that changes in the oxidation profile influence the Hall conductivity more strongly than $M_s$.

\subsection{\label{MAE}Magnetic anisotropy}
The magnetic anisotropy was characterized by measuring the effective anisotropy field $B_{\textrm{K}}^{\textrm{eff}}$. For samples with either uniaxial or easy plane anisotropy, this field is given by
\begin{equation}\label{eq:BKEFF}
B_{\textrm{K}}^{\textrm{eff}}= B_{\textrm{K}} + B_{\textrm{dem}} =\frac{2}{\mu_0M_s}K -\mu_0M_s,
\end{equation}
where $K$ is the anisotropy constant, including the effects of the magnetocrystalline and interface anisotropy, $\mu_0$ the permeability of vacuum, and the last term represents the demagnetizing field of a thin film. Following Eq.~\ref{eq:BKEFF}, samples with PMA have $B_{\textrm{K}}^{\textrm{eff}}> 0$, whereas samples with IMA have $B_{\textrm{K}}^{\textrm{eff}}< 0$.

Both Pt/Co/AlO$_{\textrm{x}}$ and Pt/CoO$_{\textrm{x}}$/Co/AlO$_{\textrm{x}}$ present PMA with a dominant uniaxial behavior.  $B_{\textrm{K}}^{\textrm{eff}}$ was measured by fitting the anomalous Hall resistance as a function of a magnetic field applied perpendicular to the easy axis, as detailed in the Appendix. Figure \ref{BK}(a) shows $B_{\textrm{K}}^{\textrm{eff}}$ as a function of $t_{\rm{Al}}$ in the Pt/Co/AlO$_{\textrm{x}}$ series. With increasing $t_{\rm{Al}}$, $B_{\textrm{K}}^{\textrm{eff}}$ decreases almost in a linear fashion from 1.7~T to 0.4~T; samples with $t_{\rm{Al}} > 18.9$~\AA~ have in-plane magnetization. This trend is attributed to the linear increase of $M_{s}$ in the over-oxidized range ($t_{\rm{Al}}$ = 13.4-16.9~\AA) [Fig.~\ref{MS}(a)]. In this range, the top and bottom interfaces do not change significantly so that $B_{\textrm{K}}^{\textrm{eff}}$ is mostly affected by the change of the demagnetizing field, which is directly proportional to $M_s$. In the optimally-oxidized and under-oxidized range ($t_{\rm{Al}}$ = 17.6-18.9~\AA), $M_{s}$ remains approximately constant, but the oxidation of the top Co interface (higher Co/Al interface anisotropy K$_i^{\textrm{Co/Al}}$) leads to an increase of the PMA~\cite{Yang2011}.

In the Pt/CoO$_{\textrm{x}}$/Co/AlO$_{\textrm{x}}$ series, $B_{\textrm{K}}^{\textrm{eff}}$ remains close to 1.7~T in the entire CoO$_{\textrm{x}}$ thickness range [Fig. \ref{BK}(b)], despite the moderate decrease of $M_s$ as a function of $t_{\rm{CoO_{\textrm{x}}}}$ [Fig. \ref{MS}(b)]. We attribute this behavior to the strong PMA induced by the CoO$_{\textrm{x}}$/Co interface, as suggested by previous studies~\cite{Chung2016,Pan2019}.

The samples belonging to the PtO$_{\textrm{x}}$/Co/AlO$_{\textrm{x}}$ series have IMA. $B_{\textrm{K}}^{\textrm{eff}}$ in this case was estimated by the magnetic field required to saturate the sample in the hard out-of-plane direction, as described in Ref.~40. Figure \ref{BK}(c) shows $B_{\textrm{K}}^{\textrm{eff}}$ as a function of the O$_2$ ratio measured 2 and 14 days after deposition. We identify two different regimes below and above 20\% of O$_2$. $B_{\textrm{K}}^{\textrm{eff}}$ decreases going from 0 to 10\% O$_2$, presumably due to the oxidation of Pt at the Pt/Co interface at this low O$_2$ concentration. However, at 20\% O$_2$ ratio $B_{\textrm{K}}^{\textrm{eff}}$ suddenly approaches zero and then decreases again for larger O$_2$ ratios. One plausible explanation of this peculiar behavior is that oxygen diffuses through Co up to the Al interface with increasing O$_2$ ratio. At 20\% O$_2$, the Co/Al interface is oxidized, leading to PMA and almost complete compensation of the demagnetizing field. Further increasing the O$_2$ ratio leads to a decrease of $B_{\textrm{K}}^{\textrm{eff}}$ due to oxygen diffusing throughout the Co layer, as shown by the more pronounced decrease of $B_{\textrm{K}}^{\textrm{eff}}$ after 14 days.

Overall, this set of data proves that the magnetic properties of Pt/Co/Al trilayers are extremely sensitive to the oxidation of both the top and bottom Co interfaces as well as to the oxygen concentration within the layers.

\begin{figure*}[t]
	\includegraphics[width=160mm]{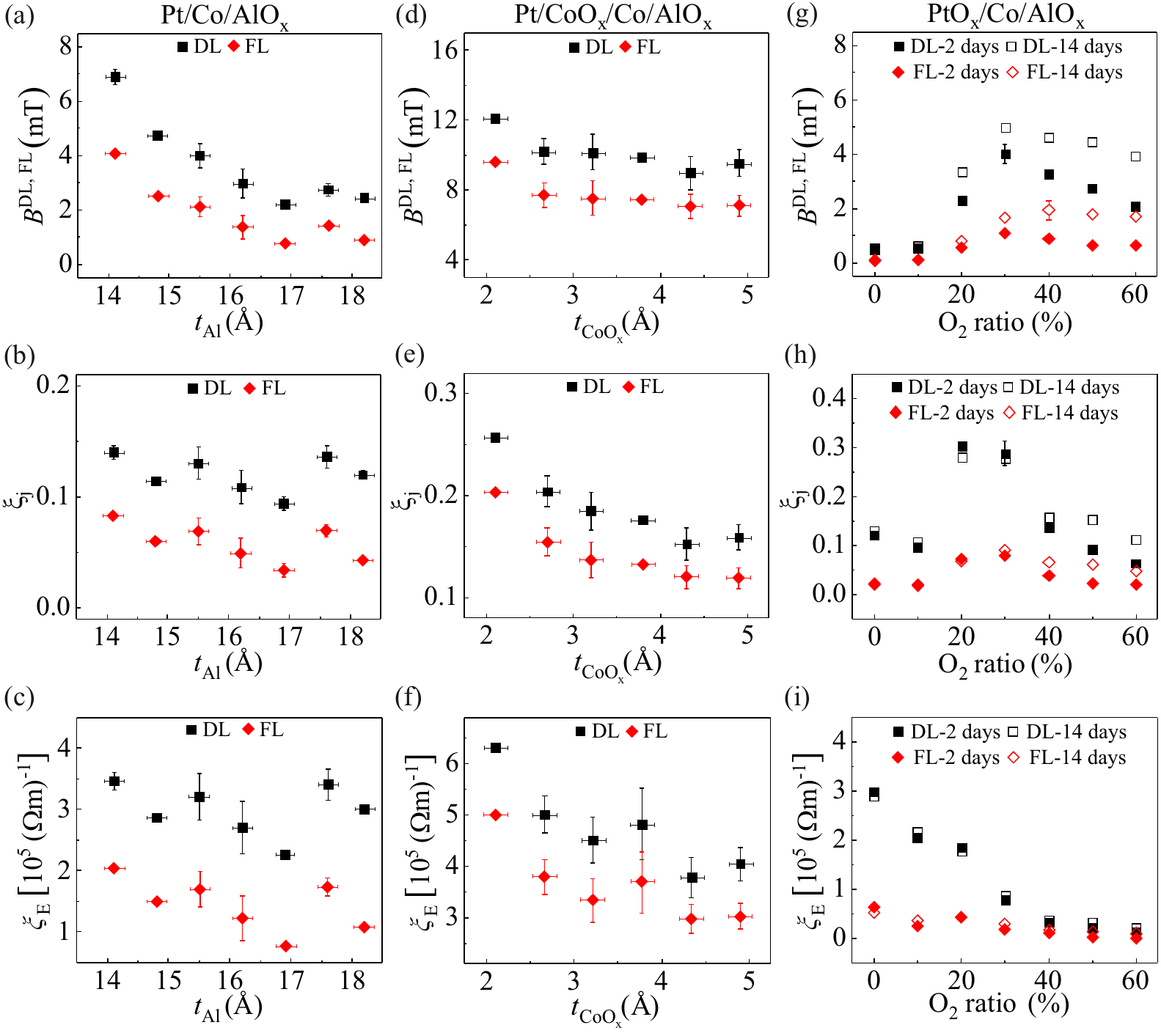}
	\caption{\label{SOT} (a) $B^{DL,FL}$, (b) $\xi_j^{DL, FL}$ and (c) $\xi_E^{DL, FL}$ as a function of $t_{\rm{Al}}$ in Pt/Co/AlO$_{\textrm{x}}$. (d-f) Same quantities as a function of $t_{\rm{CoO_{\textrm{x}}}}$ in Pt/CoO$_{\textrm{x}}$/Co/AlO$_{\textrm{x}}$ and (g-i) O$_2$:(Ar+O$_2$) ratio in PtO$_{\textrm{x}}$/Co/AlO$_{\textrm{x}}$. 
The data shown as solid and open symbols in (g-i) were obtained, respectively, 2 and 14 days after the deposition of PtO$_{\textrm{x}}$/Co/AlO$_{\textrm{x}}$. Vertical error bars represent the standard deviation of the average of different devices from the same series. Vertical error bars represent the standard deviation of the average of different devices from the same series. Horizontal error bars represent the uncertainty of the thickness calibration.}

\end{figure*}

\section{\label{SOT}Spin-orbit torques}
In our experimental geometry, and to the lowest order in the unitary magnetization $\textbf{m}=\textbf{M}/M_s$, where $\textbf{M}$ is the magnetization, the SOT is given by the sum of damping-like (DL) and field-like (FL) components~\cite{Garello-nn-2013,Manchon-rmp-2019},
\begin{eqnarray}
&\textbf{T}=T^{DL} \textbf{m} \times ( \textbf{m}\times \textbf{y})+ T^{FL} \textbf{m} \times \textbf{y},
\label{Tor}
\end{eqnarray}
where \textbf{y} represents the direction of the current-induced spin polarization.
For characterization purposes, it is convenient to work with the two effective magnetic fields corresponding to the damping-like and field-like torques,
\begin{eqnarray}
&\textbf{B}^{DL}= \textbf{T}^{DL}\times \textbf{m} = B^{DL}(\textbf{m}\times \textbf{y}),
\\
&\textbf{B}^{FL}= \textbf{T}^{FL}\times \textbf{m}=B^{FL} \textbf{y}.
\label{Bfl}
\end{eqnarray}
The advantage of the effective field formulation is that their action on the magnetization can be directly compared to that of an external field of known magnitude and direction, which allows for quantifying $B^{DL}$ and $B^{FL}$ by means of a variety of techniques~\cite{Manchon-rmp-2019}. Here we measure $B^{DL}$ and $B^{FL}$ using the harmonic Hall voltage method~\cite{Garello-nn-2013, Can-thermal, Hayashi-nm2013, Hayashi2014}. A description of the method, representative second harmonic Hall resistance measurements of the different samples as well as the analysis of thermal voltages and planar Hall effect are reported in the Appendix. For simplicity, we focus on the effects of oxidation on the lowest order SOT terms (Eq.~\ref{Tor}); a discussion of the higher order terms, i.e., of the angular dependence of the SOT, is given in Refs.~\onlinecite{Garello-nn-2013,Gweon2019,Avci2014a,Belashchenko2019}.

The fields $B^{DL}$ and $B^{FL}$ scale inversely with the thickness of the ferromagnet. Therefore, in order to compare the SOT in different systems, we consider the SOT efficiencies
\begin{eqnarray}
\xi_j^{DL,FL}&=&\frac{2e}{\hbar}M_st_{\textrm{Co}}\frac{B^{DL,FL}}{j},
\label{A9}
\end{eqnarray}
where $t_{\textrm{Co}}$ is the thickness of the Co layer, $\hbar$ the reduced Planck constant, and $j$ the injected current density obtained by dividing the current by the cross section of the conductive layers. The current density applied to study the SOT in the Pt/Co/AlO$_{\textrm{x}}$ and Pt/CoO$_{\textrm{x}}$/Co/AlO$_{\textrm{x}}$ series was $1.0 \times 10^7$~A/cm$^2$, which was calculated by dividing the total current by the width of the Hall bar and the sum of the thickness of the Pt and Co layers, respectively, 6 and 5.65~nm. For the PtO$_{\textrm{x}}$/Co/AlO$_{\textrm{x}}$ series,
we applied a current of 1~mA, which corresponds to a current density of $1.7 \times 10^6$~A/cm$^2$ when considering both PtO$_{\textrm{x}}$(5) and Co(1) as the conductive layers.

Because the current distribution is not expected to be homogeneous in samples made of different layers with varying degrees of oxidation, we also consider the SOT efficiency per unit electric field~\cite{Effciencypercurrent-PRL2016}
\begin{eqnarray}
\xi_E^{DL,FL}&=&\frac{2e}{\hbar}M_st_{\textrm{Co}}\frac{B^{DL,FL}}{E}=\frac{\xi_j^{DL,FL}}{\rho},
\label{A10}
\end{eqnarray}
where $E=\rho j$ is the electric field and $\rho$ is the resistivity. $\xi_E^{DL,FL}$ can be easily calculated by knowing the total magnetic moment measured by SQUID and is therefore independent of $t_{\textrm{Co}}$ as well as on any assumption on the current distribution and the thickness of the conductive layers. When comparing samples with similar resistivity, it is equivalent to use either $\xi_j^{DL,FL}$ or $\xi_E^{DL,FL}$, which can be considered as an effective spin Hall angle and spin Hall conductivity, respectively. However, for samples with very different resistivities such as those belonging to the PtO$_{\textrm{x}}$/Co/AlO$_{\textrm{x}}$ series, the electric field normalization provides a better means of comparison.

Figure \ref{SOT} summarizes the results obtained on the three sample series for $B^{DL,FL}$, $\xi_j^{DL,FL}$, and $\xi_E^{DL,FL}$, which we discuss in detail below.

\subsection{\label{SOT:1} Pt/Co/AlO$_{\textrm{x}}$}
Here we consider the effect of changing $t_{\rm{Al}}$, hence the oxidation of the Co/Al interface, in the Pt/Co/AlO$_{\textrm{x}}$ system. As shown in Fig. \ref{SOT}(a), the effective fields $B^{DL,FL}$ decrease monotonically as a function of $t_{\rm{Al}}$. This decrease mostly reflects the increase of $M_s$ as the Co layer changes from over- to under-oxidized (see Sec.~\ref{MS1}). Accordingly, the SOT efficiencies show a weaker dependence on $t_{\rm{Al}}$. $\xi_j^{DL, FL}$ and $\xi_E^{DL, FL}$ decrease slowly up to $t_{\rm{Al}}=16.9$~{\AA}, and then increase rather abruptly from $t_{\rm{Al}}=16.9$ to 17.6~{\AA} [Fig. \ref{SOT}(b,c)]. As the latter thickness corresponds to the optimal oxidation of the Co/AlO$_{\textrm{x}}$ interface (see Sec.~\ref{XPS}), we attribute the local maxima of $\xi_j^{DL, FL}$ and $\xi_E^{DL, FL}$ to the enhancement of the charge transfer and ensuing interfacial electric field between the Co and AlO$_{\textrm{x}}$ layers. We note that this enhancement affects both the damping-like and field-like components of the SOT, as expected from the spin current contributions originating from spin-orbit coupling and spin-dependent scattering at interfaces~\cite{Amin2016,Amin2016a, Amin2018}. Moreover, the larger SOT efficiencies found for the optimized samples are consistent with the larger torkances (torques per unit electric field and unit magnetization) calculated ab-initio for Pt/Co/O relative to Pt/Co/Al~\cite{Freimuth2014}.

The sign of the field-like SOT is such that $B^{FL}$ opposes the Oersted field generated by the current flowing the Pt layer. $\xi_j^{DL}$ ($\xi_j^{FL}$) and $\xi_E^{DL}$ ($\xi_E^{FL}$) peak at 0.14 (0.08) and 3.5 (2.1) $\times 10^5\,(\Omega m)^{-1}$, respectively, which compares well with the SOT efficiencies reported in previous studies of optimized Pt/Co/AlO$_{\textrm{x}}$~\cite{Garello-nn-2013}. Interestingly, these SOT efficiencies are consistently larger than those reported for  Pt(5)/Co(1)/MgO(2), where  $\xi_j^{DL}$ ($\xi_j^{FL}$) = 0.11 (0.024) and $\xi_E^{DL}$ ($\xi_E^{FL}$) = 2.43 (0.53)  $\times 10^5\,(\Omega m)^{-1}$~\cite{Effciencypercurrent-PRL2016}. Such a difference might be related to the different band alignment between Co and either AlO$_{\textrm{x}}$ or MgO, leading to different interfacial electrical fields and scattering properties. 

\subsection{\label{SOT:2}Pt/CoO$_{\textrm{x}}$/Co/AlO$_{\textrm{x}}$}
Inserting a 2.1~\AA~ thick CoO$_{\textrm{x}}$ layer between Pt and Co leads to an enhancement of the SOT effective fields and efficiencies by about a factor 2 compared to the optimized Pt/Co/AlO$_{\textrm{x}}$ trilayer [Fig. \ref{SOT}(d-f)]. This remarkable result indicates that the Pt/CoO$_{\textrm{x}}$ and CoO$_{\textrm{x}}$/Co interfaces are very efficient in transferring spins from Pt to Co. Given that the N\'{e}el and blocking temperatures of CoO are well below room temperature in films thinner than 10~nm~\cite{Ambrose1998}, the spin transfer across the insulating CoO$_{\textrm{x}}$ layer is likely mediated by antiferromagnetic spin fluctuations, as found for the reciprocal effect of spin pumping in Y$_3$Fe$_5$O$_{12}$/NiO/Pt and Y$_3$Fe$_5$O$_{12}$/CoO/Pt trilayers~\cite{Hahn2014,Qiu2016}. In such cases, a pronounced maximum of the pumping efficiency was reported when approaching the N\'{e}el temperature of the antiferromagnet~\cite{Lin2016}. Differently from these studies, however, here the FM layer is metallic and the spin current flows towards the FM, which is extremely promising for the efficient manipulation of the magnetization by an electric current. Additionally, we find that $\xi_j^{DL,FL}$ and $\xi_E^{DL,FL}$ decrease upon increasing the thickness of the CoO$_{\textrm{x}}$ layer, consistently with diffusive spin transport mediated by correlated spin fluctuations~\cite{Hahn2014}. Such a decrease shows that a significant fraction of the spin current giving rise to the SOT originates from either the Pt or the Pt/CoO$_{\textrm{x}}$ interface. However, we also observe that $\xi_j^{DL,FL}$ and $\xi_E^{DL,FL}$ tend to saturate towards finite values rather than tending to zero, which may indicate that the CoO$_{\textrm{x}}$/Co interface also contributes to generating a spin current~\cite{Amin2016,Amin2016a, Amin2018}. Overall, these results indicate that the insertion of a thin paramagnetic CoO$_{\textrm{x}}$ layer significantly improves the charge-spin conversion efficiency of HM/FM/Ox structures at a very small cost in terms of material engineering, consistently with recent reports~\cite{Hasegawa2018,Wang2019}.

\subsection{\label{SOT:3}PtO$_{\textrm{x}}$/Co/AlO$_{\textrm{x}}$}
The oxidation of Pt leads to a very different behavior compared to the oxidation of Co. Figures \ref{SOT}(g-i) show the SOT effective fields and efficiencies measured 2 days (full symbols) and 14 days (open symbols) after deposition of the PtO$_{\textrm{x}}$/Co/AlO$_{\textrm{x}}$ series. Note that the values of $B^{DL,FL}$ at 0\% O$_2$ ratio, i.e., for unoxidized Pt, are lower than reported for Pt/Co/AlO$_{\textrm{x}}$ [Fig. \ref{SOT}(a)] because the current density employed in these measurements was about a factor 6 smaller compared to the Pt/Co/AlO$_{\textrm{x}}$ series, whereas $\xi_j^{DL,FL}$ and $\xi_E^{DL,FL}$ are comparable. We recall also that changing the O$_2$ ratio from 0 to 30\% O$_2$ leads to about a 10-fold increase of the resistance (Sec.~\ref{ELECTRICAL}) and $R_{\textrm{AHE}}$ (Sec.~\ref{AHE}) as well as to a 3-fold decrease of $M_s$ (Sec.~\ref{MS1}). The samples of these series have IMA, the strength of which varies with the O$_2$ ratio (Sec.~\ref{MAE}).

Figure \ref{SOT}(g) shows that $B^{DL,FL}$ increase first by about a factor 10 as the O$_2$ ratio reaches 30\% and then decrease by an amount that depends on the time elapsed between the deposition and the measurements. Samples measured 20 days after deposition show the same SOT compared with the measurements performed after 14 days, suggesting that the oxidation profiles becomes stable after several days. The initial increase of $B^{DL,FL}$ is partly assigned to the reduction of $M_s$, as shown by the comparatively smaller change of $\xi_j^{DL, FL}$ [Fig. \ref{SOT}(h)]. $\xi_j^{DL}$ and $\xi_j^{FL}$ peak at 0.30 and 0.07, respectively, close to 20\% O$_2$ ratio. Our XPS measurements, however, show that Pt is not significantly oxidized at this ratio, as oxygen migrates towards the Co and Al layers (Sec.~\ref{XPS}).
A possible reason for this initial enhancement of $\xi_j^{DL,FL}$ could be the spontaneous formation of a CoO$_{\textrm{x}}$ layer next to Pt, in analogy with the results reported in Sec.~\ref{SOT:2}. Yet, it is unlikely that oxygen migration will lead to the formation of a homogenous CoO$_{\textrm{x}}$ layer between Pt and metallic Co, similar to the Pt/CoO$_{\textrm{x}}$/Co/AlO$_{\textrm{x}}$ series.
Further, measurements of the magnetic anisotropy suggest that oxygen reaches the Co/Al interface at around 20\% O$_2$ ratio (Sec.~\ref{MAE}). It is therefore likely that the peak of $\xi_j^{DL, FL}$ reflects a change of the current distribution in the PtO$_{\textrm{x}}$/Co/AlO$_{\textrm{x}}$ layers rather than a true increase of the SOT efficiency. For example, if oxygen from the topmost part of the PtO$_{\textrm{x}}$ layer migrates towards Al, most of the current will flow close to the Pt/Co interface, leading to an apparent increase of $\xi_j^{DL, FL}$. We recall that, for consistency between different samples, $\xi_j^{DL,FL}$ is calculated by assuming a constant cross-section for current flow, proportional to the total thickness of the PtO$_{\textrm{x}}$ and Co layers.

Further insight into the dependence of the SOT on the O$_2$ ratio can be gained by analyzing the behavior of $\xi_E^{DL,FL}$, which does not depend on the current distribution. Figure \ref{SOT}(i) shows that $\xi_E^{DL, FL}$ decreases monotonically with increasing O$_2$ ratio. Specifically, $\xi_E^{DL}$ starts at $2.98 \times 10^5$~$(\Omega m)^{-1}$ at 0\% O$_2$ and reaches $0.13 \times 10^5$~$(\Omega m)^{-1}$ at 60\% O$_2$; $\xi_E^{FL}$ shows a similar decreasing trend, from $0.63 \times 10^5$~$(\Omega m)^{-1}$  to $0.01 \times 10^5$~$(\Omega m)^{-1}$. According to Eq.~\ref{A10}, for measurements performed at constant current, $\xi_E^{DL,FL} \propto B^{DL,FL}M_s/R$. 
The strong decrease of $\xi_E^{DL, FL}$ is thus ascribed to the increment of $R$ and reduction of $M_s$ that occur upon augmenting the O$_2$ ratio. 

Our results are in stark contrast with previous work carried out on PtO$_{\textrm{x}}$/Ni$_{81}$Fe$_{19}$/SiO$_2$ trilayers~\cite{An2018,An2018a}, in which both $\xi_j^{DL,FL}$ and $\xi_E^{DL,FL}$ were found to increase monotonically with O$_2$ ratio, up to  $\xi_j^{DL}$ ($\xi_j^{FL}$) = 0.92 (0.19) and $\xi_E^{DL}$ ($\xi_E^{FL}$) = 8.7 (1.8)  $\times 10^5\,(\Omega m)^{-1}$ at 100\% O$_2$. In the following, we discuss possible reasons for this discrepancy. First, the FM and cap layers are different, which may result in significant differences in both oxygen diffusion and oxidation profile as well as in the generation and transmission of spin currents at interfaces. Second, the measurements of $B^{DL,FL}$ in Refs.~\onlinecite{An2018,An2018a} were performed by spin-torque ferromagnetic resonance (FMR) and by measuring the broadening of the FMR lineshape upon dc current injection. The current and field normalizations required to calculate $\xi_j^{DL,FL}$ and $\xi_E^{DL,FL}$ in Refs.~\onlinecite{An2018,An2018a} were carried out by assuming current flow in the Ni$_{81}$Fe$_{19}$ layer only and constant resistance, respectively. The current normalization affects the absolute value of $\xi_j^{DL,FL}$, but would not induce a different trend as a function of O$_2$ ratio compared to ours, since we also assume a constant cross section for the current. The electric field normalization required to calculate $\xi_E^{DL,FL}$ is more critical, since the resistance increases continuously as a function of O$_2$ ratio in PtO$_x$/Ni$_{81}$Fe$_{19}$/SiO$_2$~\citep{An2018}. However, even assuming a constant resistance, we do not observe a monotonous increase of $\xi_E^{DL,FL}$. We also note that assumptions on the distribution of the current as well as on the different contributions to the FMR lineshape can significantly affect the determination of the effective fields by spin-torque FMR~\cite{Harder-pr-2016}. Finally, no hints of post-deposition oxygen diffusion or changes of $M_s$ were reported for either PtO$_{\textrm{x}}$/Ni$_{81}$Fe$_{19}$/SiO$_2$ or PtO$_{\textrm{x}}$/CoTb/MgO~\cite{An2018,An2018a}, but also not specifically analyzed, which prevents a more stringent comparison with our results.

\begin{figure*}[t]
	\includegraphics[width=160mm]{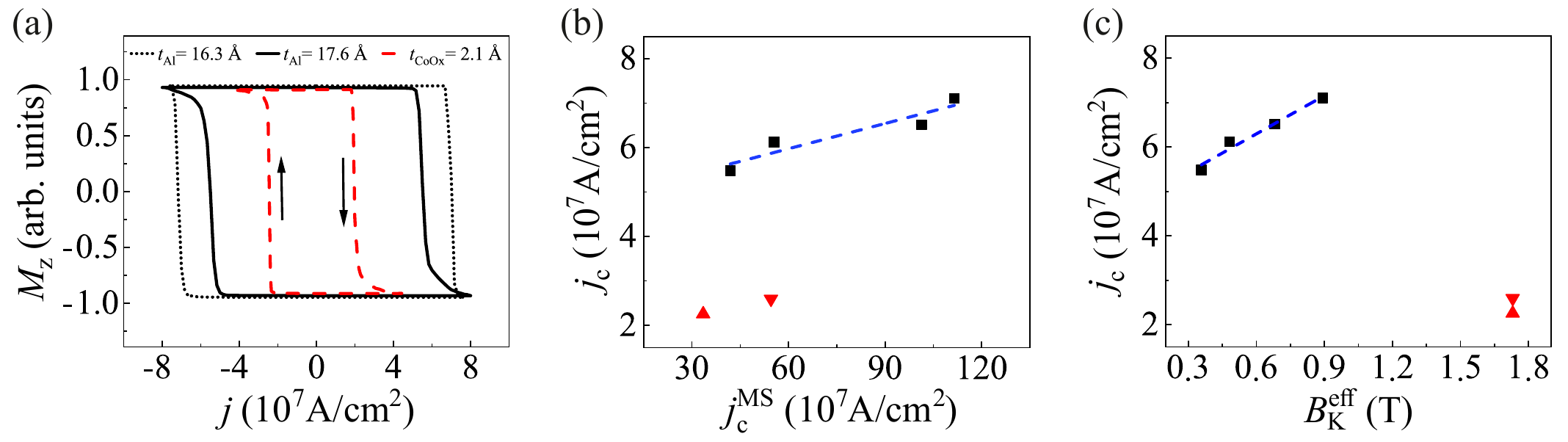}
	\caption{\label{switching} (a) Out-of-plane component of the magnetization $M_z= R^\omega_H / R_{\textrm{AHE}}$ of Pt/Co/AlO$_{\textrm{x}}$ (dotted and solid black lines) and Pt/CoO$_{\textrm{x}}$/Co/AlO$_{\textrm{x}}$ (red dashed line) as a function of dc current density during a switching loop.
(b) $j_c$ versus $j_{c}^{\textrm{MS}}$ calculated using Eq.~\ref{eq:jsw_op}. (c) $j_c$ as a function of $B_{\textrm{K}}^{\textrm{eff}}$. In (b) and (c) black squares and red triangles represent data for Pt/Co/AlO$_{\textrm{x}}$ and Pt/CoO$_{\textrm{x}}$/Co/AlO$_{\textrm{x}}$, respectively. The dashed lines represent linear fits to $j_c$ of Pt/Co/AlO$_{\textrm{x}}$. Up (down) triangles denote Pt/CoO$_{\textrm{x}}$/Co/AlO$_{\textrm{x}}$ samples with $\xi_j^{DL}=0.26$ (0.19). All data have been acquired in the presence of a constant field $B_x=+70$~mT for Pt/Co/AlO$_{\textrm{x}}$ and $B_x=+410$~mT for Pt/CoO$_{\textrm{x}}$/Co/AlO$_{\textrm{x}}$.
Note that $j_c$ could not be measured for all the samples of the Pt/CoO$_{\textrm{x}}$/Co/AlO$_{\textrm{x}}$ series as the specimens with thicker CoO$_{\textrm{x}}$ were irreversibly modified by the current.}
\end{figure*}

\section{\label{SWITCHING}Current-induced switching}
By increasing the current density beyond a critical threshold $j_c \approx 10^7- 10^8$ A/cm$^2$, the SOT become large enough to reverse the magnetization of trilayers with strong PMA, such as Pt/Co/AlO$_{\textrm{x}}$~\cite{Miron-switching2011,Garello-apl2014}, Pt/Co/MgO~\cite{Avci2012}, Ta/CoFeB/MgO~\cite{Can-thermal,Zhang-jap-2014}, and Pt/[Co/Ni]$_{\rm N}$/Al~\cite{Rs-apl-2016}. Magnetization reversal is driven by the damping-like torque~\cite{Miron-switching2011}; the field-like torque, which is equivalent to a hard axis field, facilitates the switching by lowering the energy barrier for domain nucleation and propagation~\cite{Miron-rashba-2010,Manuel-nn, Taniguchi2015}. As the damping-like torque is perpendicular to both the current direction and the normal to the plane, an in-plane external field $B_x$ parallel to the current is required in order to break the symmetry of the system and determine the polarity of the switching~\cite{Miron-switching2011,Manchon-rmp-2019}.
In the macrospin approximation, the critical switching current is given by~\cite{Lee2013}
\begin{equation}\label{eq:jsw_op}
j_{c}^{\textrm{MS}}=\frac{2e}{\hbar}\frac{M_s t_{\textrm{Co}}}{\xi_j^{DL}}\left(\frac{B_{K}^{\textrm{eff}}}{2}-\frac{B_{x}}{\sqrt{2}} \right).
\end{equation}
Although SOT-driven magnetization switching in structures with lateral dimensions larger than about 20~nm occurs by nucleation and expansion of magnetic domains~\cite{Emori2013,Ryu2013,Yu2014b,Safeer2015,Manuel-nn}, this expression provides a useful starting point to evaluate the dependence of $j_{c}$ on critical material parameters.

In realistic systems, $j_{c}$ depends on the parameters appearing in Eq.~\eqref{eq:jsw_op} as well as on the Dzyaloshinskii–Moriya interaction, domain pinning field, device geometry, size, temperature, and duration of the current pulses~\cite{Manuel-nn, Garello-apl2014, Lee2014, Zhang_2018, Pham2018, Mikuszeit2015,Laguna-marco2015,Pizzini2014}. The temperature, which is determined by the current distribution as well as by the thermal conductivity of the different materials in the stack, plays a major role, both in activating the switching as well as in changing critical parameters such as $M_s$, $B_{\textrm K}$, and $\xi_j^{DL}$ during switching. These factors vary significantly, even among nominally similar material systems reported in the literature~\cite{Manchon-rmp-2019}, so that a comparative study of the critical switching current versus magnetic anisotropy or SOT efficiency is hard to achieve. Here we use the controlled oxidation of the top Co interfaces in Pt/Co/AlO$_{\textrm{x}}$ in order to investigate such effects.

Figure \ref{switching}(a) shows the switching of the out-of-plane magnetization of Pt/Co/AlO$_{\textrm{x}}$ with $t_{\rm{Al}} = 16.3$ and 17.6~\AA~ (dotted and solid black line, respectively) measured by the change of the anomalous Hall resistance during a sweep of a dc current in the presence of a constant field $B_x=+70$~mT. The curve represents the normalized signal $M_z = R^\omega_H / R_{\textrm{AHE}}$, where $R^\omega_H$ is the first harmonic Hall resistance (see Appendix). The dashed red line shows the switching of Pt/CoO$_{\textrm{x}}$/Co/AlO$_{\textrm{x}}$ with $t_{\rm{CoO}_{\rm{x}}} = 2.1$~\AA~ in the presence of a constant field $B_x=+410$~mT. The square loops indicate full switching of the magnetic layer with current in both types of samples. However, there are noticeable differences in the threshold current at which the magnetization switches, which we discuss below. In the following, the critical switching current density is defined as the current for which $M_z = 0$.

Figure~\ref{switching}(b) shows the experimental $j_c$ versus the macrospin $j_{c}^{\textrm{MS}}$ calculated using Eq.~\ref{eq:jsw_op} for Pt/Co/AlO$_{\textrm{x}}$ (black squares) and Pt/CoO$_{\textrm{x}}$/Co/AlO$_{\textrm{x}}$ (red triangles). As seen in Sec.~\ref{MAG}, reducing the oxidation of the top Co interface of Pt/Co/AlO$_{\textrm{x}}$ increases $M_s$ and decreases $B_{\textrm{K}}^{\textrm{eff}}$ by about a factor 4, whereas $\xi_j^{DL}$ changes by about 40\% depending on $t_{\textrm{Al}}$.
Accordingly, we find that $j_c$ scales linearly with $j_{c}^{\textrm{MS}}$, which in turn is proportional to $M_s/\xi_j^{DL}$ times $B_{\textrm{K}}^{\textrm{eff}}$ reduced by the constant in-plane field $B_{x}$. The plot further shows that $j_c$ is about one order of magnitude smaller than predicted by $j_{c}^{\textrm{MS}}$ and has a nonzero intercept for $j_{c}^{\textrm{MS}}\rightarrow 0$. The strong reduction of $j_c$ compared to the macrospin threshold can be explained by the fact that magnetization reversal occurs by domain nucleation and propagation~\cite{Manuel-nn,Mikuszeit2015} and is a thermally activated process~\cite{Grimaldi2020}, whereas Eq.~\ref{eq:jsw_op} assumes coherent reversal at zero temperature. Further, the nonzero intercept of $j_c$ denotes the presence of a finite activation barrier for switching. Accordingly, we find that the factor that mostly affects $j_{c}$ is the strength of the magnetic anisotropy, as seen by the linear scaling of $j_{c}$ with $B_{\textrm{K}}^{\textrm{eff}}$ in Fig.~\ref{switching}(c).

Samples of the Pt/CoO$_{\textrm{x}}$/Co/AlO$_{\textrm{x}}$ series have larger $\xi_j^{DL}$ and $B_{\textrm{K}}^{\textrm{eff}}$ relative to Pt/Co/AlO$_{\textrm{x}}$ and comparable or smaller $M_s$. In Eq.~\ref{eq:jsw_op} these factors partially compensate, such that $j_{c}^{\textrm{MS}}$ of the Pt/CoO$_{\textrm{x}}$/Co/AlO$_{\textrm{x}}$ samples with larger $\xi_j^{DL}$ [red triangles in Fig.~\ref{switching}(b)] are close to the lower values of $j_{c}^{\textrm{MS}}$ of the Pt/Co/AlO$_{\textrm{x}}$ series [black squares in Fig.~\ref{switching}(b)]. However, the experimental $j_c$ in Pt/CoO$_{\textrm{x}}$/Co/AlO$_{\textrm{x}}$ is 30 to 40\% that of Pt/Co/AlO$_{\textrm{x}}$. The different $j_c$ in the two sample series can be possibly ascribed to the larger $B_{x}$ employed to switch the high anisotropy Pt/CoO$_{\textrm{x}}$/Co/AlO$_{\textrm{x}}$ samples. In models of spin-transfer torque switching, however, the barrier for magnetization reversal scales proportionally to $B_{\textrm{K}}^{\textrm{eff}}\left(1-B_{x}/B_{\textrm{K}}^{\textrm{eff}}\right)^2$~\cite{Koch2004,Seki2008}. This scaling does not account for the observed reduction of $j_c$ in Pt/CoO$_{\textrm{x}}$/Co/AlO$_{\textrm{x}}$. Therefore, we conclude that either $B_{x}$ has a stronger influence on SOT switching relative to spin transfer torque or there are other factors that promote switching in Pt/CoO$_{\textrm{x}}$/Co/AlO$_{\textrm{x}}$, such as the field-like torque and the Dzyaloshinskii–Moriya interaction, whose interplay cannot be discerned in our measurements.

\section{\label{CONCLUSIONS}Conclusions}
In summary, we investigated the oxidation effects on the resistivity, magnetization, magnetic anisotropy, anomalous Hall resistance, and SOT of Pt/Co/AlO$_{\textrm{x}}$, Pt/CoO$_{\textrm{x}}$/Co/AlO$_{\textrm{x}}$, and PtO$_{\textrm{x}}$/Co/AlO$_{\textrm{x}}$ heterostructures. In all samples, increasing levels of oxidation lead to a reduction of $M_s$ due to the formation of CoO$_{\textrm{x}}$. On the contrary, $R_{\textrm{AHE}}$ decreases with increasing oxidation in Pt/Co/AlO$_{\textrm{x}}$ but increases in Pt/CoO$_{\textrm{x}}$/Co/AlO$_{\textrm{x}}$ and PtO$_{\textrm{x}}$/Co/AlO$_{\textrm{x}}$, consistently with the AHE being influenced by the magnetic volume as well as by spin-dependent scattering at oxidized interfaces. 

The magnetic anisotropy is determined by the competition between the PMA of the Pt/Co and Co/Al interfaces and shape anisotropy. The latter scales as $M_s$ and is strongly influenced by the oxidation of Co. In Pt/Co/AlO$_{\textrm{x}}$, $B_{\textrm{K}}^{\textrm{eff}}$ decreases linearly with increasing $t_{\textrm{Al}}$, from 1.76~T to 0.36~T going from over-oxidized to under-oxidized Co/Al interfaces.
Interestingly, the insertion of a thin CoO$_{\textrm{x}}$ layer between Pt and Co results in very strong PMA of the Pt/CoO$_{\textrm{x}}$/Co/AlO$_{\textrm{x}}$ series, with $B_{K}^{\textrm eff}\approx 1.7$~T, which is nearly independent on $t_{\textrm{CoO}}$. Direct oxidation of Pt, on the other hand, leads to a reduction of the PMA in PtO$_{\textrm{x}}$/Co/AlO$_{\textrm{x}}$ compared to Pt/Co/AlO$_{\textrm{x}}$.

The amplitude of the damping-like and field-like SOT varies substantially depending on the oxidation profile, reflecting changes of $M_s$ and SOT efficiency. For Pt/Co/AlO$_{\textrm{x}}$, we find that $\xi_j^{DL}$ and $\xi_j^{FL}$ peak at 0.14 and 0.07, respectively, upon optimal oxidation of the Co/Al interface, which corresponds to a maximum of $M_s$ and minimum oxidation of Co, as confirmed by XPS. The insertion of a 2~\AA ~thick paramagnetic CoO$_{\textrm{x}}$ layer between Pt and Co nearly doubles $\xi_j^{DL}$ and $\xi_j^{FL}$, indicating that the overall spin conversion efficiency of Pt/CoO$_{\textrm{x}}$/Co/AlO$_{\textrm{x}}$ is significantly higher than Pt/Co/AlO$_{\textrm{x}}$. Both $\xi_j^{DL}$ and $\xi_j^{FL}$ decrease with increasing $t_{{\rm CoO}_x}$, but appear to level off around 0.15 and 0.12, respectively, for $t_{{\rm CoO}_x}>4$~\AA. This trend suggests that not only the CoO$_{\textrm{x}}$ layer enhances the transmission of the spin current generated by the Pt layer and Pt/CoO$_{\textrm{x}}$ interface~\cite{Hahn2014,Qiu2016,Hasegawa2018,Wang2019}, but also that the CoO$_{\textrm{x}}$/Co interface generates an additional spin accumulation that is independent of $t_{{\rm CoO}_x}$.

The oxidation of the Pt layer leads to significant oxygen diffusion upon deposition of Co and Al, as confirmed by XPS and rationalized by the larger enthalpy of formation of Al$_2$O$_3$ and CoO relative to PtO. The inhomogeneous oxidation profile of the PtO$_{\textrm{x}}$/Co/AlO$_{\textrm{x}}$ series prevents a clear cut determination of $\xi_j^{DL}$ and $\xi_j^{FL}$, which remain, in any case, significantly smaller than reported for PtO$_{\textrm{x}}$/Ni$_{81}$Fe$_{19}$/SiO$_2$~\cite{An2018,An2018a}. Additionally, the SOT efficiencies normalized by the electric field are found to decrease continuously with O$_2$ ratio. These results are in stark contrast with the SOT efficiency reported in Refs.~\onlinecite{An2018,An2018a}, which shows a monotonic increase with increasing concentration of O$_2$. Possible explanations for such a discrepancy are discussed in Sec.~\ref{SOT:3}.

Finally, we proved that the critical current density required to switch the magnetization of the Co layer scales linearly with the theoretical current density calculated using the macrospin approximation, but is smaller by about one order of magnitude. This difference is attributed to an activated magnetization reversal process limited by the energy barrier for domain nucleation. The critical current density of Pt/Co/AlO$_{\textrm{x}}$ is found to depend linearly on $B_{\textrm{K}}^{\textrm{eff}}$, which is determined by the degree of oxidation of the Co/Al interface. Owing to the large PMA, switching of Pt/CoO$_{\textrm{x}}$/Co/AlO$_{\textrm{x}}$ generally requires the presence of a higher in-plane static field relative to Pt/Co/AlO$_{\textrm{x}}$. However, even when accounting for the different in-plane field, the critical current density of Pt/CoO$_{\textrm{x}}$/Co/AlO$_{\textrm{x}}$ is lower than expected.

Our results demonstrate the critical impact of interfacial oxidation on the SOT and magnetotransport properties of HM/FM/Ox trilayers. Taken together with recent studies~\cite{Bauer2014, Emori2014, Bi2014, Qiu2015,ZFS-wedge-nn-2014, Yu2014b, Pai2015, Akyol2015, Hibino2017,Hasegawa2018,Wang2019}, our work shows that devices with strongly improved PMA and charge-to-spin conversion efficiency might be realized by careful control of the oxidation of the top and bottom interfaces of the FM layer.

\begin{acknowledgments}
We acknowledge funding from the the Swiss National Science Foundation
(grant No. 200020-172775). We acknowledge Prof. N. Spencer for access to the surface analytical facilities. J. Feng was supported by a scholarship from the Chinese Scholarship Council.
\end{acknowledgments}

\appendix
\section{\label{APPENDIX}Harmonic Hall voltage measurements}
The anomalous Hall resistance, magnetic anisotropy, and SOT effective fields were measured using the harmonic Hall voltage method~\cite{Garello-nn-2013, Hayashi-nm2013, Hayashi2014, Can-thermal, Manchon-rmp-2019}. The method is based on the injection of an ac current $I = I_0\cos(\omega t)$ with amplitude $I_0$ and frequency $f=\omega/2\pi$, typically about 10~Hz. As the magnetization oscillates around its equilibrium state due to the oscillating SOTs, the Hall resistance varies as $R_H(t)=R_H(\textbf{B}_0+\textbf{B}_I)$, where $\textbf{B}_0$ is the sum of the external field $\textbf{B}_{\textrm{ext}}$ and effective anisotropy field $\textbf{B}_{K}^{\textrm eff}$, and $\textbf{B}_I$ is the sum of the current-induced effective fields $\textbf{B}^{DL}$, $\textbf{B}^{FL}$, and the Oersted field $\textbf{B}^{\textrm{Oe}}$. In this situation, the Hall voltage can be written as

\begin{eqnarray}
V_{H}& = &I_0\cos(\omega t)\, R_{H}(\textbf{B}_0+\textbf{B}_I)\label{A1}\\
& = & I_0[R_H^{0} + R_H^{\omega}\cos(\omega t) + R_H^{2\omega}\cos(2\omega t)] .
 \label{A2}
\end{eqnarray}

\begin{figure*}[t]
	\includegraphics[width=160mm]{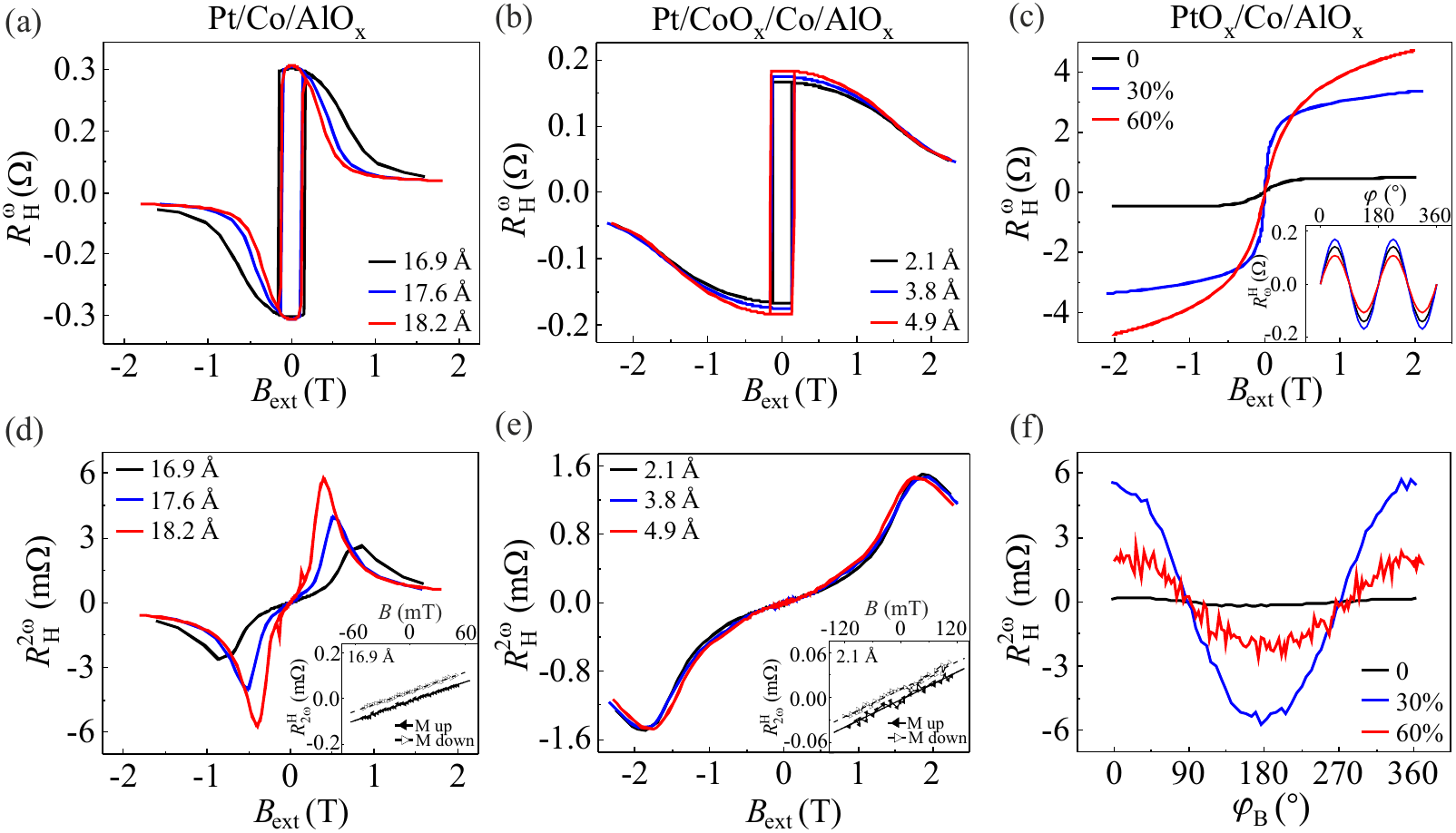}
	\caption{\label{FigS1} $R_H^{\omega}$ as a function of external field for (a) Pt/Co/AlO$_{\textrm{x}}$ with different $t_{\rm{Al}}$, (b) Pt/CoO$_{\textrm{x}}$/Co/AlO$_{\textrm{x}}$ with different $t_{\rm{CoO_{\textrm{x}}}}$, and (c) PtO$_{\textrm{x}}$/Co/AlO$_{\textrm{x}}$ with different O$_2$:(Ar+O$_2$) ratios. (d,e) $R_H^{2\omega}$ measured simultaneously with the curves shown in (a,b). Close ups of the low field behavior of $R_H^{2\omega}$ for the two systems are shown in the insets. (f) $R_H^{2\omega}$ as a function of $\varphi_B$ for PtO$_{\textrm{x}}$/Co/AlO$_{\textrm{x}}$, measured simultaneously with the angle scans of $R_H^{\omega}$ shown in the inset of (c). For Pt/Co/AlO$_{\textrm{x}}$ and Pt/CoO$_{\textrm{x}}$/AlO$_{\textrm{x}}$, the external field is applied at angles $\theta_B=86^\circ$ and $\varphi_B=0^\circ$. For PtO$_{\textrm{x}}$/Co/AlO$_{\textrm{x}}$, the field is applied at $\theta_B=0^\circ$. The in-plane angle scans in (f) are performed by rotating the samples in a constant field of 1~T.}
\end{figure*}

Here, $R_H^{0}$, $R_H^{\omega}$, and $R_H^{2\omega}$ are the zeroth, first, and second harmonic Hall resistance, respectively. $R_H^{\omega}$ corresponds to the Hall resistance measured in a dc experiment, which is determined by the anomalous Hall effect (AHE) and planar Hall effect (PHE):
\begin{equation}
R_H^{\omega}= R_{\textrm{AHE}}\cos\theta + R_{\textrm{PHE}}\sin^2\theta \sin(2\varphi),
\label{A3}
\end{equation}
where $R_{\textrm{AHE}}$ and $R_{\textrm{PHE}}$ are the AHE and PHE coefficients, respectively, and $\theta$ and $\varphi$ are the polar and azimuthal angles of the magnetization, as defined in Fig.~1(e) of the main text. The ordinary Hall resistance is negligibly small in our samples compared with $R_{\textrm{AHE}}$ and $R_{\textrm{PHE}}$, and ignored in the following. For samples with perpendicular magnetic anisotropy (PMA), namely Pt/Co/AlO$_{\textrm{x}}$ and Pt/CoO$_{\textrm{x}}$/Co/AlO$_{\textrm{x}}$, $R_{\textrm{AHE}}$ is given by half of the difference of the $R_H^{\omega}$ when the magnetization is up/down at $B_{\textrm{ext}}=0$ [Fig.~\ref{FigS1}(a,b)]. For samples with in-plane magnetic anisotropy, namely PtO$_{\textrm{x}}$/Co/AlO$_{\textrm{x}}$, we swept $B_{\textrm{ext}}$ out-of-plane up to 2.2~T magnetic field and estimated $R_{\textrm{AHE}}$ as the intercept of the linear fit of the $R_H^{\omega}$ with the $y$-axis [Fig.~\ref{FigS1}(c)]. Examples of $R_H^{\omega}$ and $R_H^{2\omega}$ as a function of applied field for samples of the three series with different oxidation levels are shown in Fig.~\ref{FigS1}.

\subsection{\label{APPENDIX:BK}Measurement of $B_{K}^{\rm eff}$}

The effective magnetic anisotropy field (Eq.~1) is determined from the field dependence of $R_H^{\omega}$ reported in Fig.~\ref{FigS1}(a-c). For the samples with PMA, $B_{K}^{\rm eff}$ is given by
\begin{eqnarray}
B_{K}^{\rm eff}&=&\left|\frac{cos\theta_B}{cos\theta}-\frac{sin\theta_B}{sin\theta}\right|B_{\textrm{ext}},
\label{A4}
\end{eqnarray}
where $\theta_B$ is the polar angle of the applied field, which is set prior to the measurement, and $\theta$ is obtained from Eq.~\ref{A3}. This equation assumes that the magnetization behaves as a macrospin in a uniaxial anisotropy field, which is approximately correct as long as $R_H^{\omega}$ has a reversible behavior versus the external field.

All the samples of the PtO$_{\textrm{x}}$/Co/AlO$_{\textrm{x}}$ series have in-plane magnetic anisotropy ($B_{K}^{\rm eff}< $0), which cannot be estimated using the method described above. An approximate estimate of $B_{K}^{\rm eff}$ in this series can be obtained by measuring the field required to saturate the magnetization along the out-of-plane hard-axis. In the macrospin approximation, such a field corresponds to the sum of the demagnetizing field and magnetocrystalline anisotropy field, as given in Eq.~1. The hard axis saturation field is obtained by the crossing point of the two lines that fit $R_H^{\omega}$ at high field and in the low field region where the magnetization coherently rotates from in-plane to out-of-plane [Fig.~\ref{FigS1}(c)].

\subsection{\label{APPENDIX:SOT}Measurement of $B^{DL}$ and $B^{FL}$}
\subsubsection{\label{APPENDIX:SOT:PMA} Samples with perpendicular magnetic anisotropy}
For samples with PMA, the dependence of $R_H^{2\omega}$ on the current is conveniently expressed as~\cite{Garello-nn-2013,Can-thermal}
\begin{eqnarray}
\begin{split}
R_H^{2\omega}=&[R_{\textrm{AHE}}-2R_{\textrm{PHE}}\cos\theta \sin(2\varphi)]\frac{d\cos\theta}{d\textbf{B}_I}\cdot\textbf{B}_I\\&+ R_{\textrm{PHE}}\sin^2\theta\frac{d\sin(2\varphi)}{d\textbf{B}_I}\cdot\textbf{B}_I + R_{\nabla T}\sin\theta \cos\varphi,
\end{split}
\label{A5}
\end{eqnarray}
where $\textbf{B}_{I}=\textbf{B}^{DL}+\textbf{B}^{FL}+\textbf{B}^{\textrm{Oe}}$ and $R_{\nabla T}$ is the thermal resistance due to the out-of-plane temperature gradient $\nabla T \propto I_0^2$ induced by Joule heating~\cite{Can-thermal}. In principle, both the anomalous Nernst effect (ANE) and the spin Seebeck effect can contribute to $R_{\nabla T}$. However, in metal bilayers, the ANE is expected to dominate. In the latter case, one has $R_{\nabla T}=\alpha \nabla T /I_0$, where $\alpha$ is the anomalous Nernst constant in units of $\mathrm{V}\mathrm{m}\mathrm{K}^{-1}$.
The angle dependent thermal resistance must be subtracted from $R_H^{2\omega}$ in order to avoid overestimating the SOTs, as described later.

In order to measure the first order contributions to the SOT, it is sufficient to consider the limit of small oscillations of the magnetization about the out-of-plane direction ($\theta \approx 0^\circ$). In such a case, ${B}^{DL}$ and ${B}^{FL}$ are obtained as follows~\citep{Hayashi-nm2013, Hayashi2014}
\begin{eqnarray}
B^{DL,FL}_{\theta\approx0^\circ}&=&-2\frac{\partial (R_H^{2\omega} - R_{\nabla T}\sin\theta \cos\varphi)}
{\partial B_{\varphi=0^\circ, 90^\circ}}/\frac{\partial ^2 R_H^{\omega}}{\partial B_{\varphi=0^\circ ,90^\circ}^2},
\label{A6}
\end{eqnarray}
where $\frac{\partial (R_H^{2\omega} - R_{\nabla T}\sin\theta \cos\varphi)}
{\partial B_{\varphi=0^\circ, 90^\circ}}$ is the derivative of $R_H^{2\omega} - R_{\nabla T}\sin\theta \cos\varphi$ as a function of external field and $\frac{\partial ^2 R_H^{\omega}}{\partial B_{\varphi=0^\circ ,90^\circ}^2}$ is the second derivative of
$R_H^{\omega}$ as a function of external field ($\theta \approx 0^\circ$, and $\varphi = 0^\circ$ for $B^{DL}$, $\varphi = 90^\circ$ for $B^{FL}$ ). The thermal component $R_{\nabla T}\sin\theta \cos\varphi$ in the numerator of  Eq.~\ref{A6} must be subtracted from $R_H^{2\omega}$ prior to evaluating the fields, as detailed in Sec.~\ref{APPENDIX:SOT:thermal}.

\begin{figure*}[t]
	\includegraphics[width=160mm]{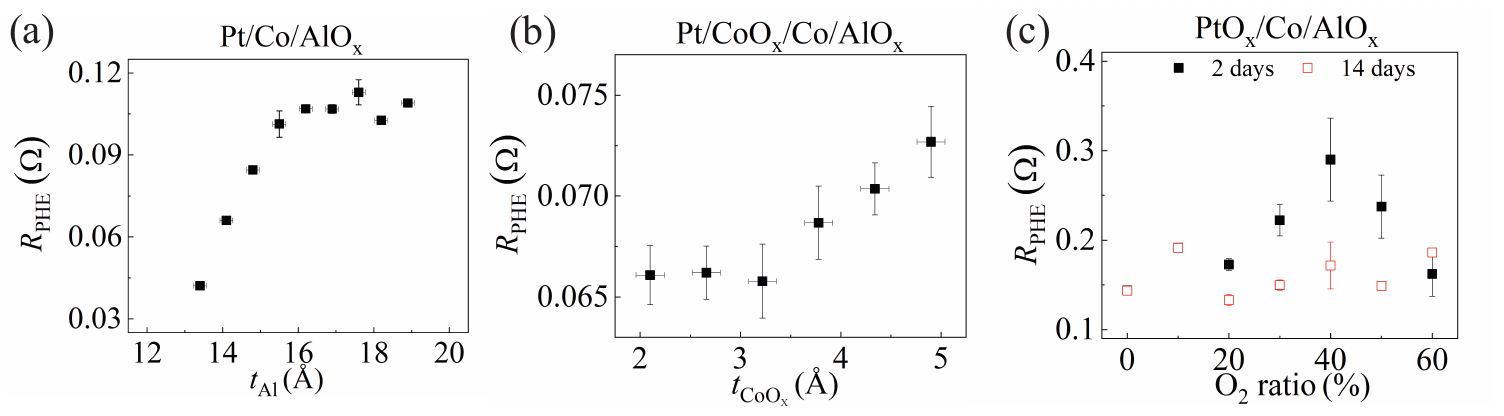}
	\caption{\label{FigS2} $R_{\textrm{PHE}}$ in (a) Pt/Co/AlO$_{\textrm{x}}$, (b) Pt/CoO$_{\textrm{x}}$/Co/AlO$_{\textrm{x}}$, and (c) PtO$_{\textrm{x}}$/Co/AlO$_{\textrm{x}}$. The data shown as solid and open symbols in (c) were obtained, respectively, 2 and 14 days after the deposition of PtO$_{\textrm{x}}$/Co/AlO$_{\textrm{x}}$.}
\end{figure*}

If $R_{\textrm{PHE}}$ $\ll$ $R_{\textrm{AHE}}$, Eq.~\ref{A6} is sufficient to quantify the SOTs effective fields, which are proportional to the ratio of the current-dependent Hall susceptibility and a normalization factor that accounts for the tendency of the magnetization to remain aligned along the easy axis. If $R_{\textrm{PHE}}$ cannot be ignored, a correction to Eq.~\ref{A6} is required~\citep{Garello-nn-2013,Hayashi2014}, which gives
\begin{eqnarray}
B^{DL,FL}&=& \frac{B^{DL,FL}_{\theta\approx0^\circ} \pm 2r B^{FL,DL}_{\theta\approx0^\circ}}{1-4r^2},
\label{A7}
\end{eqnarray}
where $r=R_{\textrm{PHE}}/R_{\textrm{AHE}}$. 
Figure~\ref{FigS2} shows that $R_{\textrm{PHE}}$ is relatively large in all the samples investigated in this work, with the larger values found in the PtO$_{\textrm{x}}$/Co/AlO$_{\textrm{x}}$ series. The overall trend of $R_{\textrm{PHE}}$ is similar to that of $R_{\textrm{AHE}}$ discussed in Sect.~III of the main text, despite the different origin of the PHE and AHE. The similar behavior is attributed to the fact that both effects depend on the volume fraction and $M_s$ of metallic Co as well as on the current distribution in the multilayer structure. In most samples, $r$ attains values between 0.3 and 0.4, such that Eq.~\ref{A7} is used to determine the fields $B^{DL}$ and $B^{FL}$.

\subsubsection{\label{APPENDIX:SOT:IMA} Samples with in-plane magnetic anisotropy}
In order to quantify the SOTs in samples with in-plane magnetic anisotropy we performed harmonic Hall voltage measurements by rotating the sample in the xy plane and applying a sufficiently large external field (with various amplitudes) to ensure the magnetic saturation of the sample is along the field direction. In this geometry, assuming that the in-plane magnetization is isotropic, $R_H^{2\omega}$ is given by \cite{Can-thermal}:
\begin{eqnarray}
\begin{split}
R_H^{2\omega} = &(R_{\textrm{AHE}}\frac{B^{DL}}{{B_{\textrm{ext}}-B_{\textrm K}^{\textrm{eff}}}} + R_{\nabla T}) \cos\varphi \\&+2R_{\textrm{PHE}}(2\cos^3\varphi-\cos\varphi)\frac{B^{FL}+B^{\textrm{Oe}}}{B_{\textrm{ext}}}.
\end{split}
\label{A8}
\end{eqnarray}
By symmetry, there are two contributions to the above signal, the first one is proportional to $\cos \varphi$ and depends on both the DL-SOT and thermal voltage, and the second one is proportional to $2R_{\textrm{PHE}}(2\cos^3 \varphi- \cos \varphi)$ and depends on the FL-SOT. The fields $B^{DL}$ and $B^{FL}$ are found using the fitting procedure described below.

\subsubsection{\label{APPENDIX:SOT:thermal} Determination of $R_{\nabla T}$}
We fit Eq.~\ref{A8} as a function of $\varphi$ by the sum of two functions proportional to $\cos \varphi$ and $(2\cos^3 \varphi- \cos \varphi)$ and obtain two coefficients, ($R_{\textrm{AHE}}\frac{B^{DL}}{{B_{\textrm{ext}}-B_{\textrm K}^{\textrm{eff}}}} + R_{\nabla T}$) and $\frac{B^{FL}+B^{\textrm{Oe}}}{B_{\textrm{ext}}}$. This procedure is repeated for each value of $B_{\textrm{ext}}$. Plotting these two coefficients as a function of $\frac{1}{B_{\textrm{ext}}-B_{\textrm K}^{\textrm{eff}}}$ and $\frac{1}{B_{\textrm{ext}}}$ yields a linear curve with slopes corresponding to $B^{DL}$ and $B^{FL}$, respectively. Finally, the intercept of ($R_{\textrm{AHE}}\frac{B^{DL}}{{B_{\textrm{ext}}-B_{\textrm K}^{\textrm{eff}}}} + R_{\nabla T}$) vs. $\frac{1}{B_{\textrm{ext}}-B_{\textrm K}^{\textrm{eff}}}$ gives the thermal resistance $R_{\nabla T}$ \cite{Can-thermal}. We note that here $B^{FL}$ is the sum of SOT and Oersted field contributions. The latter is calculated as ${B}^{\textrm{Oe}}=\mu_0j_{\textrm{Pt}}t_{\text{Pt}}/2$, where $j_{\textrm{Pt}}$ is the current density flowing in Pt and $t_{\text{Pt}}$ is the Pt thickness~\cite{Garello-nn-2013,Ghosh2017}. For a current density of $1.0 \times 10^7$~A/cm$^2$ in Pt(5)/Co(1)/AlO$_{x}$ we thus find ${B}^{\textrm{Oe}} = 0.31$~mT. Note that Fig.~6 in the main text includes ${B}^{\textrm{Oe}}$ as well.
\begin{figure}[t]
	\includegraphics[width=60mm]{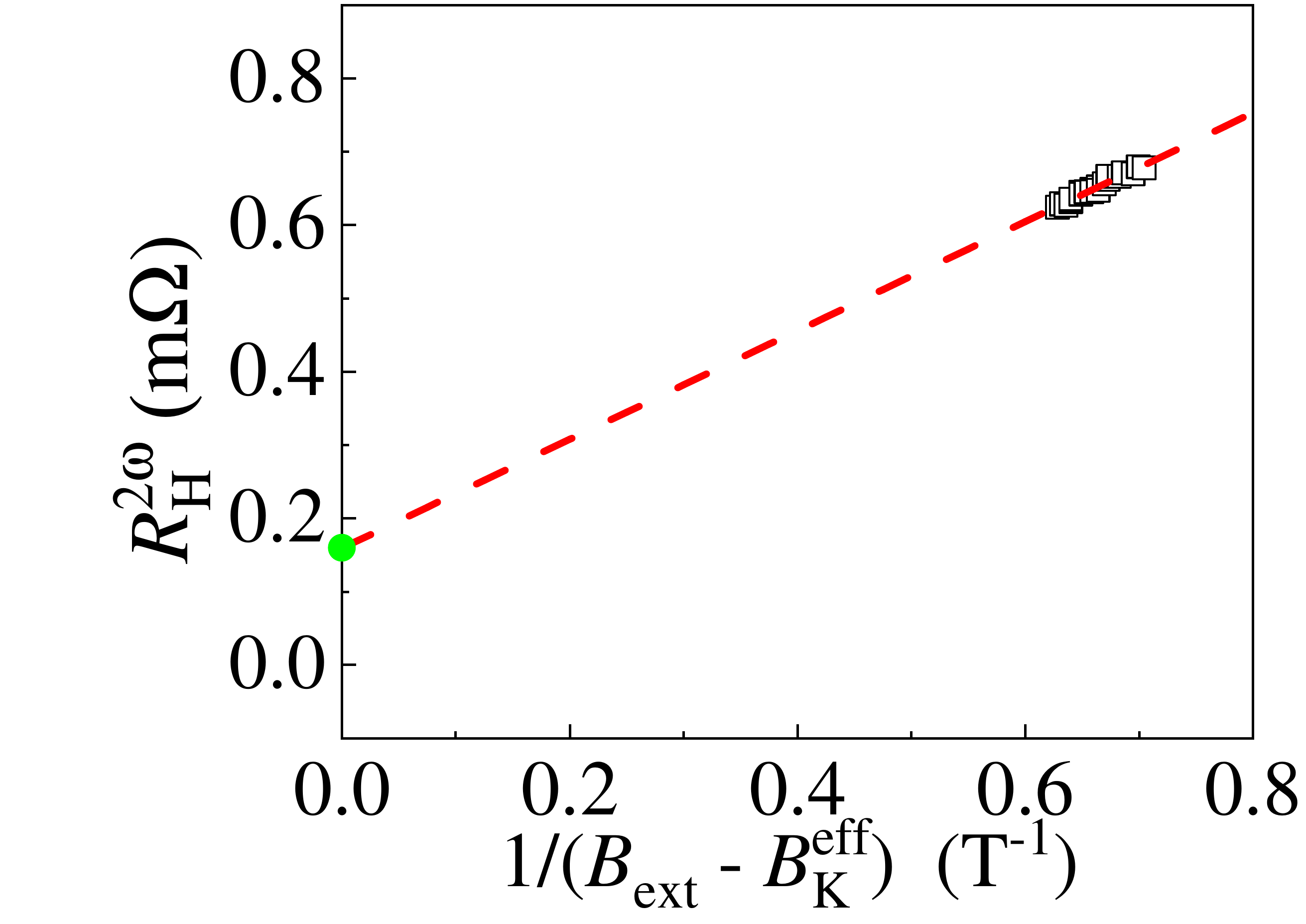}
	\caption{\label{FigS3} Separation of $B^{DL}$ and thermal signal from $R_H^{2\omega}$ for Pt(5)/Co(1)/AlO$_{\textrm{x}}$. Plot of the cosine component of $R_H^{2\omega}$ as a function of $\frac{1}{B_{\textrm{ext}}-B_{\textrm K}^{\textrm{eff}}}$ measured in a field scan performed at $\varphi=45^\circ$ with current density $1.0 \times 10^7$~A/cm$^2$. The intercept (green dot) of the linear fit is the thermal resistance $R_{\nabla T}$.}
\end{figure}

Equation~\ref{A8} and the analysis described above are equally valid for samples with PMA as long as the magnetization is fully saturated in-plane by applying an external field $B_{\textrm{ext}}$ larger than $B_{\textrm K}^{\textrm{eff}}$. Alternative to the angle scan measurements, which are time consuming, an easier way to determine the thermal resistance and $B^{DL}$ in samples with PMA is by performing an in-plane field sweep along $\varphi=45^\circ$ at magnetic saturation (i.e., $B_{\textrm{ext}} > B_{\textrm K}^{\textrm{eff}}$). In this measurement, the second term on the right hand side of Eq.~\ref{A8} vanishes by symmetry since $2\cos^3\varphi-\cos\varphi$ = 0. After subtracting a magnetically irrelevant offset, the data are plotted as a function of $\frac{1}{B_{\textrm{ext}}-B_{\textrm K}^{\textrm{eff}}}$, which allows one to obtain $B^{DL}$ and $R_{\nabla T}$ at $\varphi=45^\circ$, as shown in Fig.~\ref{FigS3}.

\begin{figure*}[h]
	\includegraphics[width=160mm]{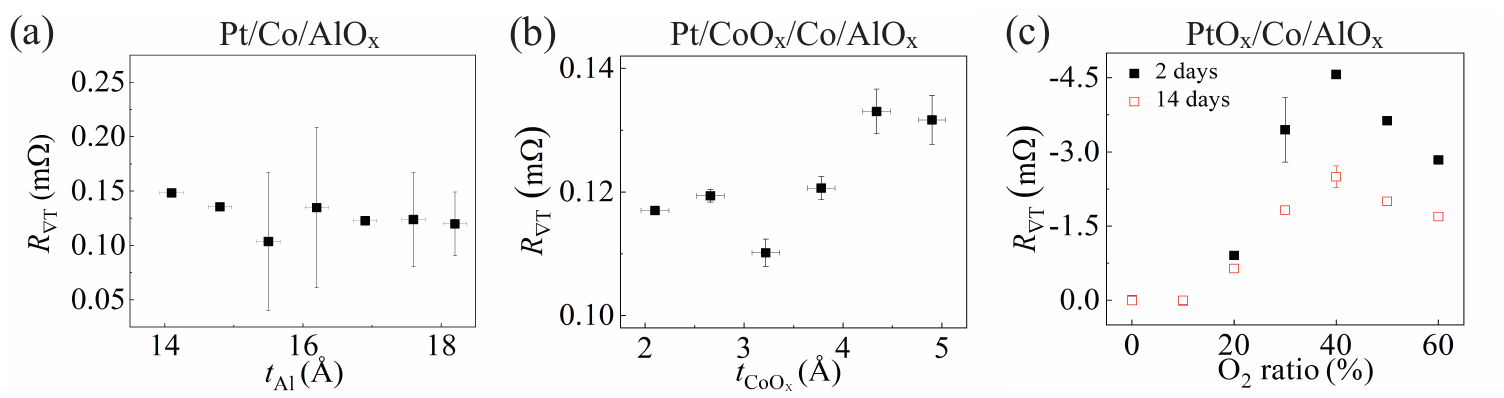}
	\caption{\label{FigS4} $R_{\nabla T}$ in (a) Pt/Co/AlO$_{\textrm{x}}$, (b) Pt/CoO$_{\textrm{x}}$/Co/AlO$_{\textrm{x}}$, and (c) PtO$_{\textrm{x}}$/Co/AlO$_{\textrm{x}}$. The measurements have been performed by using the in-plane field sweep method at $\varphi=45^\circ$ for the samples with PMA (a,b) and the angle scan method for the samples with in-plane magnetic anisotropy (c). The data shown as solid and open symbols in (c) were obtained, respectively, 2 and 14 days after the deposition of PtO$_{\textrm{x}}$/Co/AlO$_{\textrm{x}}$.}
\end{figure*}
Figure~\ref{FigS4} reports the values of $R_{\nabla T}$ in the three sample series. We observe that $R_{\nabla T}$ is approximately constant in Pt/Co/AlO$_{\textrm{x}}$, independently of $t_{\rm{Al}}$. This behavior differs from that of the AHE [see Fig.~4(d) in the main text], which is the electrical counterpart of the ANE. A possible reason for this discrepancy is the different direction of the electric field driving the AHE, which is in-plane, and the thermal gradient driving the ANE, which is out-of-plane. Moreover, since the current distribution is not significantly altered by the oxidation of the top interface, we expect that the thermal gradient remains approximately constant in this series. In Pt/CoO$_{\textrm{x}}$/Co/AlO$_{\textrm{x}}$, we find that $R_{\nabla T}$ varies by less than 20\% as a function of $t_{\rm{CoO_{\textrm{x}}}}$, similar to the AHE [see Fig.~4(e) in the main text]. This small variation likely reflects the fact that neither the current distribution nor the interfaces vary significantly in this system. On the other hand, $R_{\nabla T}$ changes strongly as a function of the O$_2$:(Ar+O$_2$) ratio in PtO$_{\textrm{x}}$/Co/AlO$_{\textrm{x}}$, closely reflecting the trend observed for the AHE [see Fig.~4(f) in the main text]. We believe that oxygen migration in this system disrupts the layered structure of the conducting elements, resulting in a more isotropic electrical and thermal conductivty compared to Pt/Co/AlO$_{\textrm{x}}$. Interestingly, the sign of $R_{\nabla T}$ is inverted in PtO$_{\textrm{x}}$/Co/AlO$_{\textrm{x}}$ relative to Pt/Co/AlO$_{\textrm{x}}$ and Pt/CoO$_{\textrm{x}}$/Co/AlO$_{\textrm{x}}$. We have no clear explanation for this effect, other than surmising the inversion of the thermal gradient in PtO$_{\textrm{x}}$/Co/AlO$_{\textrm{x}}$, which might be due to a change of the current distribution in the Pt and Co layers relative to Pt/Co/AlO$_{\textrm{x}}$ and Pt/CoO$_{\textrm{x}}$/Co/AlO$_{\textrm{x}}$.

\clearpage

\begin{thebibliography}{79}%
\makeatletter
\providecommand \@ifxundefined [1]{%
 \@ifx{#1\undefined}
}%
\providecommand \@ifnum [1]{%
 \ifnum #1\expandafter \@firstoftwo
 \else \expandafter \@secondoftwo
 \fi
}%
\providecommand \@ifx [1]{%
 \ifx #1\expandafter \@firstoftwo
 \else \expandafter \@secondoftwo
 \fi
}%
\providecommand \natexlab [1]{#1}%
\providecommand \enquote  [1]{``#1''}%
\providecommand \bibnamefont  [1]{#1}%
\providecommand \bibfnamefont [1]{#1}%
\providecommand \citenamefont [1]{#1}%
\providecommand \href@noop [0]{\@secondoftwo}%
\providecommand \href [0]{\begingroup \@sanitize@url \@href}%
\providecommand \@href[1]{\@@startlink{#1}\@@href}%
\providecommand \@@href[1]{\endgroup#1\@@endlink}%
\providecommand \@sanitize@url [0]{\catcode `\\12\catcode `\$12\catcode
  `\&12\catcode `\#12\catcode `\^12\catcode `\_12\catcode `\%12\relax}%
\providecommand \@@startlink[1]{}%
\providecommand \@@endlink[0]{}%
\providecommand \url  [0]{\begingroup\@sanitize@url \@url }%
\providecommand \@url [1]{\endgroup\@href {#1}{\urlprefix }}%
\providecommand \urlprefix  [0]{URL }%
\providecommand \Eprint [0]{\href }%
\providecommand \doibase [0]{https://doi.org/}%
\providecommand \selectlanguage [0]{\@gobble}%
\providecommand \bibinfo  [0]{\@secondoftwo}%
\providecommand \bibfield  [0]{\@secondoftwo}%
\providecommand \translation [1]{[#1]}%
\providecommand \BibitemOpen [0]{}%
\providecommand \bibitemStop [0]{}%
\providecommand \bibitemNoStop [0]{.\EOS\space}%
\providecommand \EOS [0]{\spacefactor3000\relax}%
\providecommand \BibitemShut  [1]{\csname bibitem#1\endcsname}%
\let\auto@bib@innerbib\@empty
\bibitem [{\citenamefont {Ikeda}\ \emph {et~al.}(2010)\citenamefont {Ikeda},
  \citenamefont {Miura}, \citenamefont {Yamamoto}, \citenamefont {Mizunuma},
  \citenamefont {Gan}, \citenamefont {Endo}, \citenamefont {Kanai},
  \citenamefont {Hayakawa}, \citenamefont {Matsukura},\ and\ \citenamefont
  {Ohno}}]{Mtjcofebperpen2010}%
  \BibitemOpen
  \bibfield  {author} {\bibinfo {author} {\bibfnamefont {S.}~\bibnamefont
  {Ikeda}}, \bibinfo {author} {\bibfnamefont {K.}~\bibnamefont {Miura}},
  \bibinfo {author} {\bibfnamefont {H.}~\bibnamefont {Yamamoto}}, \bibinfo
  {author} {\bibfnamefont {K.}~\bibnamefont {Mizunuma}}, \bibinfo {author}
  {\bibfnamefont {H.~D.}\ \bibnamefont {Gan}}, \bibinfo {author} {\bibfnamefont
  {M.}~\bibnamefont {Endo}}, \bibinfo {author} {\bibfnamefont {S.}~\bibnamefont
  {Kanai}}, \bibinfo {author} {\bibfnamefont {J.}~\bibnamefont {Hayakawa}},
  \bibinfo {author} {\bibfnamefont {F.}~\bibnamefont {Matsukura}},\ and\
  \bibinfo {author} {\bibfnamefont {H.}~\bibnamefont {Ohno}},\ }\bibfield
  {title} {\bibinfo {title} {A perpendicular-anisotropy $\mathrm{ CoFeB-MgO}$
  magnetic tunnel junction},\ }\href {https://doi.org/10.1038/nmat2804}
  {\bibfield  {journal} {\bibinfo  {journal} {Nat. Mater.}\ }\textbf {\bibinfo
  {volume} {9}},\ \bibinfo {pages} {721} (\bibinfo {year} {2010})}\BibitemShut
  {NoStop}%
\bibitem [{\citenamefont {Brataas}\ \emph {et~al.}(2012)\citenamefont
  {Brataas}, \citenamefont {Kent},\ and\ \citenamefont {Ohno}}]{Brataasnm2012}%
  \BibitemOpen
  \bibfield  {author} {\bibinfo {author} {\bibfnamefont {A.}~\bibnamefont
  {Brataas}}, \bibinfo {author} {\bibfnamefont {A.~D.}\ \bibnamefont {Kent}},\
  and\ \bibinfo {author} {\bibfnamefont {H.}~\bibnamefont {Ohno}},\ }\bibfield
  {title} {\bibinfo {title} {Current-induced torques in magnetic materials},\
  }\href {https://doi.org/10.1038/nmat3311} {\bibfield  {journal} {\bibinfo
  {journal} {Nat. Mater.}\ }\textbf {\bibinfo {volume} {11}},\ \bibinfo {pages}
  {372} (\bibinfo {year} {2012})}\BibitemShut {NoStop}%
\bibitem [{\citenamefont {Cubukcu}\ \emph {et~al.}(2014)\citenamefont
  {Cubukcu}, \citenamefont {Boulle}, \citenamefont {Drouard}, \citenamefont
  {Garello}, \citenamefont {O.~Avci}, \citenamefont {M.~Miron}, \citenamefont
  {Langer}, \citenamefont {Ocker}, \citenamefont {Gambardella},\ and\
  \citenamefont {Gaudin}}]{Cubukcuapl2014}%
  \BibitemOpen
  \bibfield  {author} {\bibinfo {author} {\bibfnamefont {M.}~\bibnamefont
  {Cubukcu}}, \bibinfo {author} {\bibfnamefont {O.}~\bibnamefont {Boulle}},
  \bibinfo {author} {\bibfnamefont {M.}~\bibnamefont {Drouard}}, \bibinfo
  {author} {\bibfnamefont {K.}~\bibnamefont {Garello}}, \bibinfo {author}
  {\bibfnamefont {C.}~\bibnamefont {O.~Avci}}, \bibinfo {author} {\bibfnamefont
  {I.}~\bibnamefont {M.~Miron}}, \bibinfo {author} {\bibfnamefont
  {J.}~\bibnamefont {Langer}}, \bibinfo {author} {\bibfnamefont
  {B.}~\bibnamefont {Ocker}}, \bibinfo {author} {\bibfnamefont
  {P.}~\bibnamefont {Gambardella}},\ and\ \bibinfo {author} {\bibfnamefont
  {G.}~\bibnamefont {Gaudin}},\ }\bibfield  {title} {\bibinfo {title}
  {Spin-orbit torque magnetization switching of a three-terminal perpendicular
  magnetic tunnel junction},\ }\href {https://doi.org/10.1063/1.4863407}
  {\bibfield  {journal} {\bibinfo  {journal} {Appl. Phys. Lett.}\ }\textbf
  {\bibinfo {volume} {104}},\ \bibinfo {pages} {042406} (\bibinfo {year}
  {2014})}\BibitemShut {NoStop}%
\bibitem [{\citenamefont {{Apalkov}}\ \emph {et~al.}(2016)\citenamefont
  {{Apalkov}}, \citenamefont {{Dieny}},\ and\ \citenamefont
  {{Slaughter}}}]{MRAM-IE-2016}%
  \BibitemOpen
  \bibfield  {author} {\bibinfo {author} {\bibfnamefont {D.}~\bibnamefont
  {{Apalkov}}}, \bibinfo {author} {\bibfnamefont {B.}~\bibnamefont {{Dieny}}},\
  and\ \bibinfo {author} {\bibfnamefont {J.~M.}\ \bibnamefont {{Slaughter}}},\
  }\bibfield  {title} {\bibinfo {title} {Magnetoresistive random access
  memory},\ }\href {https://doi.org/10.1109/JPROC.2016.2590142} {\bibfield
  {journal} {\bibinfo  {journal} {Proceedings of the IEEE}\ }\textbf {\bibinfo
  {volume} {104}},\ \bibinfo {pages} {1796} (\bibinfo {year}
  {2016})}\BibitemShut {NoStop}%
\bibitem [{\citenamefont {Monso}\ \emph {et~al.}(2002)\citenamefont {Monso},
  \citenamefont {Rodmacq}, \citenamefont {Auffret}, \citenamefont {Casali},
  \citenamefont {Fettar}, \citenamefont {Gilles}, \citenamefont {Dieny},\ and\
  \citenamefont {Boyer}}]{Monso-2002}%
  \BibitemOpen
  \bibfield  {author} {\bibinfo {author} {\bibfnamefont {S.}~\bibnamefont
  {Monso}}, \bibinfo {author} {\bibfnamefont {B.}~\bibnamefont {Rodmacq}},
  \bibinfo {author} {\bibfnamefont {S.}~\bibnamefont {Auffret}}, \bibinfo
  {author} {\bibfnamefont {G.}~\bibnamefont {Casali}}, \bibinfo {author}
  {\bibfnamefont {F.}~\bibnamefont {Fettar}}, \bibinfo {author} {\bibfnamefont
  {B.}~\bibnamefont {Gilles}}, \bibinfo {author} {\bibfnamefont
  {B.}~\bibnamefont {Dieny}},\ and\ \bibinfo {author} {\bibfnamefont
  {P.}~\bibnamefont {Boyer}},\ }\bibfield  {title} {\bibinfo {title} {Crossover
  from in-plane to perpendicular anisotropy in $\mathrm{Pt/CoFe/AlOx}$
  sandwiches as a function of al oxidation: A very accurate control of the
  oxidation of tunnel barriers},\ }\href {https://doi.org/10.1063/1.1483122}
  {\bibfield  {journal} {\bibinfo  {journal} {Appl. Phys. Lett.}\ }\textbf
  {\bibinfo {volume} {80}},\ \bibinfo {pages} {4157} (\bibinfo {year}
  {2002})}\BibitemShut {NoStop}%
\bibitem [{\citenamefont {{Pellegren}}\ and\ \citenamefont
  {{Sokalski}}(2015)}]{Sokalski2015}%
  \BibitemOpen
  \bibfield  {author} {\bibinfo {author} {\bibfnamefont {J.~P.}\ \bibnamefont
  {{Pellegren}}}\ and\ \bibinfo {author} {\bibfnamefont {V.~M.}\ \bibnamefont
  {{Sokalski}}},\ }\bibfield  {title} {\bibinfo {title} {Thickness and
  interface-dependent crystallization of {CoFeB} alloy thin films},\ }\href
  {https://doi.org/10.1109/TMAG.2015.2435798} {\bibfield  {journal} {\bibinfo
  {journal} {IEEE Transactions on Magnetics}\ }\textbf {\bibinfo {volume}
  {51}},\ \bibinfo {pages} {1} (\bibinfo {year} {2015})}\BibitemShut {NoStop}%
\bibitem [{\citenamefont {Baumgartner}\ \emph {et~al.}(2017)\citenamefont
  {Baumgartner}, \citenamefont {Garello}, \citenamefont {Mendil}, \citenamefont
  {Avci}, \citenamefont {Grimaldi}, \citenamefont {Murer}, \citenamefont
  {Feng}, \citenamefont {Gabureac}, \citenamefont {Stamm}, \citenamefont
  {Acremann}, \citenamefont {Finizio}, \citenamefont {Wintz}, \citenamefont
  {Raabe},\ and\ \citenamefont {Gambardella}}]{Manuel-nn}%
  \BibitemOpen
  \bibfield  {author} {\bibinfo {author} {\bibfnamefont {M.}~\bibnamefont
  {Baumgartner}}, \bibinfo {author} {\bibfnamefont {K.}~\bibnamefont
  {Garello}}, \bibinfo {author} {\bibfnamefont {J.}~\bibnamefont {Mendil}},
  \bibinfo {author} {\bibfnamefont {C.~O.}\ \bibnamefont {Avci}}, \bibinfo
  {author} {\bibfnamefont {E.}~\bibnamefont {Grimaldi}}, \bibinfo {author}
  {\bibfnamefont {C.}~\bibnamefont {Murer}}, \bibinfo {author} {\bibfnamefont
  {J.}~\bibnamefont {Feng}}, \bibinfo {author} {\bibfnamefont {M.}~\bibnamefont
  {Gabureac}}, \bibinfo {author} {\bibfnamefont {C.}~\bibnamefont {Stamm}},
  \bibinfo {author} {\bibfnamefont {Y.}~\bibnamefont {Acremann}}, \bibinfo
  {author} {\bibfnamefont {S.}~\bibnamefont {Finizio}}, \bibinfo {author}
  {\bibfnamefont {S.}~\bibnamefont {Wintz}}, \bibinfo {author} {\bibfnamefont
  {J.}~\bibnamefont {Raabe}},\ and\ \bibinfo {author} {\bibfnamefont
  {P.}~\bibnamefont {Gambardella}},\ }\bibfield  {title} {\bibinfo {title}
  {Spatially and time-resolved magnetization dynamics driven by spin-orbit
  torques},\ }\href {https://doi.org/10.1038/nnano.2017.151} {\bibfield
  {journal} {\bibinfo  {journal} {Nat. Nanotech.}\ }\textbf {\bibinfo {volume}
  {12}},\ \bibinfo {pages} {980} (\bibinfo {year} {2017})}\BibitemShut
  {NoStop}%
\bibitem [{\citenamefont {{Jan}}\ \emph {et~al.}(2014)\citenamefont {{Jan}},
  \citenamefont {{Thomas}}, \citenamefont {{Le}}, \citenamefont {{Lee}},
  \citenamefont {{Liu}}, \citenamefont {{Zhu}}, \citenamefont {{Tong}},
  \citenamefont {{Pi}}, \citenamefont {{Wang}}, \citenamefont {{Shen}},
  \citenamefont {{He}}, \citenamefont {{Haq}}, \citenamefont {{Teng}},
  \citenamefont {{Lam}}, \citenamefont {{Huang}}, \citenamefont {{Zhong}},
  \citenamefont {{Torng}},\ and\ \citenamefont {{Wang}}}]{Jan2014}%
  \BibitemOpen
  \bibfield  {author} {\bibinfo {author} {\bibfnamefont {G.}~\bibnamefont
  {{Jan}}}, \bibinfo {author} {\bibfnamefont {L.}~\bibnamefont {{Thomas}}},
  \bibinfo {author} {\bibfnamefont {S.}~\bibnamefont {{Le}}}, \bibinfo {author}
  {\bibfnamefont {Y.}~\bibnamefont {{Lee}}}, \bibinfo {author} {\bibfnamefont
  {H.}~\bibnamefont {{Liu}}}, \bibinfo {author} {\bibfnamefont
  {J.}~\bibnamefont {{Zhu}}}, \bibinfo {author} {\bibfnamefont
  {R.}~\bibnamefont {{Tong}}}, \bibinfo {author} {\bibfnamefont
  {K.}~\bibnamefont {{Pi}}}, \bibinfo {author} {\bibfnamefont {Y.}~\bibnamefont
  {{Wang}}}, \bibinfo {author} {\bibfnamefont {D.}~\bibnamefont {{Shen}}},
  \bibinfo {author} {\bibfnamefont {R.}~\bibnamefont {{He}}}, \bibinfo {author}
  {\bibfnamefont {J.}~\bibnamefont {{Haq}}}, \bibinfo {author} {\bibfnamefont
  {J.}~\bibnamefont {{Teng}}}, \bibinfo {author} {\bibfnamefont
  {V.}~\bibnamefont {{Lam}}}, \bibinfo {author} {\bibfnamefont
  {K.}~\bibnamefont {{Huang}}}, \bibinfo {author} {\bibfnamefont
  {T.}~\bibnamefont {{Zhong}}}, \bibinfo {author} {\bibfnamefont
  {T.}~\bibnamefont {{Torng}}},\ and\ \bibinfo {author} {\bibfnamefont
  {P.}~\bibnamefont {{Wang}}},\ }\bibfield  {title} {\bibinfo {title}
  {Demonstration of fully functional 8mb perpendicular stt-mram chips with
  sub-5ns writing for non-volatile embedded memories},\ }in\ \href
  {https://doi.org/10.1109/VLSIT.2014.6894357} {\emph {\bibinfo {booktitle}
  {2014 Symposium on VLSI Technology (VLSI-Technology): Digest of Technical
  Papers}}}\ (\bibinfo {year} {2014})\ pp.\ \bibinfo {pages} {1--2}\BibitemShut
  {NoStop}%
\bibitem [{\citenamefont {{Garello}}\ \emph {et~al.}(2018)\citenamefont
  {{Garello}}, \citenamefont {{Yasin}}, \citenamefont {{Couet}}, \citenamefont
  {{Souriau}}, \citenamefont {{Swerts}}, \citenamefont {{Rao}}, \citenamefont
  {{Van Beek}}, \citenamefont {{Kim}}, \citenamefont {{Liu}}, \citenamefont
  {{Kundu}}, \citenamefont {{Tsvetanova}}, \citenamefont {{Croes}},
  \citenamefont {{Jossart}}, \citenamefont {{Grimaldi}}, \citenamefont
  {{Baumgartner}}, \citenamefont {{Crotti}}, \citenamefont {{Fumémont}},
  \citenamefont {{Gambardella}},\ and\ \citenamefont {{Kar}}}]{Garello-ie2018}%
  \BibitemOpen
  \bibfield  {author} {\bibinfo {author} {\bibfnamefont {K.}~\bibnamefont
  {{Garello}}}, \bibinfo {author} {\bibfnamefont {F.}~\bibnamefont {{Yasin}}},
  \bibinfo {author} {\bibfnamefont {S.}~\bibnamefont {{Couet}}}, \bibinfo
  {author} {\bibfnamefont {L.}~\bibnamefont {{Souriau}}}, \bibinfo {author}
  {\bibfnamefont {J.}~\bibnamefont {{Swerts}}}, \bibinfo {author}
  {\bibfnamefont {S.}~\bibnamefont {{Rao}}}, \bibinfo {author} {\bibfnamefont
  {S.}~\bibnamefont {{Van Beek}}}, \bibinfo {author} {\bibfnamefont
  {W.}~\bibnamefont {{Kim}}}, \bibinfo {author} {\bibfnamefont
  {E.}~\bibnamefont {{Liu}}}, \bibinfo {author} {\bibfnamefont
  {S.}~\bibnamefont {{Kundu}}}, \bibinfo {author} {\bibfnamefont
  {D.}~\bibnamefont {{Tsvetanova}}}, \bibinfo {author} {\bibfnamefont
  {K.}~\bibnamefont {{Croes}}}, \bibinfo {author} {\bibfnamefont
  {N.}~\bibnamefont {{Jossart}}}, \bibinfo {author} {\bibfnamefont
  {E.}~\bibnamefont {{Grimaldi}}}, \bibinfo {author} {\bibfnamefont
  {M.}~\bibnamefont {{Baumgartner}}}, \bibinfo {author} {\bibfnamefont
  {D.}~\bibnamefont {{Crotti}}}, \bibinfo {author} {\bibfnamefont
  {A.}~\bibnamefont {{Fumémont}}}, \bibinfo {author} {\bibfnamefont
  {P.}~\bibnamefont {{Gambardella}}},\ and\ \bibinfo {author} {\bibfnamefont
  {G.~S.}\ \bibnamefont {{Kar}}},\ }\bibfield  {title} {\bibinfo {title}
  {Sot-mram 300mm integration for low power and ultrafast embedded memories},\
  }in\ \href {https://doi.org/10.1109/VLSIC.2018.8502269} {\emph {\bibinfo
  {booktitle} {2018 IEEE Symposium on VLSI Circuits}}}\ (\bibinfo {year}
  {2018})\ pp.\ \bibinfo {pages} {81--82}\BibitemShut {NoStop}%
\bibitem [{\citenamefont {Miron}\ \emph {et~al.}(2010)\citenamefont {Miron},
  \citenamefont {Gaudin}, \citenamefont {Auffret}, \citenamefont {Rodmacq},
  \citenamefont {Schuhl}, \citenamefont {Pizzini}, \citenamefont {Vogel},\ and\
  \citenamefont {Gambardella}}]{Miron-rashba-2010}%
  \BibitemOpen
  \bibfield  {author} {\bibinfo {author} {\bibfnamefont {I.~M.}\ \bibnamefont
  {Miron}}, \bibinfo {author} {\bibfnamefont {G.}~\bibnamefont {Gaudin}},
  \bibinfo {author} {\bibfnamefont {S.}~\bibnamefont {Auffret}}, \bibinfo
  {author} {\bibfnamefont {B.}~\bibnamefont {Rodmacq}}, \bibinfo {author}
  {\bibfnamefont {A.}~\bibnamefont {Schuhl}}, \bibinfo {author} {\bibfnamefont
  {S.}~\bibnamefont {Pizzini}}, \bibinfo {author} {\bibfnamefont
  {J.}~\bibnamefont {Vogel}},\ and\ \bibinfo {author} {\bibfnamefont
  {P.}~\bibnamefont {Gambardella}},\ }\bibfield  {title} {\bibinfo {title}
  {Current-driven spin torque induced by the rashba effect in a ferromagnetic
  metal layer},\ }\href {https://doi.org/10.1038/nmat2613} {\bibfield
  {journal} {\bibinfo  {journal} {Nat. Mater.}\ }\textbf {\bibinfo {volume}
  {9}},\ \bibinfo {pages} {230} (\bibinfo {year} {2010})}\BibitemShut {NoStop}%
\bibitem [{\citenamefont {Miron}\ \emph {et~al.}(2011)\citenamefont {Miron},
  \citenamefont {Garello}, \citenamefont {Gaudin}, \citenamefont {Zermatten},
  \citenamefont {Costache}, \citenamefont {Auffret}, \citenamefont {Bandiera},
  \citenamefont {Rodmacq}, \citenamefont {Schuhl},\ and\ \citenamefont
  {Gambardella}}]{Miron-switching2011}%
  \BibitemOpen
  \bibfield  {author} {\bibinfo {author} {\bibfnamefont {I.~M.}\ \bibnamefont
  {Miron}}, \bibinfo {author} {\bibfnamefont {K.}~\bibnamefont {Garello}},
  \bibinfo {author} {\bibfnamefont {G.}~\bibnamefont {Gaudin}}, \bibinfo
  {author} {\bibfnamefont {P.~J.}\ \bibnamefont {Zermatten}}, \bibinfo {author}
  {\bibfnamefont {M.~V.}\ \bibnamefont {Costache}}, \bibinfo {author}
  {\bibfnamefont {S.}~\bibnamefont {Auffret}}, \bibinfo {author} {\bibfnamefont
  {S.}~\bibnamefont {Bandiera}}, \bibinfo {author} {\bibfnamefont
  {B.}~\bibnamefont {Rodmacq}}, \bibinfo {author} {\bibfnamefont
  {A.}~\bibnamefont {Schuhl}},\ and\ \bibinfo {author} {\bibfnamefont
  {P.}~\bibnamefont {Gambardella}},\ }\bibfield  {title} {\bibinfo {title}
  {Perpendicular switching of a single ferromagnetic layer induced by in-plane
  current injection},\ }\href {https://doi.org/10.1038/nature10309} {\bibfield
  {journal} {\bibinfo  {journal} {Nature}\ }\textbf {\bibinfo {volume} {476}},\
  \bibinfo {pages} {189} (\bibinfo {year} {2011})}\BibitemShut {NoStop}%
\bibitem [{\citenamefont {O.~Avci}\ \emph {et~al.}(2012)\citenamefont
  {O.~Avci}, \citenamefont {Garello}, \citenamefont {M.~Miron}, \citenamefont
  {Gaudin}, \citenamefont {Auffret}, \citenamefont {Boulle},\ and\
  \citenamefont {Gambardella}}]{Avci2012}%
  \BibitemOpen
  \bibfield  {author} {\bibinfo {author} {\bibfnamefont {C.}~\bibnamefont
  {O.~Avci}}, \bibinfo {author} {\bibfnamefont {K.}~\bibnamefont {Garello}},
  \bibinfo {author} {\bibfnamefont {I.}~\bibnamefont {M.~Miron}}, \bibinfo
  {author} {\bibfnamefont {G.}~\bibnamefont {Gaudin}}, \bibinfo {author}
  {\bibfnamefont {S.}~\bibnamefont {Auffret}}, \bibinfo {author} {\bibfnamefont
  {O.}~\bibnamefont {Boulle}},\ and\ \bibinfo {author} {\bibfnamefont
  {P.}~\bibnamefont {Gambardella}},\ }\bibfield  {title} {\bibinfo {title}
  {Magnetization switching of an $\mathrm{MgO/Co/Pt}$ layer by in-plane current
  injection},\ }\href {https://doi.org/10.1063/1.4719677} {\bibfield  {journal}
  {\bibinfo  {journal} {Appl. Phys. Lett.}\ }\textbf {\bibinfo {volume}
  {100}},\ \bibinfo {pages} {212404} (\bibinfo {year} {2012})}\BibitemShut
  {NoStop}%
\bibitem [{\citenamefont {Liu}\ \emph {et~al.}(2012)\citenamefont {Liu},
  \citenamefont {Pai}, \citenamefont {Li}, \citenamefont {Tseng}, \citenamefont
  {Ralph},\ and\ \citenamefont {Buhrman}}]{Liu-science2012}%
  \BibitemOpen
  \bibfield  {author} {\bibinfo {author} {\bibfnamefont {L.}~\bibnamefont
  {Liu}}, \bibinfo {author} {\bibfnamefont {C.-F.}\ \bibnamefont {Pai}},
  \bibinfo {author} {\bibfnamefont {Y.}~\bibnamefont {Li}}, \bibinfo {author}
  {\bibfnamefont {H.~W.}\ \bibnamefont {Tseng}}, \bibinfo {author}
  {\bibfnamefont {D.~C.}\ \bibnamefont {Ralph}},\ and\ \bibinfo {author}
  {\bibfnamefont {R.~A.}\ \bibnamefont {Buhrman}},\ }\bibfield  {title}
  {\bibinfo {title} {Spin-torque switching with the giant spin hall effect of
  tantalum},\ }\href {https://doi.org/10.1126/science.1218197} {\bibfield
  {journal} {\bibinfo  {journal} {Science}\ }\textbf {\bibinfo {volume}
  {336}},\ \bibinfo {pages} {555} (\bibinfo {year} {2012})}\BibitemShut
  {NoStop}%
\bibitem [{\citenamefont {Garello}\ \emph {et~al.}(2013)\citenamefont
  {Garello}, \citenamefont {Miron}, \citenamefont {Avci}, \citenamefont
  {Freimuth}, \citenamefont {Mokrousov}, \citenamefont {Blugel}, \citenamefont
  {Auffret}, \citenamefont {Boulle}, \citenamefont {Gaudin},\ and\
  \citenamefont {Gambardella}}]{Garello-nn-2013}%
  \BibitemOpen
  \bibfield  {author} {\bibinfo {author} {\bibfnamefont {K.}~\bibnamefont
  {Garello}}, \bibinfo {author} {\bibfnamefont {I.~M.}\ \bibnamefont {Miron}},
  \bibinfo {author} {\bibfnamefont {C.~O.}\ \bibnamefont {Avci}}, \bibinfo
  {author} {\bibfnamefont {F.}~\bibnamefont {Freimuth}}, \bibinfo {author}
  {\bibfnamefont {Y.}~\bibnamefont {Mokrousov}}, \bibinfo {author}
  {\bibfnamefont {S.}~\bibnamefont {Blugel}}, \bibinfo {author} {\bibfnamefont
  {S.}~\bibnamefont {Auffret}}, \bibinfo {author} {\bibfnamefont
  {O.}~\bibnamefont {Boulle}}, \bibinfo {author} {\bibfnamefont
  {G.}~\bibnamefont {Gaudin}},\ and\ \bibinfo {author} {\bibfnamefont
  {P.}~\bibnamefont {Gambardella}},\ }\bibfield  {title} {\bibinfo {title}
  {Symmetry and magnitude of spin-orbit torques in ferromagnetic
  heterostructures},\ }\href {https://doi.org/10.1038/nnano.2013.145}
  {\bibfield  {journal} {\bibinfo  {journal} {Nat. Nanotech.}\ }\textbf
  {\bibinfo {volume} {8}},\ \bibinfo {pages} {587} (\bibinfo {year}
  {2013})}\BibitemShut {NoStop}%
\bibitem [{\citenamefont {Kim}\ \emph {et~al.}(2013)\citenamefont {Kim},
  \citenamefont {Sinha}, \citenamefont {Hayashi}, \citenamefont {Yamanouchi},
  \citenamefont {Fukami}, \citenamefont {Suzuki}, \citenamefont {Mitani},\ and\
  \citenamefont {Ohno}}]{Hayashi-nm2013}%
  \BibitemOpen
  \bibfield  {author} {\bibinfo {author} {\bibfnamefont {J.}~\bibnamefont
  {Kim}}, \bibinfo {author} {\bibfnamefont {J.}~\bibnamefont {Sinha}}, \bibinfo
  {author} {\bibfnamefont {M.}~\bibnamefont {Hayashi}}, \bibinfo {author}
  {\bibfnamefont {M.}~\bibnamefont {Yamanouchi}}, \bibinfo {author}
  {\bibfnamefont {S.}~\bibnamefont {Fukami}}, \bibinfo {author} {\bibfnamefont
  {T.}~\bibnamefont {Suzuki}}, \bibinfo {author} {\bibfnamefont
  {S.}~\bibnamefont {Mitani}},\ and\ \bibinfo {author} {\bibfnamefont
  {H.}~\bibnamefont {Ohno}},\ }\bibfield  {title} {\bibinfo {title} {Layer
  thickness dependence of the current-induced effective field vector in
  $\mathrm{Ta|CoFeB|MgO}$},\ }\href {https://doi.org/10.1038/nmat3522}
  {\bibfield  {journal} {\bibinfo  {journal} {Nat. Mater.}\ }\textbf {\bibinfo
  {volume} {12}},\ \bibinfo {pages} {240} (\bibinfo {year} {2013})}\BibitemShut
  {NoStop}%
\bibitem [{\citenamefont {{Cubukcu}}\ \emph {et~al.}(2018)\citenamefont
  {{Cubukcu}}, \citenamefont {{Boulle}}, \citenamefont {{Mikuszeit}},
  \citenamefont {{Hamelin}}, \citenamefont {{Brächer}}, \citenamefont
  {{Lamard}}, \citenamefont {{Cyrille}}, \citenamefont {{Buda-Prejbeanu}},
  \citenamefont {{Garello}}, \citenamefont {{Miron}}, \citenamefont {{Klein}},
  \citenamefont {{de Loubens}}, \citenamefont {{Naletov}}, \citenamefont
  {{Langer}}, \citenamefont {{Ocker}}, \citenamefont {{Gambardella}},\ and\
  \citenamefont {{Gaudin}}}]{Cubukcu2018}%
  \BibitemOpen
  \bibfield  {author} {\bibinfo {author} {\bibfnamefont {M.}~\bibnamefont
  {{Cubukcu}}}, \bibinfo {author} {\bibfnamefont {O.}~\bibnamefont {{Boulle}}},
  \bibinfo {author} {\bibfnamefont {N.}~\bibnamefont {{Mikuszeit}}}, \bibinfo
  {author} {\bibfnamefont {C.}~\bibnamefont {{Hamelin}}}, \bibinfo {author}
  {\bibfnamefont {T.}~\bibnamefont {{Brächer}}}, \bibinfo {author}
  {\bibfnamefont {N.}~\bibnamefont {{Lamard}}}, \bibinfo {author}
  {\bibfnamefont {M.}~\bibnamefont {{Cyrille}}}, \bibinfo {author}
  {\bibfnamefont {L.}~\bibnamefont {{Buda-Prejbeanu}}}, \bibinfo {author}
  {\bibfnamefont {K.}~\bibnamefont {{Garello}}}, \bibinfo {author}
  {\bibfnamefont {I.~M.}\ \bibnamefont {{Miron}}}, \bibinfo {author}
  {\bibfnamefont {O.}~\bibnamefont {{Klein}}}, \bibinfo {author} {\bibfnamefont
  {G.}~\bibnamefont {{de Loubens}}}, \bibinfo {author} {\bibfnamefont {V.~V.}\
  \bibnamefont {{Naletov}}}, \bibinfo {author} {\bibfnamefont {J.}~\bibnamefont
  {{Langer}}}, \bibinfo {author} {\bibfnamefont {B.}~\bibnamefont {{Ocker}}},
  \bibinfo {author} {\bibfnamefont {P.}~\bibnamefont {{Gambardella}}},\ and\
  \bibinfo {author} {\bibfnamefont {G.}~\bibnamefont {{Gaudin}}},\ }\bibfield
  {title} {\bibinfo {title} {Ultra-fast perpendicular spin–orbit torque
  mram},\ }\href {https://doi.org/10.1109/TMAG.2017.2772185} {\bibfield
  {journal} {\bibinfo  {journal} {IEEE Transactions on Magnetics}\ }\textbf
  {\bibinfo {volume} {54}},\ \bibinfo {pages} {1} (\bibinfo {year}
  {2018})}\BibitemShut {NoStop}%
\bibitem [{\citenamefont {Manchon}\ \emph {et~al.}(2019)\citenamefont
  {Manchon}, \citenamefont {\ifmmode~\check{Z}\else \v{Z}\fi{}elezn\'y},
  \citenamefont {Miron}, \citenamefont {Jungwirth}, \citenamefont {Sinova},
  \citenamefont {Thiaville}, \citenamefont {Garello},\ and\ \citenamefont
  {Gambardella}}]{Manchon-rmp-2019}%
  \BibitemOpen
  \bibfield  {author} {\bibinfo {author} {\bibfnamefont {A.}~\bibnamefont
  {Manchon}}, \bibinfo {author} {\bibfnamefont {J.}~\bibnamefont
  {\ifmmode~\check{Z}\else \v{Z}\fi{}elezn\'y}}, \bibinfo {author}
  {\bibfnamefont {I.~M.}\ \bibnamefont {Miron}}, \bibinfo {author}
  {\bibfnamefont {T.}~\bibnamefont {Jungwirth}}, \bibinfo {author}
  {\bibfnamefont {J.}~\bibnamefont {Sinova}}, \bibinfo {author} {\bibfnamefont
  {A.}~\bibnamefont {Thiaville}}, \bibinfo {author} {\bibfnamefont
  {K.}~\bibnamefont {Garello}},\ and\ \bibinfo {author} {\bibfnamefont
  {P.}~\bibnamefont {Gambardella}},\ }\bibfield  {title} {\bibinfo {title}
  {Current-induced spin-orbit torques in ferromagnetic and antiferromagnetic
  systems},\ }\href {https://doi.org/10.1103/RevModPhys.91.035004} {\bibfield
  {journal} {\bibinfo  {journal} {Rev. Mod. Phys.}\ }\textbf {\bibinfo {volume}
  {91}},\ \bibinfo {pages} {035004} (\bibinfo {year} {2019})}\BibitemShut
  {NoStop}%
\bibitem [{\citenamefont {Zhang}\ \emph {et~al.}(2002)\citenamefont {Zhang},
  \citenamefont {Levy},\ and\ \citenamefont {Fert}}]{Zhang2002a}%
  \BibitemOpen
  \bibfield  {author} {\bibinfo {author} {\bibfnamefont {S.}~\bibnamefont
  {Zhang}}, \bibinfo {author} {\bibfnamefont {P.~M.}\ \bibnamefont {Levy}},\
  and\ \bibinfo {author} {\bibfnamefont {A.}~\bibnamefont {Fert}},\ }\bibfield
  {title} {\bibinfo {title} {Mechanisms of spin-polarized current-driven
  magnetization switching},\ }\href
  {https://doi.org/10.1103/PhysRevLett.88.236601} {\bibfield  {journal}
  {\bibinfo  {journal} {Phys. Rev. Lett.}\ }\textbf {\bibinfo {volume} {88}},\
  \bibinfo {pages} {236601} (\bibinfo {year} {2002})}\BibitemShut {NoStop}%
\bibitem [{\citenamefont {Manchon}\ \emph {et~al.}(2008)\citenamefont
  {Manchon}, \citenamefont {Ducruet}, \citenamefont {Lombard}, \citenamefont
  {Auffret}, \citenamefont {Rodmacq}, \citenamefont {Dieny}, \citenamefont
  {Pizzini}, \citenamefont {Vogel}, \citenamefont {Uhlíř}, \citenamefont
  {Hochstrasser},\ and\ \citenamefont {Panaccione}}]{Manchon-jap-2008}%
  \BibitemOpen
  \bibfield  {author} {\bibinfo {author} {\bibfnamefont {A.}~\bibnamefont
  {Manchon}}, \bibinfo {author} {\bibfnamefont {C.}~\bibnamefont {Ducruet}},
  \bibinfo {author} {\bibfnamefont {L.}~\bibnamefont {Lombard}}, \bibinfo
  {author} {\bibfnamefont {S.}~\bibnamefont {Auffret}}, \bibinfo {author}
  {\bibfnamefont {B.}~\bibnamefont {Rodmacq}}, \bibinfo {author} {\bibfnamefont
  {B.}~\bibnamefont {Dieny}}, \bibinfo {author} {\bibfnamefont
  {S.}~\bibnamefont {Pizzini}}, \bibinfo {author} {\bibfnamefont
  {J.}~\bibnamefont {Vogel}}, \bibinfo {author} {\bibfnamefont
  {V.}~\bibnamefont {Uhlíř}}, \bibinfo {author} {\bibfnamefont
  {M.}~\bibnamefont {Hochstrasser}},\ and\ \bibinfo {author} {\bibfnamefont
  {G.}~\bibnamefont {Panaccione}},\ }\bibfield  {title} {\bibinfo {title}
  {Analysis of oxygen induced anisotropy crossover in trilayers
  $\mathrm{Pt/Co/MO_x}$},\ }\href {https://doi.org/10.1063/1.2969711}
  {\bibfield  {journal} {\bibinfo  {journal} {J. Appl. Phys.}\ }\textbf
  {\bibinfo {volume} {104}},\ \bibinfo {pages} {043914} (\bibinfo {year}
  {2008})}\BibitemShut {NoStop}%
\bibitem [{\citenamefont {Rodmacq}\ \emph {et~al.}(2009)\citenamefont
  {Rodmacq}, \citenamefont {Manchon}, \citenamefont {Ducruet}, \citenamefont
  {Auffret},\ and\ \citenamefont {Dieny}}]{Rodmacq2009}%
  \BibitemOpen
  \bibfield  {author} {\bibinfo {author} {\bibfnamefont {B.}~\bibnamefont
  {Rodmacq}}, \bibinfo {author} {\bibfnamefont {A.}~\bibnamefont {Manchon}},
  \bibinfo {author} {\bibfnamefont {C.}~\bibnamefont {Ducruet}}, \bibinfo
  {author} {\bibfnamefont {S.}~\bibnamefont {Auffret}},\ and\ \bibinfo {author}
  {\bibfnamefont {B.}~\bibnamefont {Dieny}},\ }\bibfield  {title} {\bibinfo
  {title} {Influence of thermal annealing on the perpendicular magnetic
  anisotropy of $\mathrm{Pt/Co/AlOx}$ trilayers},\ }\href
  {https://doi.org/10.1103/PhysRevB.79.024423} {\bibfield  {journal} {\bibinfo
  {journal} {Phys. Rev. B}\ }\textbf {\bibinfo {volume} {79}},\ \bibinfo
  {pages} {024423} (\bibinfo {year} {2009})}\BibitemShut {NoStop}%
\bibitem [{\citenamefont {Chiba}\ \emph {et~al.}(2011)\citenamefont {Chiba},
  \citenamefont {Fukami}, \citenamefont {Shimamura}, \citenamefont {Ishiwata},
  \citenamefont {Kobayashi},\ and\ \citenamefont {Ono}}]{Chiba2011}%
  \BibitemOpen
  \bibfield  {author} {\bibinfo {author} {\bibfnamefont {D.}~\bibnamefont
  {Chiba}}, \bibinfo {author} {\bibfnamefont {S.}~\bibnamefont {Fukami}},
  \bibinfo {author} {\bibfnamefont {K.}~\bibnamefont {Shimamura}}, \bibinfo
  {author} {\bibfnamefont {N.}~\bibnamefont {Ishiwata}}, \bibinfo {author}
  {\bibfnamefont {K.}~\bibnamefont {Kobayashi}},\ and\ \bibinfo {author}
  {\bibfnamefont {T.}~\bibnamefont {Ono}},\ }\bibfield  {title} {\bibinfo
  {title} {Electrical control of the ferromagnetic phase transition in cobalt
  at room temperature},\ }\href {https://doi.org/10.1038/nmat3130} {\bibfield
  {journal} {\bibinfo  {journal} {Nat. Mater.}\ }\textbf {\bibinfo {volume}
  {10}},\ \bibinfo {pages} {853} (\bibinfo {year} {2011})}\BibitemShut
  {NoStop}%
\bibitem [{\citenamefont {Garad}\ \emph {et~al.}(2013)\citenamefont {Garad},
  \citenamefont {Ortega}, \citenamefont {Ramos}, \citenamefont {Joly},
  \citenamefont {Fettar}, \citenamefont {Auffret}, \citenamefont {Rodmacq},
  \citenamefont {Diény}, \citenamefont {Proux},\ and\ \citenamefont
  {Erko}}]{Garad2013}%
  \BibitemOpen
  \bibfield  {author} {\bibinfo {author} {\bibfnamefont {H.}~\bibnamefont
  {Garad}}, \bibinfo {author} {\bibfnamefont {L.}~\bibnamefont {Ortega}},
  \bibinfo {author} {\bibfnamefont {A.~Y.}\ \bibnamefont {Ramos}}, \bibinfo
  {author} {\bibfnamefont {Y.}~\bibnamefont {Joly}}, \bibinfo {author}
  {\bibfnamefont {F.}~\bibnamefont {Fettar}}, \bibinfo {author} {\bibfnamefont
  {S.}~\bibnamefont {Auffret}}, \bibinfo {author} {\bibfnamefont
  {B.}~\bibnamefont {Rodmacq}}, \bibinfo {author} {\bibfnamefont
  {B.}~\bibnamefont {Diény}}, \bibinfo {author} {\bibfnamefont
  {O.}~\bibnamefont {Proux}},\ and\ \bibinfo {author} {\bibfnamefont {A.~I.}\
  \bibnamefont {Erko}},\ }\bibfield  {title} {\bibinfo {title} {Competition
  between $\mathrm{CoOx}$ and $\mathrm{CoPt}$ phases in $\mathrm{Pt/Co/AlOx}$
  semi tunnel junctions},\ }\href {https://doi.org/10.1063/1.4816620}
  {\bibfield  {journal} {\bibinfo  {journal} {J. Appl. Phys.}\ }\textbf
  {\bibinfo {volume} {114}},\ \bibinfo {pages} {053508} (\bibinfo {year}
  {2013})}\BibitemShut {NoStop}%
\bibitem [{\citenamefont {Yu}\ \emph {et~al.}(2014{\natexlab{a}})\citenamefont
  {Yu}, \citenamefont {Upadhyaya}, \citenamefont {Fan}, \citenamefont {Alzate},
  \citenamefont {Jiang}, \citenamefont {Wong}, \citenamefont {Takei},
  \citenamefont {Bender}, \citenamefont {Chang}, \citenamefont {Jiang},
  \citenamefont {Lang}, \citenamefont {Tang}, \citenamefont {Wang},
  \citenamefont {Tserkovnyak}, \citenamefont {Amiri},\ and\ \citenamefont
  {Wang}}]{ZFS-wedge-nn-2014}%
  \BibitemOpen
  \bibfield  {author} {\bibinfo {author} {\bibfnamefont {G.}~\bibnamefont
  {Yu}}, \bibinfo {author} {\bibfnamefont {P.}~\bibnamefont {Upadhyaya}},
  \bibinfo {author} {\bibfnamefont {Y.}~\bibnamefont {Fan}}, \bibinfo {author}
  {\bibfnamefont {J.~G.}\ \bibnamefont {Alzate}}, \bibinfo {author}
  {\bibfnamefont {W.}~\bibnamefont {Jiang}}, \bibinfo {author} {\bibfnamefont
  {K.~L.}\ \bibnamefont {Wong}}, \bibinfo {author} {\bibfnamefont
  {S.}~\bibnamefont {Takei}}, \bibinfo {author} {\bibfnamefont {S.~A.}\
  \bibnamefont {Bender}}, \bibinfo {author} {\bibfnamefont {L.~T.}\
  \bibnamefont {Chang}}, \bibinfo {author} {\bibfnamefont {Y.}~\bibnamefont
  {Jiang}}, \bibinfo {author} {\bibfnamefont {M.}~\bibnamefont {Lang}},
  \bibinfo {author} {\bibfnamefont {J.}~\bibnamefont {Tang}}, \bibinfo {author}
  {\bibfnamefont {Y.}~\bibnamefont {Wang}}, \bibinfo {author} {\bibfnamefont
  {Y.}~\bibnamefont {Tserkovnyak}}, \bibinfo {author} {\bibfnamefont {P.~K.}\
  \bibnamefont {Amiri}},\ and\ \bibinfo {author} {\bibfnamefont {K.~L.}\
  \bibnamefont {Wang}},\ }\bibfield  {title} {\bibinfo {title} {Switching of
  perpendicular magnetization by spin-orbit torques in the absence of external
  magnetic fields},\ }\href {https://doi.org/10.1038/nnano.2014.94} {\bibfield
  {journal} {\bibinfo  {journal} {Nat. Nanotech.}\ }\textbf {\bibinfo {volume}
  {9}},\ \bibinfo {pages} {548} (\bibinfo {year}
  {2014}{\natexlab{a}})}\BibitemShut {NoStop}%
\bibitem [{\citenamefont {Yu}\ \emph {et~al.}(2014{\natexlab{b}})\citenamefont
  {Yu}, \citenamefont {Chang}, \citenamefont {Akyol}, \citenamefont
  {Upadhyaya}, \citenamefont {He}, \citenamefont {Li}, \citenamefont {Wong},
  \citenamefont {Amiri},\ and\ \citenamefont {Wang}}]{Yu2014b}%
  \BibitemOpen
  \bibfield  {author} {\bibinfo {author} {\bibfnamefont {G.}~\bibnamefont
  {Yu}}, \bibinfo {author} {\bibfnamefont {L.-T.}\ \bibnamefont {Chang}},
  \bibinfo {author} {\bibfnamefont {M.}~\bibnamefont {Akyol}}, \bibinfo
  {author} {\bibfnamefont {P.}~\bibnamefont {Upadhyaya}}, \bibinfo {author}
  {\bibfnamefont {C.}~\bibnamefont {He}}, \bibinfo {author} {\bibfnamefont
  {X.}~\bibnamefont {Li}}, \bibinfo {author} {\bibfnamefont {K.~L.}\
  \bibnamefont {Wong}}, \bibinfo {author} {\bibfnamefont {P.~K.}\ \bibnamefont
  {Amiri}},\ and\ \bibinfo {author} {\bibfnamefont {K.~L.}\ \bibnamefont
  {Wang}},\ }\bibfield  {title} {\bibinfo {title} {Current-driven perpendicular
  magnetization switching in $\mathrm{Ta}/\mathrm{CoFeB}/\mathrm{[{TaO}_x}$ or
  $\mathrm{MgO}/\mathrm{{TaO}_x]}$ films with lateral structural asymmetry},\
  }\href {https://doi.org/10.1063/1.4895735} {\bibfield  {journal} {\bibinfo
  {journal} {Appl. Phys. Lett.}\ }\textbf {\bibinfo {volume} {105}},\ \bibinfo
  {pages} {102411} (\bibinfo {year} {2014}{\natexlab{b}})}\BibitemShut
  {NoStop}%
\bibitem [{\citenamefont {Qiu}\ \emph {et~al.}(2015)\citenamefont {Qiu},
  \citenamefont {Narayanapillai}, \citenamefont {Wu}, \citenamefont {Deorani},
  \citenamefont {Yang}, \citenamefont {Noh}, \citenamefont {Park},
  \citenamefont {Lee}, \citenamefont {Lee},\ and\ \citenamefont
  {Yang}}]{Qiu2015}%
  \BibitemOpen
  \bibfield  {author} {\bibinfo {author} {\bibfnamefont {X.}~\bibnamefont
  {Qiu}}, \bibinfo {author} {\bibfnamefont {K.}~\bibnamefont {Narayanapillai}},
  \bibinfo {author} {\bibfnamefont {Y.}~\bibnamefont {Wu}}, \bibinfo {author}
  {\bibfnamefont {P.}~\bibnamefont {Deorani}}, \bibinfo {author} {\bibfnamefont
  {D.-H.}\ \bibnamefont {Yang}}, \bibinfo {author} {\bibfnamefont {W.-S.}\
  \bibnamefont {Noh}}, \bibinfo {author} {\bibfnamefont {J.-H.}\ \bibnamefont
  {Park}}, \bibinfo {author} {\bibfnamefont {K.-J.}\ \bibnamefont {Lee}},
  \bibinfo {author} {\bibfnamefont {H.-W.}\ \bibnamefont {Lee}},\ and\ \bibinfo
  {author} {\bibfnamefont {H.}~\bibnamefont {Yang}},\ }\bibfield  {title}
  {\bibinfo {title} {Spin-orbit-torque engineering via oxygen manipulation},\
  }\href {https://doi.org/10.1038/nnano.2015.18} {\bibfield  {journal}
  {\bibinfo  {journal} {Nat. Nanotech.}\ }\textbf {\bibinfo {volume} {10}},\
  \bibinfo {pages} {333} (\bibinfo {year} {2015})}\BibitemShut {NoStop}%
\bibitem [{\citenamefont {Pai}\ \emph {et~al.}(2015)\citenamefont {Pai},
  \citenamefont {Ou}, \citenamefont {Vilela-Le\~ao}, \citenamefont {Ralph},\
  and\ \citenamefont {Buhrman}}]{Pai2015}%
  \BibitemOpen
  \bibfield  {author} {\bibinfo {author} {\bibfnamefont {C.-F.}\ \bibnamefont
  {Pai}}, \bibinfo {author} {\bibfnamefont {Y.}~\bibnamefont {Ou}}, \bibinfo
  {author} {\bibfnamefont {L.~H.}\ \bibnamefont {Vilela-Le\~ao}}, \bibinfo
  {author} {\bibfnamefont {D.~C.}\ \bibnamefont {Ralph}},\ and\ \bibinfo
  {author} {\bibfnamefont {R.~A.}\ \bibnamefont {Buhrman}},\ }\bibfield
  {title} {\bibinfo {title} {Dependence of the efficiency of spin hall torque
  on the transparency of {P}t/ferromagnetic layer interfaces},\ }\href
  {https://doi.org/10.1103/PhysRevB.92.064426} {\bibfield  {journal} {\bibinfo
  {journal} {Phys. Rev. B}\ }\textbf {\bibinfo {volume} {92}},\ \bibinfo
  {pages} {064426} (\bibinfo {year} {2015})}\BibitemShut {NoStop}%
\bibitem [{\citenamefont {Akyol}\ \emph {et~al.}(2015)\citenamefont {Akyol},
  \citenamefont {Alzate}, \citenamefont {Yu}, \citenamefont {Upadhyaya},
  \citenamefont {Wong}, \citenamefont {Ekicibil}, \citenamefont
  {Khalili~Amiri},\ and\ \citenamefont {Wang}}]{Akyol2015}%
  \BibitemOpen
  \bibfield  {author} {\bibinfo {author} {\bibfnamefont {M.}~\bibnamefont
  {Akyol}}, \bibinfo {author} {\bibfnamefont {J.~G.}\ \bibnamefont {Alzate}},
  \bibinfo {author} {\bibfnamefont {G.}~\bibnamefont {Yu}}, \bibinfo {author}
  {\bibfnamefont {P.}~\bibnamefont {Upadhyaya}}, \bibinfo {author}
  {\bibfnamefont {K.~L.}\ \bibnamefont {Wong}}, \bibinfo {author}
  {\bibfnamefont {A.}~\bibnamefont {Ekicibil}}, \bibinfo {author}
  {\bibfnamefont {P.}~\bibnamefont {Khalili~Amiri}},\ and\ \bibinfo {author}
  {\bibfnamefont {K.~L.}\ \bibnamefont {Wang}},\ }\bibfield  {title} {\bibinfo
  {title} {Effect of the oxide layer on current-induced spin-orbit torques in
  $\mathrm{Hf|CoFeB|MgO}$ and $\mathrm{Hf|CoFeB|TaOx}$ structures},\ }\href
  {https://doi.org/10.1063/1.4906352} {\bibfield  {journal} {\bibinfo
  {journal} {Appl. Phys. Lett.}\ }\textbf {\bibinfo {volume} {106}},\ \bibinfo
  {pages} {032406} (\bibinfo {year} {2015})}\BibitemShut {NoStop}%
\bibitem [{\citenamefont {Hibino}\ \emph {et~al.}(2017)\citenamefont {Hibino},
  \citenamefont {Hirai}, \citenamefont {Hasegawa}, \citenamefont {Koyama},\
  and\ \citenamefont {Chiba}}]{Hibino2017}%
  \BibitemOpen
  \bibfield  {author} {\bibinfo {author} {\bibfnamefont {Y.}~\bibnamefont
  {Hibino}}, \bibinfo {author} {\bibfnamefont {T.}~\bibnamefont {Hirai}},
  \bibinfo {author} {\bibfnamefont {K.}~\bibnamefont {Hasegawa}}, \bibinfo
  {author} {\bibfnamefont {T.}~\bibnamefont {Koyama}},\ and\ \bibinfo {author}
  {\bibfnamefont {D.}~\bibnamefont {Chiba}},\ }\bibfield  {title} {\bibinfo
  {title} {Enhancement of the spin-orbit torque in a $\mathrm{Pt/Co}$ system
  with a naturally oxidized $\mathrm{Co}$ layer},\ }\href
  {https://doi.org/10.1063/1.4995292} {\bibfield  {journal} {\bibinfo
  {journal} {Appl. Phys. Lett.}\ }\textbf {\bibinfo {volume} {111}},\ \bibinfo
  {pages} {132404} (\bibinfo {year} {2017})}\BibitemShut {NoStop}%
\bibitem [{\citenamefont {Gweon}\ \emph {et~al.}(2019)\citenamefont {Gweon},
  \citenamefont {Lee},\ and\ \citenamefont {Lim}}]{Gweon2019}%
  \BibitemOpen
  \bibfield  {author} {\bibinfo {author} {\bibfnamefont {H.~K.}\ \bibnamefont
  {Gweon}}, \bibinfo {author} {\bibfnamefont {K.-J.}\ \bibnamefont {Lee}},\
  and\ \bibinfo {author} {\bibfnamefont {S.~H.}\ \bibnamefont {Lim}},\
  }\bibfield  {title} {\bibinfo {title} {Influence of $\mathrm{Mg}\mathrm{O}$
  sputtering power and post annealing on strength and angular dependence of
  spin-orbit torques in $\mathrm{Pt}/\mathrm{Co}/\mathrm{Mg}\mathrm{O}$
  trilayers},\ }\href {https://doi.org/10.1103/PhysRevApplied.11.014034}
  {\bibfield  {journal} {\bibinfo  {journal} {Phys. Rev. Applied}\ }\textbf
  {\bibinfo {volume} {11}},\ \bibinfo {pages} {014034} (\bibinfo {year}
  {2019})}\BibitemShut {NoStop}%
\bibitem [{\citenamefont {Bauer}\ \emph {et~al.}(2014)\citenamefont {Bauer},
  \citenamefont {Yao}, \citenamefont {Tan}, \citenamefont {Agrawal},
  \citenamefont {Emori}, \citenamefont {Tuller}, \citenamefont {van Dijken},\
  and\ \citenamefont {Beach}}]{Bauer2014}%
  \BibitemOpen
  \bibfield  {author} {\bibinfo {author} {\bibfnamefont {U.}~\bibnamefont
  {Bauer}}, \bibinfo {author} {\bibfnamefont {L.}~\bibnamefont {Yao}}, \bibinfo
  {author} {\bibfnamefont {A.~J.}\ \bibnamefont {Tan}}, \bibinfo {author}
  {\bibfnamefont {P.}~\bibnamefont {Agrawal}}, \bibinfo {author} {\bibfnamefont
  {S.}~\bibnamefont {Emori}}, \bibinfo {author} {\bibfnamefont {H.~L.}\
  \bibnamefont {Tuller}}, \bibinfo {author} {\bibfnamefont {S.}~\bibnamefont
  {van Dijken}},\ and\ \bibinfo {author} {\bibfnamefont {G.~S.~D.}\
  \bibnamefont {Beach}},\ }\bibfield  {title} {\bibinfo {title} {Magneto-ionic
  control of interfacial magnetism},\ }\href {https://doi.org/10.1038/nmat4134}
  {\bibfield  {journal} {\bibinfo  {journal} {Nat. Mater.}\ }\textbf {\bibinfo
  {volume} {14}},\ \bibinfo {pages} {174} (\bibinfo {year} {2014})}\BibitemShut
  {NoStop}%
\bibitem [{\citenamefont {Emori}\ \emph {et~al.}(2014)\citenamefont {Emori},
  \citenamefont {Bauer}, \citenamefont {Woo},\ and\ \citenamefont
  {Beach}}]{Emori2014}%
  \BibitemOpen
  \bibfield  {author} {\bibinfo {author} {\bibfnamefont {S.}~\bibnamefont
  {Emori}}, \bibinfo {author} {\bibfnamefont {U.}~\bibnamefont {Bauer}},
  \bibinfo {author} {\bibfnamefont {S.}~\bibnamefont {Woo}},\ and\ \bibinfo
  {author} {\bibfnamefont {G.~S.~D.}\ \bibnamefont {Beach}},\ }\bibfield
  {title} {\bibinfo {title} {Large voltage-induced modification of spin-orbit
  torques in $\mathrm{Pt/Co/GdOx}$},\ }\href
  {https://doi.org/10.1063/1.4903041} {\bibfield  {journal} {\bibinfo
  {journal} {Appl. Phys. Lett.}\ }\textbf {\bibinfo {volume} {105}},\ \bibinfo
  {pages} {222401} (\bibinfo {year} {2014})}\BibitemShut {NoStop}%
\bibitem [{\citenamefont {Bi}\ \emph {et~al.}(2014)\citenamefont {Bi},
  \citenamefont {Liu}, \citenamefont {Newhouse-Illige}, \citenamefont {Xu},
  \citenamefont {Rosales}, \citenamefont {Freeland}, \citenamefont {Mryasov},
  \citenamefont {Zhang}, \citenamefont {te~Velthuis},\ and\ \citenamefont
  {Wang}}]{Bi2014}%
  \BibitemOpen
  \bibfield  {author} {\bibinfo {author} {\bibfnamefont {C.}~\bibnamefont
  {Bi}}, \bibinfo {author} {\bibfnamefont {Y.}~\bibnamefont {Liu}}, \bibinfo
  {author} {\bibfnamefont {T.}~\bibnamefont {Newhouse-Illige}}, \bibinfo
  {author} {\bibfnamefont {M.}~\bibnamefont {Xu}}, \bibinfo {author}
  {\bibfnamefont {M.}~\bibnamefont {Rosales}}, \bibinfo {author} {\bibfnamefont
  {J.~W.}\ \bibnamefont {Freeland}}, \bibinfo {author} {\bibfnamefont
  {O.}~\bibnamefont {Mryasov}}, \bibinfo {author} {\bibfnamefont
  {S.}~\bibnamefont {Zhang}}, \bibinfo {author} {\bibfnamefont {S.~G.~E.}\
  \bibnamefont {te~Velthuis}},\ and\ \bibinfo {author} {\bibfnamefont {W.~G.}\
  \bibnamefont {Wang}},\ }\bibfield  {title} {\bibinfo {title} {Reversible
  control of $\mathrm{Co}$ magnetism by voltage-induced oxidation},\ }\href
  {https://doi.org/10.1103/PhysRevLett.113.267202} {\bibfield  {journal}
  {\bibinfo  {journal} {Phys. Rev. Lett.}\ }\textbf {\bibinfo {volume} {113}},\
  \bibinfo {pages} {267202} (\bibinfo {year} {2014})}\BibitemShut {NoStop}%
\bibitem [{\citenamefont {Demasius}\ \emph {et~al.}(2016)\citenamefont
  {Demasius}, \citenamefont {Phung}, \citenamefont {Zhang}, \citenamefont
  {Hughes}, \citenamefont {Yang}, \citenamefont {Kellock}, \citenamefont {Han},
  \citenamefont {Pushp},\ and\ \citenamefont {Parkin}}]{Demasius2016}%
  \BibitemOpen
  \bibfield  {author} {\bibinfo {author} {\bibfnamefont {K.-U.}\ \bibnamefont
  {Demasius}}, \bibinfo {author} {\bibfnamefont {T.}~\bibnamefont {Phung}},
  \bibinfo {author} {\bibfnamefont {W.}~\bibnamefont {Zhang}}, \bibinfo
  {author} {\bibfnamefont {B.~P.}\ \bibnamefont {Hughes}}, \bibinfo {author}
  {\bibfnamefont {S.-H.}\ \bibnamefont {Yang}}, \bibinfo {author}
  {\bibfnamefont {A.}~\bibnamefont {Kellock}}, \bibinfo {author} {\bibfnamefont
  {W.}~\bibnamefont {Han}}, \bibinfo {author} {\bibfnamefont {A.}~\bibnamefont
  {Pushp}},\ and\ \bibinfo {author} {\bibfnamefont {S.~S.~P.}\ \bibnamefont
  {Parkin}},\ }\bibfield  {title} {\bibinfo {title} {Enhanced spin-orbit
  torques by oxygen incorporation in tungsten films},\ }\href
  {https://doi.org/10.1038/ncomms10644} {\bibfield  {journal} {\bibinfo
  {journal} {Nat. Commun.}\ }\textbf {\bibinfo {volume} {7}},\ \bibinfo {pages}
  {10644} (\bibinfo {year} {2016})}\BibitemShut {NoStop}%
\bibitem [{\citenamefont {An}\ \emph {et~al.}(2018{\natexlab{a}})\citenamefont
  {An}, \citenamefont {Kanno}, \citenamefont {Asami},\ and\ \citenamefont
  {Ando}}]{An2018}%
  \BibitemOpen
  \bibfield  {author} {\bibinfo {author} {\bibfnamefont {H.}~\bibnamefont
  {An}}, \bibinfo {author} {\bibfnamefont {Y.}~\bibnamefont {Kanno}}, \bibinfo
  {author} {\bibfnamefont {A.}~\bibnamefont {Asami}},\ and\ \bibinfo {author}
  {\bibfnamefont {K.}~\bibnamefont {Ando}},\ }\bibfield  {title} {\bibinfo
  {title} {Giant spin-torque generation by heavily oxidized $\mathrm{Pt}$},\
  }\href {https://doi.org/10.1103/PhysRevB.98.014401} {\bibfield  {journal}
  {\bibinfo  {journal} {Phys. Rev. B}\ }\textbf {\bibinfo {volume} {98}},\
  \bibinfo {pages} {014401} (\bibinfo {year} {2018}{\natexlab{a}})}\BibitemShut
  {NoStop}%
\bibitem [{\citenamefont {An}\ \emph {et~al.}(2018{\natexlab{b}})\citenamefont
  {An}, \citenamefont {Ohno}, \citenamefont {Kanno}, \citenamefont {Kageyama},
  \citenamefont {Monnai}, \citenamefont {Maki}, \citenamefont {Shi},\ and\
  \citenamefont {Ando}}]{An2018a}%
  \BibitemOpen
  \bibfield  {author} {\bibinfo {author} {\bibfnamefont {H.}~\bibnamefont
  {An}}, \bibinfo {author} {\bibfnamefont {T.}~\bibnamefont {Ohno}}, \bibinfo
  {author} {\bibfnamefont {Y.}~\bibnamefont {Kanno}}, \bibinfo {author}
  {\bibfnamefont {Y.}~\bibnamefont {Kageyama}}, \bibinfo {author}
  {\bibfnamefont {Y.}~\bibnamefont {Monnai}}, \bibinfo {author} {\bibfnamefont
  {H.}~\bibnamefont {Maki}}, \bibinfo {author} {\bibfnamefont {J.}~\bibnamefont
  {Shi}},\ and\ \bibinfo {author} {\bibfnamefont {K.}~\bibnamefont {Ando}},\
  }\bibfield  {title} {\bibinfo {title} {Current-induced magnetization
  switching using an electrically insulating spin-torque generator},\ }\href
  {https://doi.org/10.1126/sciadv.aar2250} {\bibfield  {journal} {\bibinfo
  {journal} {Sci. Adv.}\ }\textbf {\bibinfo {volume} {4}},\ \bibinfo {pages}
  {eaar2250} (\bibinfo {year} {2018}{\natexlab{b}})}\BibitemShut {NoStop}%
\bibitem [{\citenamefont {Dua}\ \emph {et~al.}(1988)\citenamefont {Dua},
  \citenamefont {George},\ and\ \citenamefont {Agarwala}}]{Dua1988163}%
  \BibitemOpen
  \bibfield  {author} {\bibinfo {author} {\bibfnamefont {A.}~\bibnamefont
  {Dua}}, \bibinfo {author} {\bibfnamefont {V.}~\bibnamefont {George}},\ and\
  \bibinfo {author} {\bibfnamefont {R.}~\bibnamefont {Agarwala}},\ }\bibfield
  {title} {\bibinfo {title} {Characterization and microhardness measurement of
  electron-beam-evaporated alumina coatings},\ }\href
  {https://doi.org/https://doi.org/10.1016/0040-6090(88)90687-6} {\bibfield
  {journal} {\bibinfo  {journal} {Thin Solid Films}\ }\textbf {\bibinfo
  {volume} {165}},\ \bibinfo {pages} {163 } (\bibinfo {year}
  {1988})}\BibitemShut {NoStop}%
\bibitem [{\citenamefont {Nagano}(2002)}]{Nagano2002}%
  \BibitemOpen
  \bibfield  {author} {\bibinfo {author} {\bibfnamefont {Y.}~\bibnamefont
  {Nagano}},\ }\bibfield  {title} {\bibinfo {title} {Standard enthalpy of
  formation of platinum hydrous oxide},\ }\href
  {https://doi.org/10.1023/A:1020651805170} {\bibfield  {journal} {\bibinfo
  {journal} {J. Therm. Anal. Calorim.}\ }\textbf {\bibinfo {volume} {69}},\
  \bibinfo {pages} {831} (\bibinfo {year} {2002})}\BibitemShut {NoStop}%
\bibitem [{\citenamefont {Wagman}\ \emph {et~al.}(1989)\citenamefont {Wagman},
  \citenamefont {Cox},\ and\ \citenamefont {Medvedev}}]{enthalpy}%
  \BibitemOpen
  \bibfield  {author} {\bibinfo {author} {\bibfnamefont {D.~D.}\ \bibnamefont
  {Wagman}}, \bibinfo {author} {\bibfnamefont {J.~D.}\ \bibnamefont {Cox}},\
  and\ \bibinfo {author} {\bibfnamefont {V.~A.}\ \bibnamefont {Medvedev}},\
  }\href@noop {} {\emph {\bibinfo {title} {CODATA key values for
  thermodynamics}}}\ (\bibinfo  {publisher} {Hemisphere Pub. Corp.},\ \bibinfo
  {year} {1989})\BibitemShut {NoStop}%
\bibitem [{\citenamefont {Holmes}\ \emph {et~al.}(1986)\citenamefont {Holmes},
  \citenamefont {O'Neill},\ and\ \citenamefont {Arculus}}]{Holmes1986}%
  \BibitemOpen
  \bibfield  {author} {\bibinfo {author} {\bibfnamefont {R.~D.}\ \bibnamefont
  {Holmes}}, \bibinfo {author} {\bibfnamefont {H.~S.}\ \bibnamefont
  {O'Neill}},\ and\ \bibinfo {author} {\bibfnamefont {R.~J.}\ \bibnamefont
  {Arculus}},\ }\bibfield  {title} {\bibinfo {title} {Standard gibbs free
  energy of formation for $\mathrm{Cu_2O}$, $\mathrm{NiO}$, $\mathrm{CoO}$, and
  $\mathrm{Fe_xO}$: High resolution electrochemical measurements using zirconia
  solid electrolytes from 900–1400 {K}},\ }\href
  {https://doi.org/https://doi.org/10.1016/0016-7037(86)90027-X} {\bibfield
  {journal} {\bibinfo  {journal} {Geochim. Cosmochim. Acta}\ }\textbf {\bibinfo
  {volume} {50}},\ \bibinfo {pages} {2439 } (\bibinfo {year}
  {1986})}\BibitemShut {NoStop}%
\bibitem [{\citenamefont {Avci}\ \emph {et~al.}(2019)\citenamefont {Avci},
  \citenamefont {Beach},\ and\ \citenamefont {Gambardella}}]{Avci2019}%
  \BibitemOpen
  \bibfield  {author} {\bibinfo {author} {\bibfnamefont {C.~O.}\ \bibnamefont
  {Avci}}, \bibinfo {author} {\bibfnamefont {G.~S.~D.}\ \bibnamefont {Beach}},\
  and\ \bibinfo {author} {\bibfnamefont {P.}~\bibnamefont {Gambardella}},\
  }\bibfield  {title} {\bibinfo {title} {Effects of transition metal spacers on
  spin-orbit torques, spin hall magnetoresistance, and magnetic anisotropy of
  pt/co bilayers},\ }\href {https://doi.org/10.1103/PhysRevB.100.235454}
  {\bibfield  {journal} {\bibinfo  {journal} {Phys. Rev. B}\ }\textbf {\bibinfo
  {volume} {100}},\ \bibinfo {pages} {235454} (\bibinfo {year}
  {2019})}\BibitemShut {NoStop}%
\bibitem [{\citenamefont {Avci}\ \emph
  {et~al.}(2014{\natexlab{a}})\citenamefont {Avci}, \citenamefont {Garello},
  \citenamefont {Gabureac}, \citenamefont {Ghosh}, \citenamefont {Fuhrer},
  \citenamefont {Alvarado},\ and\ \citenamefont {Gambardella}}]{Can-thermal}%
  \BibitemOpen
  \bibfield  {author} {\bibinfo {author} {\bibfnamefont {C.~O.}\ \bibnamefont
  {Avci}}, \bibinfo {author} {\bibfnamefont {K.}~\bibnamefont {Garello}},
  \bibinfo {author} {\bibfnamefont {M.}~\bibnamefont {Gabureac}}, \bibinfo
  {author} {\bibfnamefont {A.}~\bibnamefont {Ghosh}}, \bibinfo {author}
  {\bibfnamefont {A.}~\bibnamefont {Fuhrer}}, \bibinfo {author} {\bibfnamefont
  {S.~F.}\ \bibnamefont {Alvarado}},\ and\ \bibinfo {author} {\bibfnamefont
  {P.}~\bibnamefont {Gambardella}},\ }\bibfield  {title} {\bibinfo {title}
  {Interplay of spin-orbit torque and thermoelectric effects in
  ferromagnet/normal-metal bilayers},\ }\href
  {https://doi.org/10.1103/PhysRevB.90.224427} {\bibfield  {journal} {\bibinfo
  {journal} {Phys. Rev. B}\ }\textbf {\bibinfo {volume} {90}},\ \bibinfo
  {pages} {224427} (\bibinfo {year} {2014}{\natexlab{a}})}\BibitemShut
  {NoStop}%
\bibitem [{SI()}]{SI}%
  \BibitemOpen
  \href@noop {} {}\bibinfo {note} {See supplementary materials for more details
  on the determination of the Hall voltage, magnetic anisotropy, and SOT
  effective fields.}\BibitemShut {Stop}%
\bibitem [{\citenamefont {Zhang}\ \emph {et~al.}(2010)\citenamefont {Zhang},
  \citenamefont {Teng}, \citenamefont {Zhang}, \citenamefont {Liu},
  \citenamefont {Li}, \citenamefont {Yu},\ and\ \citenamefont
  {Wang}}]{Zhang2010}%
  \BibitemOpen
  \bibfield  {author} {\bibinfo {author} {\bibfnamefont {S.~L.}\ \bibnamefont
  {Zhang}}, \bibinfo {author} {\bibfnamefont {J.}~\bibnamefont {Teng}},
  \bibinfo {author} {\bibfnamefont {J.~Y.}\ \bibnamefont {Zhang}}, \bibinfo
  {author} {\bibfnamefont {Y.}~\bibnamefont {Liu}}, \bibinfo {author}
  {\bibfnamefont {J.~W.}\ \bibnamefont {Li}}, \bibinfo {author} {\bibfnamefont
  {G.~H.}\ \bibnamefont {Yu}},\ and\ \bibinfo {author} {\bibfnamefont {S.~G.}\
  \bibnamefont {Wang}},\ }\bibfield  {title} {\bibinfo {title} {Large
  enhancement of the anomalous hall effect in $\mathrm{Co/Pt}$ multilayers
  sandwiched by $\mathrm{MgO}$ layers},\ }\href
  {https://doi.org/10.1063/1.3522653} {\bibfield  {journal} {\bibinfo
  {journal} {Appl. Phys. Lett.}\ }\textbf {\bibinfo {volume} {97}},\ \bibinfo
  {pages} {222504} (\bibinfo {year} {2010})}\BibitemShut {NoStop}%
\bibitem [{\citenamefont {Zhang}\ \emph {et~al.}(2013)\citenamefont {Zhang},
  \citenamefont {Wu}, \citenamefont {Wang}, \citenamefont {Zhao}, \citenamefont
  {Yang}, \citenamefont {Zhang}, \citenamefont {Liu}, \citenamefont {Liu},
  \citenamefont {Teng},\ and\ \citenamefont {Yu}}]{Zhang2013}%
  \BibitemOpen
  \bibfield  {author} {\bibinfo {author} {\bibfnamefont {J.~Y.}\ \bibnamefont
  {Zhang}}, \bibinfo {author} {\bibfnamefont {Z.~L.}\ \bibnamefont {Wu}},
  \bibinfo {author} {\bibfnamefont {S.~G.}\ \bibnamefont {Wang}}, \bibinfo
  {author} {\bibfnamefont {C.~J.}\ \bibnamefont {Zhao}}, \bibinfo {author}
  {\bibfnamefont {G.}~\bibnamefont {Yang}}, \bibinfo {author} {\bibfnamefont
  {S.~L.}\ \bibnamefont {Zhang}}, \bibinfo {author} {\bibfnamefont
  {Y.}~\bibnamefont {Liu}}, \bibinfo {author} {\bibfnamefont {S.}~\bibnamefont
  {Liu}}, \bibinfo {author} {\bibfnamefont {J.}~\bibnamefont {Teng}},\ and\
  \bibinfo {author} {\bibfnamefont {G.~H.}\ \bibnamefont {Yu}},\ }\bibfield
  {title} {\bibinfo {title} {Effect of interfacial structures on anomalous hall
  behavior in perpendicular $\mathrm{Co/Pt}$ multilayers},\ }\href
  {https://doi.org/10.1063/1.4795331} {\bibfield  {journal} {\bibinfo
  {journal} {Appl. Phys. Lett.}\ }\textbf {\bibinfo {volume} {102}},\ \bibinfo
  {pages} {102404} (\bibinfo {year} {2013})}\BibitemShut {NoStop}%
\bibitem [{\citenamefont {Yang}\ \emph {et~al.}(2011)\citenamefont {Yang},
  \citenamefont {Chshiev}, \citenamefont {Dieny}, \citenamefont {Lee},
  \citenamefont {Manchon},\ and\ \citenamefont {Shin}}]{Yang2011}%
  \BibitemOpen
  \bibfield  {author} {\bibinfo {author} {\bibfnamefont {H.~X.}\ \bibnamefont
  {Yang}}, \bibinfo {author} {\bibfnamefont {M.}~\bibnamefont {Chshiev}},
  \bibinfo {author} {\bibfnamefont {B.}~\bibnamefont {Dieny}}, \bibinfo
  {author} {\bibfnamefont {J.~H.}\ \bibnamefont {Lee}}, \bibinfo {author}
  {\bibfnamefont {A.}~\bibnamefont {Manchon}},\ and\ \bibinfo {author}
  {\bibfnamefont {K.~H.}\ \bibnamefont {Shin}},\ }\bibfield  {title} {\bibinfo
  {title} {First-principles investigation of the very large perpendicular
  magnetic anisotropy at fe$|$mgo and co$|$mgo interfaces},\ }\href
  {https://doi.org/10.1103/PhysRevB.84.054401} {\bibfield  {journal} {\bibinfo
  {journal} {Phys. Rev. B}\ }\textbf {\bibinfo {volume} {84}},\ \bibinfo
  {pages} {054401} (\bibinfo {year} {2011})}\BibitemShut {NoStop}%
\bibitem [{\citenamefont {Chung}\ \emph {et~al.}(2016)\citenamefont {Chung},
  \citenamefont {Yang}, \citenamefont {Kim},\ and\ \citenamefont
  {Hong}}]{Chung2016}%
  \BibitemOpen
  \bibfield  {author} {\bibinfo {author} {\bibfnamefont {W.~S.}\ \bibnamefont
  {Chung}}, \bibinfo {author} {\bibfnamefont {S.~M.}\ \bibnamefont {Yang}},
  \bibinfo {author} {\bibfnamefont {T.~W.}\ \bibnamefont {Kim}},\ and\ \bibinfo
  {author} {\bibfnamefont {J.~P.}\ \bibnamefont {Hong}},\ }\bibfield  {title}
  {\bibinfo {title} {Ultrathin {Co-O} oxide layer-driven perpendicular magnetic
  anisotropy in a {CoO/[Co/Pd]m} multilayer matrix upon annealing},\ }\href
  {https://doi.org/10.1038/srep37503} {\bibfield  {journal} {\bibinfo
  {journal} {Sci. Rep.}\ }\textbf {\bibinfo {volume} {6}},\ \bibinfo {pages}
  {37503} (\bibinfo {year} {2016})}\BibitemShut {NoStop}%
\bibitem [{\citenamefont {Pan}\ \emph {et~al.}(2019)\citenamefont {Pan},
  \citenamefont {An}, \citenamefont {Harumoto}, \citenamefont {Zhang},
  \citenamefont {Nakamura},\ and\ \citenamefont {Shi}}]{Pan2019}%
  \BibitemOpen
  \bibfield  {author} {\bibinfo {author} {\bibfnamefont {C.}~\bibnamefont
  {Pan}}, \bibinfo {author} {\bibfnamefont {H.}~\bibnamefont {An}}, \bibinfo
  {author} {\bibfnamefont {T.}~\bibnamefont {Harumoto}}, \bibinfo {author}
  {\bibfnamefont {Z.}~\bibnamefont {Zhang}}, \bibinfo {author} {\bibfnamefont
  {Y.}~\bibnamefont {Nakamura}},\ and\ \bibinfo {author} {\bibfnamefont
  {J.}~\bibnamefont {Shi}},\ }\bibfield  {title} {\bibinfo {title} {Control of
  the perpendicular magnetic anisotropy and perpendicular exchange bias in
  {CoPt/CoOx} thin films},\ }\href
  {https://doi.org/https://doi.org/10.1016/j.jmmm.2019.04.050} {\bibfield
  {journal} {\bibinfo  {journal} {J. Magn. Magn. Mater.}\ }\textbf {\bibinfo
  {volume} {484}},\ \bibinfo {pages} {320 } (\bibinfo {year}
  {2019})}\BibitemShut {NoStop}%
\bibitem [{\citenamefont {Hayashi}\ \emph {et~al.}(2014)\citenamefont
  {Hayashi}, \citenamefont {Kim}, \citenamefont {Yamanouchi},\ and\
  \citenamefont {Ohno}}]{Hayashi2014}%
  \BibitemOpen
  \bibfield  {author} {\bibinfo {author} {\bibfnamefont {M.}~\bibnamefont
  {Hayashi}}, \bibinfo {author} {\bibfnamefont {J.}~\bibnamefont {Kim}},
  \bibinfo {author} {\bibfnamefont {M.}~\bibnamefont {Yamanouchi}},\ and\
  \bibinfo {author} {\bibfnamefont {H.}~\bibnamefont {Ohno}},\ }\bibfield
  {title} {\bibinfo {title} {Quantitative characterization of the spin-orbit
  torque using harmonic hall voltage measurements},\ }\href
  {https://doi.org/10.1103/PhysRevB.89.144425} {\bibfield  {journal} {\bibinfo
  {journal} {Phys. Rev. B}\ }\textbf {\bibinfo {volume} {89}},\ \bibinfo
  {pages} {144425} (\bibinfo {year} {2014})}\BibitemShut {NoStop}%
\bibitem [{\citenamefont {Avci}\ \emph
  {et~al.}(2014{\natexlab{b}})\citenamefont {Avci}, \citenamefont {Garello},
  \citenamefont {Nistor}, \citenamefont {Godey}, \citenamefont {Ballesteros},
  \citenamefont {Mugarza}, \citenamefont {Barla}, \citenamefont {Valvidares},
  \citenamefont {Pellegrin}, \citenamefont {Ghosh}, \citenamefont {Miron},
  \citenamefont {Boulle}, \citenamefont {Auffret}, \citenamefont {Gaudin},\
  and\ \citenamefont {Gambardella}}]{Avci2014a}%
  \BibitemOpen
  \bibfield  {author} {\bibinfo {author} {\bibfnamefont {C.~O.}\ \bibnamefont
  {Avci}}, \bibinfo {author} {\bibfnamefont {K.}~\bibnamefont {Garello}},
  \bibinfo {author} {\bibfnamefont {C.}~\bibnamefont {Nistor}}, \bibinfo
  {author} {\bibfnamefont {S.}~\bibnamefont {Godey}}, \bibinfo {author}
  {\bibfnamefont {B.}~\bibnamefont {Ballesteros}}, \bibinfo {author}
  {\bibfnamefont {A.}~\bibnamefont {Mugarza}}, \bibinfo {author} {\bibfnamefont
  {A.}~\bibnamefont {Barla}}, \bibinfo {author} {\bibfnamefont
  {M.}~\bibnamefont {Valvidares}}, \bibinfo {author} {\bibfnamefont
  {E.}~\bibnamefont {Pellegrin}}, \bibinfo {author} {\bibfnamefont
  {A.}~\bibnamefont {Ghosh}}, \bibinfo {author} {\bibfnamefont {I.~M.}\
  \bibnamefont {Miron}}, \bibinfo {author} {\bibfnamefont {O.}~\bibnamefont
  {Boulle}}, \bibinfo {author} {\bibfnamefont {S.}~\bibnamefont {Auffret}},
  \bibinfo {author} {\bibfnamefont {G.}~\bibnamefont {Gaudin}},\ and\ \bibinfo
  {author} {\bibfnamefont {P.}~\bibnamefont {Gambardella}},\ }\bibfield
  {title} {\bibinfo {title} {Fieldlike and antidamping spin-orbit torques in
  as-grown and annealed ta/cofeb/mgo layers},\ }\href
  {https://doi.org/10.1103/PhysRevB.89.214419} {\bibfield  {journal} {\bibinfo
  {journal} {Phys. Rev. B}\ }\textbf {\bibinfo {volume} {89}},\ \bibinfo
  {pages} {214419} (\bibinfo {year} {2014}{\natexlab{b}})}\BibitemShut
  {NoStop}%
\bibitem [{\citenamefont {Belashchenko}\ \emph {et~al.}(2019)\citenamefont
  {Belashchenko}, \citenamefont {Kovalev},\ and\ \citenamefont {van
  Schilfgaarde}}]{Belashchenko2019}%
  \BibitemOpen
  \bibfield  {author} {\bibinfo {author} {\bibfnamefont {K.~D.}\ \bibnamefont
  {Belashchenko}}, \bibinfo {author} {\bibfnamefont {A.~A.}\ \bibnamefont
  {Kovalev}},\ and\ \bibinfo {author} {\bibfnamefont {M.}~\bibnamefont {van
  Schilfgaarde}},\ }\bibfield  {title} {\bibinfo {title} {First-principles
  calculation of spin-orbit torque in a co/pt bilayer},\ }\href
  {https://doi.org/10.1103/PhysRevMaterials.3.011401} {\bibfield  {journal}
  {\bibinfo  {journal} {Phys. Rev. Materials}\ }\textbf {\bibinfo {volume}
  {3}},\ \bibinfo {pages} {011401} (\bibinfo {year} {2019})}\BibitemShut
  {NoStop}%
\bibitem [{\citenamefont {Nguyen}\ \emph {et~al.}(2016)\citenamefont {Nguyen},
  \citenamefont {Ralph},\ and\ \citenamefont
  {Buhrman}}]{Effciencypercurrent-PRL2016}%
  \BibitemOpen
  \bibfield  {author} {\bibinfo {author} {\bibfnamefont {M.~H.}\ \bibnamefont
  {Nguyen}}, \bibinfo {author} {\bibfnamefont {D.~C.}\ \bibnamefont {Ralph}},\
  and\ \bibinfo {author} {\bibfnamefont {R.~A.}\ \bibnamefont {Buhrman}},\
  }\bibfield  {title} {\bibinfo {title} {Spin torque study of the spin $h$all
  conductivity and spin diffusion length in platinum thin films with varying
  resistivity},\ }\href {https://doi.org/10.1103/PhysRevLett.116.126601}
  {\bibfield  {journal} {\bibinfo  {journal} {Phys. Rev. Lett.}\ }\textbf
  {\bibinfo {volume} {116}},\ \bibinfo {pages} {126601} (\bibinfo {year}
  {2016})}\BibitemShut {NoStop}%
\bibitem [{\citenamefont {Amin}\ and\ \citenamefont
  {Stiles}(2016{\natexlab{a}})}]{Amin2016}%
  \BibitemOpen
  \bibfield  {author} {\bibinfo {author} {\bibfnamefont {V.~P.}\ \bibnamefont
  {Amin}}\ and\ \bibinfo {author} {\bibfnamefont {M.~D.}\ \bibnamefont
  {Stiles}},\ }\bibfield  {title} {\bibinfo {title} {Spin transport at
  interfaces with spin-orbit coupling: Phenomenology},\ }\href
  {https://doi.org/10.1103/PhysRevB.94.104420} {\bibfield  {journal} {\bibinfo
  {journal} {Phys. Rev. B}\ }\textbf {\bibinfo {volume} {94}},\ \bibinfo
  {pages} {104420} (\bibinfo {year} {2016}{\natexlab{a}})}\BibitemShut
  {NoStop}%
\bibitem [{\citenamefont {Amin}\ and\ \citenamefont
  {Stiles}(2016{\natexlab{b}})}]{Amin2016a}%
  \BibitemOpen
  \bibfield  {author} {\bibinfo {author} {\bibfnamefont {V.~P.}\ \bibnamefont
  {Amin}}\ and\ \bibinfo {author} {\bibfnamefont {M.~D.}\ \bibnamefont
  {Stiles}},\ }\bibfield  {title} {\bibinfo {title} {Spin transport at
  interfaces with spin-orbit coupling: Formalism},\ }\href
  {https://doi.org/10.1103/PhysRevB.94.104419} {\bibfield  {journal} {\bibinfo
  {journal} {Phys. Rev. B}\ }\textbf {\bibinfo {volume} {94}},\ \bibinfo
  {pages} {104419} (\bibinfo {year} {2016}{\natexlab{b}})}\BibitemShut
  {NoStop}%
\bibitem [{\citenamefont {Amin}\ \emph {et~al.}(2018)\citenamefont {Amin},
  \citenamefont {Zemen},\ and\ \citenamefont {Stiles}}]{Amin2018}%
  \BibitemOpen
  \bibfield  {author} {\bibinfo {author} {\bibfnamefont {V.~P.}\ \bibnamefont
  {Amin}}, \bibinfo {author} {\bibfnamefont {J.}~\bibnamefont {Zemen}},\ and\
  \bibinfo {author} {\bibfnamefont {M.~D.}\ \bibnamefont {Stiles}},\ }\bibfield
   {title} {\bibinfo {title} {Interface-generated spin currents},\ }\href
  {https://doi.org/10.1103/PhysRevLett.121.136805} {\bibfield  {journal}
  {\bibinfo  {journal} {Phys. Rev. Lett.}\ }\textbf {\bibinfo {volume} {121}},\
  \bibinfo {pages} {136805} (\bibinfo {year} {2018})}\BibitemShut {NoStop}%
\bibitem [{\citenamefont {Freimuth}\ \emph {et~al.}(2014)\citenamefont
  {Freimuth}, \citenamefont {Bl\"ugel},\ and\ \citenamefont
  {Mokrousov}}]{Freimuth2014}%
  \BibitemOpen
  \bibfield  {author} {\bibinfo {author} {\bibfnamefont {F.}~\bibnamefont
  {Freimuth}}, \bibinfo {author} {\bibfnamefont {S.}~\bibnamefont {Bl\"ugel}},\
  and\ \bibinfo {author} {\bibfnamefont {Y.}~\bibnamefont {Mokrousov}},\
  }\bibfield  {title} {\bibinfo {title} {Spin-orbit torques in
  $\mathrm{Co/Pt(111)}$ and $\mathrm{Mn/W(001)}$ magnetic bilayers from first
  principles},\ }\href {https://doi.org/10.1103/PhysRevB.90.174423} {\bibfield
  {journal} {\bibinfo  {journal} {Phys. Rev. B}\ }\textbf {\bibinfo {volume}
  {90}},\ \bibinfo {pages} {174423} (\bibinfo {year} {2014})}\BibitemShut
  {NoStop}%
\bibitem [{\citenamefont {Ambrose}\ and\ \citenamefont
  {Chien}(1998)}]{Ambrose1998}%
  \BibitemOpen
  \bibfield  {author} {\bibinfo {author} {\bibfnamefont {T.}~\bibnamefont
  {Ambrose}}\ and\ \bibinfo {author} {\bibfnamefont {C.~L.}\ \bibnamefont
  {Chien}},\ }\bibfield  {title} {\bibinfo {title} {Dependence of exchange
  coupling on antiferromagnetic layer thickness in $\mathrm{NiFe/CoO}$
  bilayers},\ }\href {https://doi.org/10.1063/1.367863} {\bibfield  {journal}
  {\bibinfo  {journal} {J. Appl. Phys.}\ }\textbf {\bibinfo {volume} {83}},\
  \bibinfo {pages} {6822} (\bibinfo {year} {1998})}\BibitemShut {NoStop}%
\bibitem [{\citenamefont {Hahn}\ \emph {et~al.}(2014)\citenamefont {Hahn},
  \citenamefont {de~Loubens}, \citenamefont {Naletov}, \citenamefont {Youssef},
  \citenamefont {Klein},\ and\ \citenamefont {Viret}}]{Hahn2014}%
  \BibitemOpen
  \bibfield  {author} {\bibinfo {author} {\bibfnamefont {C.}~\bibnamefont
  {Hahn}}, \bibinfo {author} {\bibfnamefont {G.}~\bibnamefont {de~Loubens}},
  \bibinfo {author} {\bibfnamefont {V.~V.}\ \bibnamefont {Naletov}}, \bibinfo
  {author} {\bibfnamefont {J.~B.}\ \bibnamefont {Youssef}}, \bibinfo {author}
  {\bibfnamefont {O.}~\bibnamefont {Klein}},\ and\ \bibinfo {author}
  {\bibfnamefont {M.}~\bibnamefont {Viret}},\ }\bibfield  {title} {\bibinfo
  {title} {Conduction of spin currents through insulating antiferromagnetic
  oxides},\ }\href {http://stacks.iop.org/0295-5075/108/i=5/a=57005} {\bibfield
   {journal} {\bibinfo  {journal} {Europhys. Lett.}\ }\textbf {\bibinfo
  {volume} {108}},\ \bibinfo {pages} {57005} (\bibinfo {year}
  {2014})}\BibitemShut {NoStop}%
\bibitem [{\citenamefont {Qiu}\ \emph {et~al.}(2016)\citenamefont {Qiu},
  \citenamefont {Li}, \citenamefont {Hou}, \citenamefont {Arenholz},
  \citenamefont {N’Diaye}, \citenamefont {Tan}, \citenamefont {Uchida},
  \citenamefont {Sato}, \citenamefont {Okamoto}, \citenamefont {Tserkovnyak},
  \citenamefont {Qiu},\ and\ \citenamefont {Saitoh}}]{Qiu2016}%
  \BibitemOpen
  \bibfield  {author} {\bibinfo {author} {\bibfnamefont {Z.}~\bibnamefont
  {Qiu}}, \bibinfo {author} {\bibfnamefont {J.}~\bibnamefont {Li}}, \bibinfo
  {author} {\bibfnamefont {D.}~\bibnamefont {Hou}}, \bibinfo {author}
  {\bibfnamefont {E.}~\bibnamefont {Arenholz}}, \bibinfo {author}
  {\bibfnamefont {A.~T.}\ \bibnamefont {N’Diaye}}, \bibinfo {author}
  {\bibfnamefont {A.}~\bibnamefont {Tan}}, \bibinfo {author} {\bibfnamefont
  {K.-i.}\ \bibnamefont {Uchida}}, \bibinfo {author} {\bibfnamefont
  {K.}~\bibnamefont {Sato}}, \bibinfo {author} {\bibfnamefont {S.}~\bibnamefont
  {Okamoto}}, \bibinfo {author} {\bibfnamefont {Y.}~\bibnamefont
  {Tserkovnyak}}, \bibinfo {author} {\bibfnamefont {Z.~Q.}\ \bibnamefont
  {Qiu}},\ and\ \bibinfo {author} {\bibfnamefont {E.}~\bibnamefont {Saitoh}},\
  }\bibfield  {title} {\bibinfo {title} {Spin-current probe for phase
  transition in an insulator},\ }\href {https://doi.org/10.1038/ncomms12670}
  {\bibfield  {journal} {\bibinfo  {journal} {Nat. Commun.}\ }\textbf {\bibinfo
  {volume} {7}},\ \bibinfo {pages} {12670} (\bibinfo {year}
  {2016})}\BibitemShut {NoStop}%
\bibitem [{\citenamefont {Lin}\ \emph {et~al.}(2016)\citenamefont {Lin},
  \citenamefont {Chen}, \citenamefont {Zhang},\ and\ \citenamefont
  {Chien}}]{Lin2016}%
  \BibitemOpen
  \bibfield  {author} {\bibinfo {author} {\bibfnamefont {W.}~\bibnamefont
  {Lin}}, \bibinfo {author} {\bibfnamefont {K.}~\bibnamefont {Chen}}, \bibinfo
  {author} {\bibfnamefont {S.}~\bibnamefont {Zhang}},\ and\ \bibinfo {author}
  {\bibfnamefont {C.~L.}\ \bibnamefont {Chien}},\ }\bibfield  {title} {\bibinfo
  {title} {Enhancement of thermally injected spin current through an
  antiferromagnetic insulator},\ }\href
  {https://doi.org/10.1103/PhysRevLett.116.186601} {\bibfield  {journal}
  {\bibinfo  {journal} {Phys. Rev. Lett.}\ }\textbf {\bibinfo {volume} {116}},\
  \bibinfo {pages} {186601} (\bibinfo {year} {2016})}\BibitemShut {NoStop}%
\bibitem [{\citenamefont {Hasegawa}\ \emph {et~al.}(2018)\citenamefont
  {Hasegawa}, \citenamefont {Hibino}, \citenamefont {Suzuki}, \citenamefont
  {Koyama},\ and\ \citenamefont {Chiba}}]{Hasegawa2018}%
  \BibitemOpen
  \bibfield  {author} {\bibinfo {author} {\bibfnamefont {K.}~\bibnamefont
  {Hasegawa}}, \bibinfo {author} {\bibfnamefont {Y.}~\bibnamefont {Hibino}},
  \bibinfo {author} {\bibfnamefont {M.}~\bibnamefont {Suzuki}}, \bibinfo
  {author} {\bibfnamefont {T.}~\bibnamefont {Koyama}},\ and\ \bibinfo {author}
  {\bibfnamefont {D.}~\bibnamefont {Chiba}},\ }\bibfield  {title} {\bibinfo
  {title} {Enhancement of spin-orbit torque by inserting $\mathrm{{CoO_x}}$
  layer into $\mathrm{{Co/Pt}}$ interface},\ }\href
  {https://doi.org/10.1103/PhysRevB.98.020405} {\bibfield  {journal} {\bibinfo
  {journal} {Phys. Rev. B}\ }\textbf {\bibinfo {volume} {98}},\ \bibinfo
  {pages} {020405} (\bibinfo {year} {2018})}\BibitemShut {NoStop}%
\bibitem [{\citenamefont {Wang}\ \emph {et~al.}(2019)\citenamefont {Wang},
  \citenamefont {Finley}, \citenamefont {Zhang}, \citenamefont {Han},
  \citenamefont {Hou},\ and\ \citenamefont {Liu}}]{Wang2019}%
  \BibitemOpen
  \bibfield  {author} {\bibinfo {author} {\bibfnamefont {H.}~\bibnamefont
  {Wang}}, \bibinfo {author} {\bibfnamefont {J.}~\bibnamefont {Finley}},
  \bibinfo {author} {\bibfnamefont {P.}~\bibnamefont {Zhang}}, \bibinfo
  {author} {\bibfnamefont {J.}~\bibnamefont {Han}}, \bibinfo {author}
  {\bibfnamefont {J.~T.}\ \bibnamefont {Hou}},\ and\ \bibinfo {author}
  {\bibfnamefont {L.}~\bibnamefont {Liu}},\ }\bibfield  {title} {\bibinfo
  {title} {Spin-orbit-torque switching mediated by an antiferromagnetic
  insulator},\ }\href {https://doi.org/10.1103/PhysRevApplied.11.044070}
  {\bibfield  {journal} {\bibinfo  {journal} {Phys. Rev. Applied}\ }\textbf
  {\bibinfo {volume} {11}},\ \bibinfo {pages} {044070} (\bibinfo {year}
  {2019})}\BibitemShut {NoStop}%
\bibitem [{\citenamefont {Harder}\ \emph {et~al.}(2016)\citenamefont {Harder},
  \citenamefont {Gui},\ and\ \citenamefont {Hu}}]{Harder-pr-2016}%
  \BibitemOpen
  \bibfield  {author} {\bibinfo {author} {\bibfnamefont {M.}~\bibnamefont
  {Harder}}, \bibinfo {author} {\bibfnamefont {Y.}~\bibnamefont {Gui}},\ and\
  \bibinfo {author} {\bibfnamefont {C.-M.}\ \bibnamefont {Hu}},\ }\bibfield
  {title} {\bibinfo {title} {Electrical detection of magnetization dynamics via
  spin rectification effects},\ }\href
  {https://doi.org/https://doi.org/10.1016/j.physrep.2016.10.002} {\bibfield
  {journal} {\bibinfo  {journal} {Phys. Rep.}\ }\textbf {\bibinfo {volume}
  {661}},\ \bibinfo {pages} {1 } (\bibinfo {year} {2016})}\BibitemShut
  {NoStop}%
\bibitem [{\citenamefont {Garello}\ \emph {et~al.}(2014)\citenamefont
  {Garello}, \citenamefont {Avci}, \citenamefont {Miron}, \citenamefont
  {Baumgartner}, \citenamefont {Ghosh}, \citenamefont {Auffret}, \citenamefont
  {Boulle}, \citenamefont {Gaudin},\ and\ \citenamefont
  {Gambardella}}]{Garello-apl2014}%
  \BibitemOpen
  \bibfield  {author} {\bibinfo {author} {\bibfnamefont {K.}~\bibnamefont
  {Garello}}, \bibinfo {author} {\bibfnamefont {C.~O.}\ \bibnamefont {Avci}},
  \bibinfo {author} {\bibfnamefont {I.~M.}\ \bibnamefont {Miron}}, \bibinfo
  {author} {\bibfnamefont {M.}~\bibnamefont {Baumgartner}}, \bibinfo {author}
  {\bibfnamefont {A.}~\bibnamefont {Ghosh}}, \bibinfo {author} {\bibfnamefont
  {S.}~\bibnamefont {Auffret}}, \bibinfo {author} {\bibfnamefont
  {O.}~\bibnamefont {Boulle}}, \bibinfo {author} {\bibfnamefont
  {G.}~\bibnamefont {Gaudin}},\ and\ \bibinfo {author} {\bibfnamefont
  {P.}~\bibnamefont {Gambardella}},\ }\bibfield  {title} {\bibinfo {title}
  {Ultrafast magnetization switching by spin-orbit torques},\ }\href
  {https://doi.org/10.1063/1.4902443} {\bibfield  {journal} {\bibinfo
  {journal} {Appl. Phys. Lett.}\ }\textbf {\bibinfo {volume} {105}},\ \bibinfo
  {pages} {212402} (\bibinfo {year} {2014})}\BibitemShut {NoStop}%
\bibitem [{\citenamefont {Zhang}\ \emph {et~al.}(2014)\citenamefont {Zhang},
  \citenamefont {Yamanouchi}, \citenamefont {Sato}, \citenamefont {Fukami},
  \citenamefont {Ikeda}, \citenamefont {Matsukura},\ and\ \citenamefont
  {Ohno}}]{Zhang-jap-2014}%
  \BibitemOpen
  \bibfield  {author} {\bibinfo {author} {\bibfnamefont {C.}~\bibnamefont
  {Zhang}}, \bibinfo {author} {\bibfnamefont {M.}~\bibnamefont {Yamanouchi}},
  \bibinfo {author} {\bibfnamefont {H.}~\bibnamefont {Sato}}, \bibinfo {author}
  {\bibfnamefont {S.}~\bibnamefont {Fukami}}, \bibinfo {author} {\bibfnamefont
  {S.}~\bibnamefont {Ikeda}}, \bibinfo {author} {\bibfnamefont
  {F.}~\bibnamefont {Matsukura}},\ and\ \bibinfo {author} {\bibfnamefont
  {H.}~\bibnamefont {Ohno}},\ }\bibfield  {title} {\bibinfo {title}
  {Magnetization reversal induced by in-plane current in {Ta/CoFeB/MgO}
  structures with perpendicular magnetic easy axis},\ }\href
  {https://doi.org/10.1063/1.4863260} {\bibfield  {journal} {\bibinfo
  {journal} {J. Appl. Phys.}\ }\textbf {\bibinfo {volume} {115}},\ \bibinfo
  {pages} {17C714} (\bibinfo {year} {2014})}\BibitemShut {NoStop}%
\bibitem [{\citenamefont {Rojas-Sánchez}\ \emph {et~al.}(2016)\citenamefont
  {Rojas-Sánchez}, \citenamefont {Laczkowski}, \citenamefont {Sampaio},
  \citenamefont {Collin}, \citenamefont {Bouzehouane}, \citenamefont {Reyren},
  \citenamefont {Jaffrès}, \citenamefont {Mougin},\ and\ \citenamefont
  {George}}]{Rs-apl-2016}%
  \BibitemOpen
  \bibfield  {author} {\bibinfo {author} {\bibfnamefont {J.-C.}\ \bibnamefont
  {Rojas-Sánchez}}, \bibinfo {author} {\bibfnamefont {P.}~\bibnamefont
  {Laczkowski}}, \bibinfo {author} {\bibfnamefont {J.}~\bibnamefont {Sampaio}},
  \bibinfo {author} {\bibfnamefont {S.}~\bibnamefont {Collin}}, \bibinfo
  {author} {\bibfnamefont {K.}~\bibnamefont {Bouzehouane}}, \bibinfo {author}
  {\bibfnamefont {N.}~\bibnamefont {Reyren}}, \bibinfo {author} {\bibfnamefont
  {H.}~\bibnamefont {Jaffrès}}, \bibinfo {author} {\bibfnamefont
  {A.}~\bibnamefont {Mougin}},\ and\ \bibinfo {author} {\bibfnamefont {J.-M.}\
  \bibnamefont {George}},\ }\bibfield  {title} {\bibinfo {title} {Perpendicular
  magnetization reversal in $\mathrm{Pt/[Co/Ni]3/Al}$ multilayers via the spin
  hall effect of $\mathrm{Pt}$},\ }\href {https://doi.org/10.1063/1.4942672}
  {\bibfield  {journal} {\bibinfo  {journal} {Appl. Phys. Lett.}\ }\textbf
  {\bibinfo {volume} {108}},\ \bibinfo {pages} {082406} (\bibinfo {year}
  {2016})}\BibitemShut {NoStop}%
\bibitem [{\citenamefont {Taniguchi}\ \emph {et~al.}(2015)\citenamefont
  {Taniguchi}, \citenamefont {Mitani},\ and\ \citenamefont
  {Hayashi}}]{Taniguchi2015}%
  \BibitemOpen
  \bibfield  {author} {\bibinfo {author} {\bibfnamefont {T.}~\bibnamefont
  {Taniguchi}}, \bibinfo {author} {\bibfnamefont {S.}~\bibnamefont {Mitani}},\
  and\ \bibinfo {author} {\bibfnamefont {M.}~\bibnamefont {Hayashi}},\
  }\bibfield  {title} {\bibinfo {title} {Critical current destabilizing
  perpendicular magnetization by the spin hall effect},\ }\href
  {https://doi.org/10.1103/PhysRevB.92.024428} {\bibfield  {journal} {\bibinfo
  {journal} {Phys. Rev. B}\ }\textbf {\bibinfo {volume} {92}},\ \bibinfo
  {pages} {024428} (\bibinfo {year} {2015})}\BibitemShut {NoStop}%
\bibitem [{\citenamefont {Lee}\ \emph {et~al.}(2013)\citenamefont {Lee},
  \citenamefont {Lee}, \citenamefont {Min},\ and\ \citenamefont
  {Lee}}]{Lee2013}%
  \BibitemOpen
  \bibfield  {author} {\bibinfo {author} {\bibfnamefont {K.-S.}\ \bibnamefont
  {Lee}}, \bibinfo {author} {\bibfnamefont {S.-W.}\ \bibnamefont {Lee}},
  \bibinfo {author} {\bibfnamefont {B.-C.}\ \bibnamefont {Min}},\ and\ \bibinfo
  {author} {\bibfnamefont {K.-J.}\ \bibnamefont {Lee}},\ }\bibfield  {title}
  {\bibinfo {title} {Threshold current for switching of a perpendicular
  magnetic layer induced by spin $\mathrm{Hall}$ effect},\ }\href
  {https://doi.org/10.1063/1.4798288} {\bibfield  {journal} {\bibinfo
  {journal} {Appl. Phys. Lett.}\ }\textbf {\bibinfo {volume} {102}},\ \bibinfo
  {pages} {112410} (\bibinfo {year} {2013})}\BibitemShut {NoStop}%
\bibitem [{\citenamefont {Emori}\ \emph {et~al.}(2013)\citenamefont {Emori},
  \citenamefont {Bauer}, \citenamefont {Ahn}, \citenamefont {Martinez},\ and\
  \citenamefont {Beach}}]{Emori2013}%
  \BibitemOpen
  \bibfield  {author} {\bibinfo {author} {\bibfnamefont {S.}~\bibnamefont
  {Emori}}, \bibinfo {author} {\bibfnamefont {U.}~\bibnamefont {Bauer}},
  \bibinfo {author} {\bibfnamefont {S.-M.}\ \bibnamefont {Ahn}}, \bibinfo
  {author} {\bibfnamefont {E.}~\bibnamefont {Martinez}},\ and\ \bibinfo
  {author} {\bibfnamefont {G.~S.~D.}\ \bibnamefont {Beach}},\ }\bibfield
  {title} {\bibinfo {title} {Current-driven dynamics of chiral ferromagnetic
  domain walls},\ }\href {https://doi.org/10.1038/nmat3675} {\bibfield
  {journal} {\bibinfo  {journal} {Nat. Mater.}\ }\textbf {\bibinfo {volume}
  {12}},\ \bibinfo {pages} {611} (\bibinfo {year} {2013})}\BibitemShut
  {NoStop}%
\bibitem [{\citenamefont {Ryu}\ \emph {et~al.}(2013)\citenamefont {Ryu},
  \citenamefont {Thomas}, \citenamefont {Yang},\ and\ \citenamefont
  {Parkin}}]{Ryu2013}%
  \BibitemOpen
  \bibfield  {author} {\bibinfo {author} {\bibfnamefont {K.-S.}\ \bibnamefont
  {Ryu}}, \bibinfo {author} {\bibfnamefont {L.}~\bibnamefont {Thomas}},
  \bibinfo {author} {\bibfnamefont {S.-H.}\ \bibnamefont {Yang}},\ and\
  \bibinfo {author} {\bibfnamefont {S.}~\bibnamefont {Parkin}},\ }\bibfield
  {title} {\bibinfo {title} {Chiral spin torque at magnetic domain walls},\
  }\href {https://doi.org/10.1038/nnano.2013.102} {\bibfield  {journal}
  {\bibinfo  {journal} {Nat. Nanotech.}\ }\textbf {\bibinfo {volume} {8}},\
  \bibinfo {pages} {527} (\bibinfo {year} {2013})}\BibitemShut {NoStop}%
\bibitem [{\citenamefont {Safeer}\ \emph {et~al.}(2015)\citenamefont {Safeer},
  \citenamefont {Jué}, \citenamefont {Lopez}, \citenamefont {Buda-Prejbeanu},
  \citenamefont {Auffret}, \citenamefont {Pizzini}, \citenamefont {Boulle},
  \citenamefont {Miron},\ and\ \citenamefont {Gaudin}}]{Safeer2015}%
  \BibitemOpen
  \bibfield  {author} {\bibinfo {author} {\bibfnamefont {C.~K.}\ \bibnamefont
  {Safeer}}, \bibinfo {author} {\bibfnamefont {E.}~\bibnamefont {Jué}},
  \bibinfo {author} {\bibfnamefont {A.}~\bibnamefont {Lopez}}, \bibinfo
  {author} {\bibfnamefont {L.}~\bibnamefont {Buda-Prejbeanu}}, \bibinfo
  {author} {\bibfnamefont {S.}~\bibnamefont {Auffret}}, \bibinfo {author}
  {\bibfnamefont {S.}~\bibnamefont {Pizzini}}, \bibinfo {author} {\bibfnamefont
  {O.}~\bibnamefont {Boulle}}, \bibinfo {author} {\bibfnamefont {I.~M.}\
  \bibnamefont {Miron}},\ and\ \bibinfo {author} {\bibfnamefont
  {G.}~\bibnamefont {Gaudin}},\ }\bibfield  {title} {\bibinfo {title}
  {Spin-orbit torque magnetization switching controlled by geometry},\ }\href
  {https://doi.org/10.1038/nnano.2015.252} {\bibfield  {journal} {\bibinfo
  {journal} {Nat. Nanotech.}\ }\textbf {\bibinfo {volume} {11}},\ \bibinfo
  {pages} {143} (\bibinfo {year} {2015})}\BibitemShut {NoStop}%
\bibitem [{\citenamefont {Lee}\ \emph {et~al.}(2014)\citenamefont {Lee},
  \citenamefont {Liu}, \citenamefont {Pai}, \citenamefont {Li}, \citenamefont
  {Tseng}, \citenamefont {Gowtham}, \citenamefont {Park}, \citenamefont
  {Ralph},\ and\ \citenamefont {Buhrman}}]{Lee2014}%
  \BibitemOpen
  \bibfield  {author} {\bibinfo {author} {\bibfnamefont {O.~J.}\ \bibnamefont
  {Lee}}, \bibinfo {author} {\bibfnamefont {L.~Q.}\ \bibnamefont {Liu}},
  \bibinfo {author} {\bibfnamefont {C.~F.}\ \bibnamefont {Pai}}, \bibinfo
  {author} {\bibfnamefont {Y.}~\bibnamefont {Li}}, \bibinfo {author}
  {\bibfnamefont {H.~W.}\ \bibnamefont {Tseng}}, \bibinfo {author}
  {\bibfnamefont {P.~G.}\ \bibnamefont {Gowtham}}, \bibinfo {author}
  {\bibfnamefont {J.~P.}\ \bibnamefont {Park}}, \bibinfo {author}
  {\bibfnamefont {D.~C.}\ \bibnamefont {Ralph}},\ and\ \bibinfo {author}
  {\bibfnamefont {R.~A.}\ \bibnamefont {Buhrman}},\ }\bibfield  {title}
  {\bibinfo {title} {Central role of domain wall depinning for perpendicular
  magnetization switching driven by spin torque from the spin hall effect},\
  }\href {https://doi.org/10.1103/PhysRevB.89.024418} {\bibfield  {journal}
  {\bibinfo  {journal} {Phys. Rev. B}\ }\textbf {\bibinfo {volume} {89}},\
  \bibinfo {pages} {024418} (\bibinfo {year} {2014})}\BibitemShut {NoStop}%
\bibitem [{\citenamefont {Zhang}\ \emph {et~al.}(2018)\citenamefont {Zhang},
  \citenamefont {Fukami}, \citenamefont {DuttaGupta}, \citenamefont {Sato},\
  and\ \citenamefont {Ohno}}]{Zhang_2018}%
  \BibitemOpen
  \bibfield  {author} {\bibinfo {author} {\bibfnamefont {C.}~\bibnamefont
  {Zhang}}, \bibinfo {author} {\bibfnamefont {S.}~\bibnamefont {Fukami}},
  \bibinfo {author} {\bibfnamefont {S.}~\bibnamefont {DuttaGupta}}, \bibinfo
  {author} {\bibfnamefont {H.}~\bibnamefont {Sato}},\ and\ \bibinfo {author}
  {\bibfnamefont {H.}~\bibnamefont {Ohno}},\ }\bibfield  {title} {\bibinfo
  {title} {Time and spatial evolution of spin{\textendash}orbit torque-induced
  magnetization switching in {W/CoFeB/MgO} structures with various sizes},\
  }\href {https://doi.org/10.7567/jjap.57.04fn02} {\bibfield  {journal}
  {\bibinfo  {journal} {Jpn. J. Appl. Phys.}\ }\textbf {\bibinfo {volume}
  {57}},\ \bibinfo {pages} {04FN02} (\bibinfo {year} {2018})}\BibitemShut
  {NoStop}%
\bibitem [{\citenamefont {Pham}\ \emph {et~al.}(2018)\citenamefont {Pham},
  \citenamefont {Je}, \citenamefont {Vallobra}, \citenamefont {Fache},
  \citenamefont {Lacour}, \citenamefont {Malinowski}, \citenamefont {Cyrille},
  \citenamefont {Gaudin}, \citenamefont {Boulle}, \citenamefont {Hehn},
  \citenamefont {Rojas-S\'anchez},\ and\ \citenamefont {Mangin}}]{Pham2018}%
  \BibitemOpen
  \bibfield  {author} {\bibinfo {author} {\bibfnamefont {T.~H.}\ \bibnamefont
  {Pham}}, \bibinfo {author} {\bibfnamefont {S.-G.}\ \bibnamefont {Je}},
  \bibinfo {author} {\bibfnamefont {P.}~\bibnamefont {Vallobra}}, \bibinfo
  {author} {\bibfnamefont {T.}~\bibnamefont {Fache}}, \bibinfo {author}
  {\bibfnamefont {D.}~\bibnamefont {Lacour}}, \bibinfo {author} {\bibfnamefont
  {G.}~\bibnamefont {Malinowski}}, \bibinfo {author} {\bibfnamefont {M.~C.}\
  \bibnamefont {Cyrille}}, \bibinfo {author} {\bibfnamefont {G.}~\bibnamefont
  {Gaudin}}, \bibinfo {author} {\bibfnamefont {O.}~\bibnamefont {Boulle}},
  \bibinfo {author} {\bibfnamefont {M.}~\bibnamefont {Hehn}}, \bibinfo {author}
  {\bibfnamefont {J.-C.}\ \bibnamefont {Rojas-S\'anchez}},\ and\ \bibinfo
  {author} {\bibfnamefont {S.}~\bibnamefont {Mangin}},\ }\bibfield  {title}
  {\bibinfo {title} {Thermal contribution to the spin-orbit torque in
  metallic-ferrimagnetic systems},\ }\href
  {https://doi.org/10.1103/PhysRevApplied.9.064032} {\bibfield  {journal}
  {\bibinfo  {journal} {Phys. Rev. Applied}\ }\textbf {\bibinfo {volume} {9}},\
  \bibinfo {pages} {064032} (\bibinfo {year} {2018})}\BibitemShut {NoStop}%
\bibitem [{\citenamefont {Mikuszeit}\ \emph {et~al.}(2015)\citenamefont
  {Mikuszeit}, \citenamefont {Boulle}, \citenamefont {Miron}, \citenamefont
  {Garello}, \citenamefont {Gambardella}, \citenamefont {Gaudin},\ and\
  \citenamefont {Buda-Prejbeanu}}]{Mikuszeit2015}%
  \BibitemOpen
  \bibfield  {author} {\bibinfo {author} {\bibfnamefont {N.}~\bibnamefont
  {Mikuszeit}}, \bibinfo {author} {\bibfnamefont {O.}~\bibnamefont {Boulle}},
  \bibinfo {author} {\bibfnamefont {I.~M.}\ \bibnamefont {Miron}}, \bibinfo
  {author} {\bibfnamefont {K.}~\bibnamefont {Garello}}, \bibinfo {author}
  {\bibfnamefont {P.}~\bibnamefont {Gambardella}}, \bibinfo {author}
  {\bibfnamefont {G.}~\bibnamefont {Gaudin}},\ and\ \bibinfo {author}
  {\bibfnamefont {L.~D.}\ \bibnamefont {Buda-Prejbeanu}},\ }\bibfield  {title}
  {\bibinfo {title} {Spin-orbit torque driven chiral magnetization reversal in
  ultrathin nanostructures},\ }\href
  {https://doi.org/10.1103/PhysRevB.92.144424} {\bibfield  {journal} {\bibinfo
  {journal} {Phys. Rev. B}\ }\textbf {\bibinfo {volume} {92}},\ \bibinfo
  {pages} {144424} (\bibinfo {year} {2015})}\BibitemShut {NoStop}%
\bibitem [{\citenamefont {Laguna-Marco}\ \emph {et~al.}(2015)\citenamefont
  {Laguna-Marco}, \citenamefont {Kayser}, \citenamefont {Alonso}, \citenamefont
  {Mart\'{\i}nez-Lope}, \citenamefont {van Veenendaal}, \citenamefont {Choi},\
  and\ \citenamefont {Haskel}}]{Laguna-marco2015}%
  \BibitemOpen
  \bibfield  {author} {\bibinfo {author} {\bibfnamefont {M.~A.}\ \bibnamefont
  {Laguna-Marco}}, \bibinfo {author} {\bibfnamefont {P.}~\bibnamefont
  {Kayser}}, \bibinfo {author} {\bibfnamefont {J.~A.}\ \bibnamefont {Alonso}},
  \bibinfo {author} {\bibfnamefont {M.~J.}\ \bibnamefont {Mart\'{\i}nez-Lope}},
  \bibinfo {author} {\bibfnamefont {M.}~\bibnamefont {van Veenendaal}},
  \bibinfo {author} {\bibfnamefont {Y.}~\bibnamefont {Choi}},\ and\ \bibinfo
  {author} {\bibfnamefont {D.}~\bibnamefont {Haskel}},\ }\bibfield  {title}
  {\bibinfo {title} {Electronic structure, local magnetism, and spin-orbit
  effects of {Ir(IV)-}, {Ir(V)-}, and {Ir(VI)-} based compounds},\ }\href
  {https://doi.org/10.1103/PhysRevB.91.214433} {\bibfield  {journal} {\bibinfo
  {journal} {Phys. Rev. B}\ }\textbf {\bibinfo {volume} {91}},\ \bibinfo
  {pages} {214433} (\bibinfo {year} {2015})}\BibitemShut {NoStop}%
\bibitem [{\citenamefont {Pizzini}\ \emph {et~al.}(2014)\citenamefont
  {Pizzini}, \citenamefont {Vogel}, \citenamefont {Rohart}, \citenamefont
  {Buda-Prejbeanu}, \citenamefont {Ju\'e}, \citenamefont {Boulle},
  \citenamefont {Miron}, \citenamefont {Safeer}, \citenamefont {Auffret},
  \citenamefont {Gaudin},\ and\ \citenamefont {Thiaville}}]{Pizzini2014}%
  \BibitemOpen
  \bibfield  {author} {\bibinfo {author} {\bibfnamefont {S.}~\bibnamefont
  {Pizzini}}, \bibinfo {author} {\bibfnamefont {J.}~\bibnamefont {Vogel}},
  \bibinfo {author} {\bibfnamefont {S.}~\bibnamefont {Rohart}}, \bibinfo
  {author} {\bibfnamefont {L.~D.}\ \bibnamefont {Buda-Prejbeanu}}, \bibinfo
  {author} {\bibfnamefont {E.}~\bibnamefont {Ju\'e}}, \bibinfo {author}
  {\bibfnamefont {O.}~\bibnamefont {Boulle}}, \bibinfo {author} {\bibfnamefont
  {I.~M.}\ \bibnamefont {Miron}}, \bibinfo {author} {\bibfnamefont {C.~K.}\
  \bibnamefont {Safeer}}, \bibinfo {author} {\bibfnamefont {S.}~\bibnamefont
  {Auffret}}, \bibinfo {author} {\bibfnamefont {G.}~\bibnamefont {Gaudin}},\
  and\ \bibinfo {author} {\bibfnamefont {A.}~\bibnamefont {Thiaville}},\
  }\bibfield  {title} {\bibinfo {title} {Chirality-induced asymmetric magnetic
  nucleation in $\mathrm{Pt}/\mathrm{Co}/\mathrm{AlO}$\textsubscript{x}
  ultrathin microstructures},\ }\href
  {https://doi.org/10.1103/PhysRevLett.113.047203} {\bibfield  {journal}
  {\bibinfo  {journal} {Phys. Rev. Lett.}\ }\textbf {\bibinfo {volume} {113}},\
  \bibinfo {pages} {047203} (\bibinfo {year} {2014})}\BibitemShut {NoStop}%
\bibitem [{\citenamefont {Grimaldi}\ \emph {et~al.}(2020)\citenamefont
  {Grimaldi}, \citenamefont {Krizakova}, \citenamefont {Sala}, \citenamefont
  {Yasin}, \citenamefont {Couet}, \citenamefont {Kar}, \citenamefont
  {Garello},\ and\ \citenamefont {Gambardella}}]{Grimaldi2020}%
  \BibitemOpen
  \bibfield  {author} {\bibinfo {author} {\bibfnamefont {E.}~\bibnamefont
  {Grimaldi}}, \bibinfo {author} {\bibfnamefont {V.}~\bibnamefont {Krizakova}},
  \bibinfo {author} {\bibfnamefont {G.}~\bibnamefont {Sala}}, \bibinfo {author}
  {\bibfnamefont {F.}~\bibnamefont {Yasin}}, \bibinfo {author} {\bibfnamefont
  {S.}~\bibnamefont {Couet}}, \bibinfo {author} {\bibfnamefont
  {G.}~\bibnamefont {Kar}}, \bibinfo {author} {\bibfnamefont {K.}~\bibnamefont
  {Garello}},\ and\ \bibinfo {author} {\bibfnamefont {P.}~\bibnamefont
  {Gambardella}},\ }\bibfield  {title} {\bibinfo {title} {Single-shot dynamics
  of spin–orbit torque and spin transfer torque switching in three-terminal
  magnetic tunnel junctions},\ }\href
  {https://doi.org/10.1038/s41565-019-0607-7} {\bibfield  {journal} {\bibinfo
  {journal} {Nat. Nanotechnol.}\ } (\bibinfo {year} {2020})}\BibitemShut
  {NoStop}%
\bibitem [{\citenamefont {Koch}\ \emph {et~al.}(2004)\citenamefont {Koch},
  \citenamefont {Katine},\ and\ \citenamefont {Sun}}]{Koch2004}%
  \BibitemOpen
  \bibfield  {author} {\bibinfo {author} {\bibfnamefont {R.~H.}\ \bibnamefont
  {Koch}}, \bibinfo {author} {\bibfnamefont {J.~A.}\ \bibnamefont {Katine}},\
  and\ \bibinfo {author} {\bibfnamefont {J.~Z.}\ \bibnamefont {Sun}},\
  }\bibfield  {title} {\bibinfo {title} {Time-resolved reversal of
  spin-transfer switching in a nanomagnet},\ }\href
  {https://doi.org/10.1103/PhysRevLett.92.088302} {\bibfield  {journal}
  {\bibinfo  {journal} {Phys. Rev. Lett.}\ }\textbf {\bibinfo {volume} {92}},\
  \bibinfo {pages} {088302} (\bibinfo {year} {2004})}\BibitemShut {NoStop}%
\bibitem [{\citenamefont {Seki}\ \emph {et~al.}(2008)\citenamefont {Seki},
  \citenamefont {Mitani},\ and\ \citenamefont {Takanashi}}]{Seki2008}%
  \BibitemOpen
  \bibfield  {author} {\bibinfo {author} {\bibfnamefont {T.}~\bibnamefont
  {Seki}}, \bibinfo {author} {\bibfnamefont {S.}~\bibnamefont {Mitani}},\ and\
  \bibinfo {author} {\bibfnamefont {K.}~\bibnamefont {Takanashi}},\ }\bibfield
  {title} {\bibinfo {title} {Nucleation-type magnetization reversal by
  spin-polarized current in perpendicularly magnetized fept layers},\ }\href
  {https://doi.org/10.1103/PhysRevB.77.214414} {\bibfield  {journal} {\bibinfo
  {journal} {Phys. Rev. B}\ }\textbf {\bibinfo {volume} {77}},\ \bibinfo
  {pages} {214414} (\bibinfo {year} {2008})}\BibitemShut {NoStop}%
  \bibitem [{\citenamefont {Ghosh}\ \emph {et~al.}(2017)\citenamefont
  {Ghosh}, \citenamefont {Garello}, \citenamefont {Avci}, \citenamefont
  {Gabureac}, \and\citenamefont {Gambardella}}]{Ghosh2017}%
  \BibitemOpen
  \bibfield  {author} {\bibinfo {author} {\bibfnamefont {A.}~\bibnamefont
  {Ghosh}}, \bibinfo {author} {\bibfnamefont {K.}\ \bibnamefont {Garello}},
  \bibinfo {author} {\bibfnamefont {C.~O.}\ \bibnamefont {Avci}}, \bibinfo
  {author} {\bibfnamefont {M.}~\bibnamefont {Gabureac}}, \and\bibinfo {author} {\bibfnamefont
  {P.}~\bibnamefont {Gambardella}},\ }\bibfield  {title} {\bibinfo {title}
  { Interface-Enhanced Spin-Orbit Torques and Current-Induced Magnetization Switching of {Pd}/{Co}/{A}l{O}\textsubscript{x} Layers },\ }\href { https://doi.org/10.1103/PhysRevApplied.7.014004}
  {\bibfield  {journal} {\bibinfo  {journal} { Phys. Rev. Appl.}\ }\textbf
  {\bibinfo {volume} {7}},\ \bibinfo {pages} {014004} (\bibinfo {year}
  {2017})}\BibitemShut {NoStop}%
\end{thebibliography}

%

\end{document}